\documentclass[journal]{vgtc}                     

\onlineid{0}

\vgtccategory{Research}

\vgtcpapertype{please specify}

\title{F-Hash: Feature-Based Hash Design for Time-Varying Volume Visualization via Multi-Resolution Tesseract Encoding}

\author{%
  \authororcid{Jianxin Sun}{0000-0002-9627-9397},
  \authororcid{David Lenz}{0000-0002-2587-2783}, 
  \authororcid{Hongfeng Yu}{0000-0002-0596-8227}, and
  \authororcid{Tom Peterka}{0000-0002-0525-3205}
}

\authorfooter{
  \item
  	Jianxin Sun and Hongfeng Yu are with the University of Nebraska-Lincoln.
  	E-mail: jianxin.sun@huskers.unl.edu, yu@cse.unl.edu.
  \item
  	David Lenz and Tom Peterka are with Argonne National Laboratory.
  	E-mail: dlenz@anl.gov, tpeterka@mcs.anl.gov.
}

\abstract{%
Interactive time-varying volume visualization is challenging due to its complex spatiotemporal features and sheer size of the dataset. Recent works transform the original discrete time-varying volumetric data into continuous Implicit Neural Representations (INR) to address the issues of compression, rendering, and super-resolution in both spatial and temporal domains. However, training the INR takes a long time to converge, especially when handling large-scale time-varying volumetric datasets. In this work, we proposed F-Hash, a novel feature-based multi-resolution Tesseract encoding architecture to greatly enhance the convergence speed compared with existing input encoding methods for modeling time-varying volumetric data. The proposed design incorporates multi-level collision-free hash functions that map dynamic 4D multi-resolution embedding grids without bucket waste, achieving high encoding capacity with compact encoding parameters. Our encoding method is agnostic to time-varying feature detection methods, making it a unified encoding solution for feature tracking and evolution visualization. Experiments show the F-Hash achieves state-of-the-art convergence speed in training various time-varying volumetric datasets for diverse features. We also proposed an adaptive ray marching algorithm to optimize the sample streaming for faster rendering of the time-varying neural representation.
}


\teaser{
  \centering
  \includegraphics[trim=2 2 2 2,clip,width=0.98\linewidth, alt={A view of a city with buildings peeking out of the clouds.}]{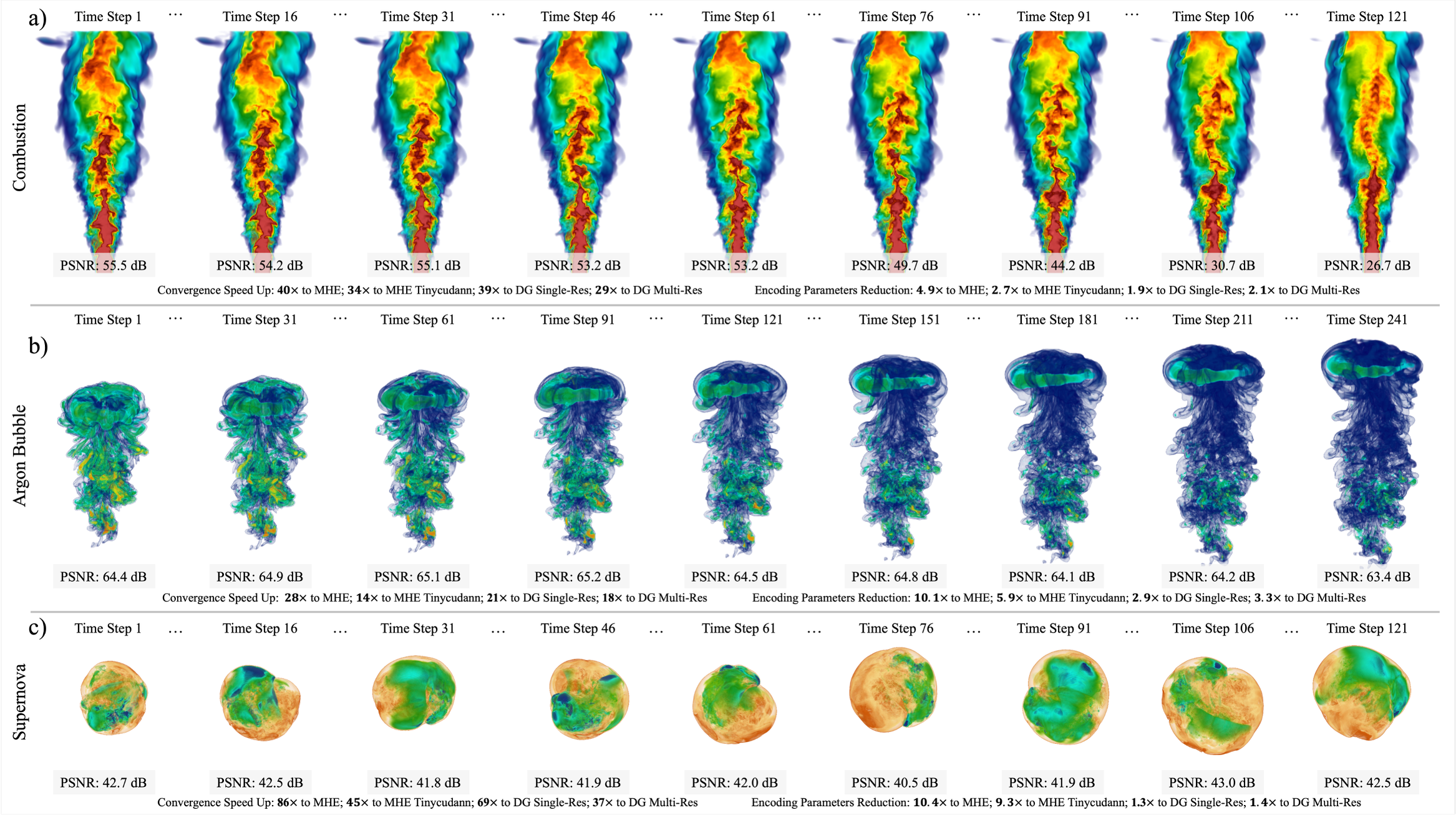}
  \caption{Time-varying volume visualization via proposed F-Hash multi-resolution tesseract encoding on Combustion(a), Argon Bubble (b), and Supernova (c) datasets. F-Hash achieves high reconstruction accuracy with shorter convergence time by order of magnitude ($10\times$--$100\times$) and fewer encoding parameters compared to existing input encoding approaches.}
  \label{fig:teaser}
}

\graphicspath{{figs/}{figures/}{pictures/}{images/}{./}} 

\usepackage{tabu}                      
\usepackage{booktabs}                  
\usepackage{lipsum}                    
\usepackage{mwe}                       

\usepackage{mathptmx}                  

\usepackage{adjustbox}
\usepackage{amsmath}
\usepackage{algorithm}
\usepackage{algpseudocode}
\usepackage{multirow}
\begin{document}




\firstsection{Introduction}

\maketitle
Numerous scientific disciplines involve intrinsically time-dependent phenomena. A substantial proportion of data produced by scientific simulations consists of time-varying volumetric data spanning fields such as computational fluid dynamics, combustion modeling, meteorological simulations, cosmological studies, and biomolecular simulations. However, the high resolution in both spatial and temporal domains results in extremely large-scale time-varying volumetric data, making interactive visualization significantly challenging.

Recent works leverage the neural network to model the discrete time-varying volumetric data into a continuous implicit neural representation (INR), which can be accelerated on modern GPU, achieving state-of-the-art performance in compression, rendering, and super-resolution tasks. Multilayer perception (MLP)-based INR, as a universal approximator, provides an efficient way of representing high-dimensional data, enabling efficient compression~\cite{10371224, tang2023ecnr} of large-scale time-varying data for I/O intensive operations. The continuous nature of INR provides infinite super-resolution~\cite{TANG2024103874, 9852325, 9229162, HAN2022168, 9552857} on both spatial and temporal domains. INR can also be trained for specific visualization tasks as generative models for fast rendering through novel view synthesis~\cite{yao2025visnerf, sun2025make, 9903564} or efficient out-of-core data management~\cite{10549835}.

Although INR gives outstanding performance in various visualization tasks, its core limitation lies in slow convergence, often requiring hours or even days of training for complex datasets. This makes it even more infeasible to directly model a large-scale dataset like time-varying volume. Numerous works have been proposed to address this issue. SIREN (Sinusoidal Representation Network)~\cite{NEURIPS2020_53c04118, Mehta_2021_ICCV} uses periodic activation functions instead of traditional activations like ReLU or tanh to improve the convergence time. Meta-learning~\cite{yang2025meta, 9428530} pretrains a warm-start INR with optimal initial weights trained from a sparse but representative subset of the whole training samples. Input encoding methods, which encode the input of INR into a higher-dimensional space, show advantages in both efficacy and simplicity. Frequency encoding~\cite{NEURIPS2020_55053683, Barron_2021_ICCV} was first proposed as a positional encoding method utilizing Fourier features mapping to better capture the high-frequency details in constructing neural radiance fields (NeRF)~\cite{10.1145/3503250}. Parametric encoding~\cite{Sun_2022_CVPR, Fridovich-Keil_2022_CVPR} was later introduced, which arranges additional trainable parameters (beyond weights and biases) in an auxiliary grid or tree data structure. Recent Multi-resolution Hash Encoding (MHE)~\cite{10.1145/3528223.3530127} gives state-of-the-art convergence speed by combining the embedding grid, multi-resolution, and a hash function to enable ``instant'' reconstruction of various types of neural graphic primitives including steady scientific volumetric data~\cite{10175377}.

However, MHE suffers from three main problems. 1) Since hash collisions were not handled, different entries in the embedding grid were forced to use the same embedding vector, adversely affecting learning precision and limiting the fitting capacity of the INR. Detailed supporting quantitative results can be found in the Appendix. 2) The embedding grid of each resolution level has the same size, leading to unused buckets in lower resolutions, causing inefficient use of the encoding parameters. This also produces an unnecessarily large INR with redundant weights. Although the first problem can be mitigated by increasing the hash table size, it will further intensify the second bucket waste problem. 3) The architecture of MHE can only process 3D spatial data. These limitations prevent MHE from effectively encoding and representing large-scale time-varying data at high resolutions. In this work, we proposed F-Hash, a novel feature-based hash design for multi-resolution input encoding, which greatly enhances the convergence speed compared to existing parametric encoding methods for modeling time-varying volumetric data. Our design consists of three components: A feature-based coreset selection method for Meta-learning, adaptive Tesseract embedding grids for spatiotemporal interpolation, and collision-free bijective hash functions with 100\% bucket utilization. F-Hash is dynamically adjusted according to the feature of interest for the best convergence speed and encoding parameter reduction. F-Hash is also compatible with various visualization tasks, from feature tracking to evolution visualization. Our work is the first attempt to apply multi-resolution hash encoding to time-varying volumetric data with state-of-the-art convergence speed. While the convergence speed remains insufficient for online training demands and the compression ratio is suboptimal compared to recently proposed volume compressors~\cite{10371224, lu2021compressive}, our method achieves a substantial reduction in INR training time for large-scale time-varying volumetric data. We also introduce a rendering framework capable of directly visualizing time-varying neural representations encoded via F-Hash and the Adaptive Ray Marching (ARM) algorithm to dynamically optimize sample streaming to the GPU, enabling interactive visualization. The main contributions of this work include: 

\setlist{nolistsep}
\begin{itemize}[leftmargin=*]
  \item A novel feature-based multi-resolution hash encoding, F-Hash, with state-of-the-art convergence speed for training neural representation from time-varying volumetric data.
  \item A minimal perfect hash function (MPHF) to efficiently map multi-resolution Tesseract embedding grid to hash table without collision and bucket waste, enabling effective while compact input encoding.
  \item A feature-based coreset selection method to select only informative samples as training data from the time-varying volumetric data.
  \item An adaptive ray marching algorithm to dynamically stream samples for interactive visualization of the time-varying neural representation.
\end{itemize}

\begin{figure*}[t]
    \centering
        \includegraphics[trim=0 0 0 0,clip,width=\textwidth]{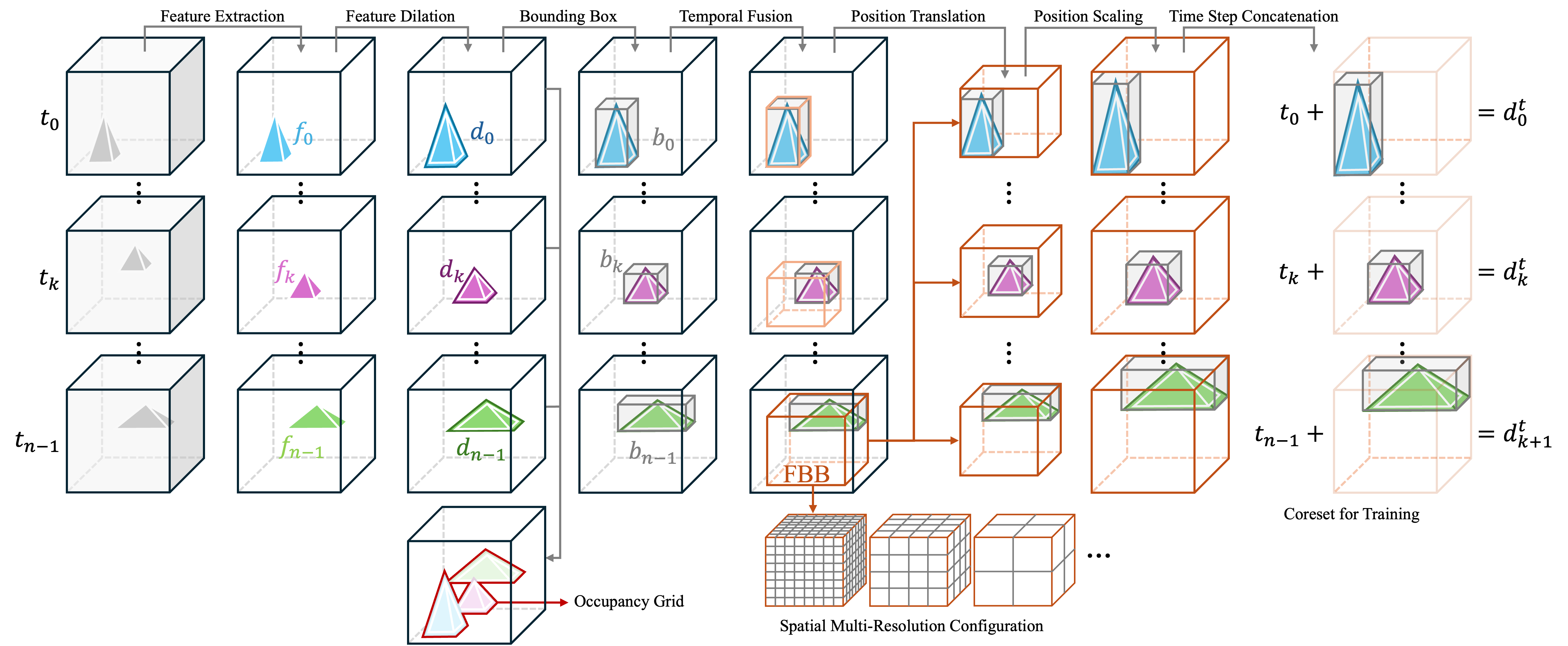}
    \vspace{-3mm}
    \caption{Coreset selection process to retrieve the coreset for training, occupancy grid, and the FBB.}
    \label{fig:coreset}
\end{figure*}

\section{Related Work} 
\label{related work}

\subsection{Time-Varying Volume Visualization}
For the past two decades, time-varying volume visualization has remained a vibrant research domain, driven by its critical role in revealing temporal dynamics and evolving patterns~\cite{bai2020time, 1182960}. Time-varying volume visualization can be systematically classified into three aspects: feature tracking, evolution visualization, and rendering. 1) Feature tracking aims to acquire the temporal evolution of features, which is essential for understanding various phenomena. Widanagamaachchi et al. introduce an interactive system~\cite{8031584} to track pressure perturbations and a merge tree-based method~\cite{7348066} to track extinction holes in turbulent combustion simulations. Kumpf et al.~\cite{8440839} presents a framework that tracks and evaluates ensemble forecast sensitivities using clustering and optical flow. Lukasczyk et al.~\cite{lukasczyk2017} present a topological framework to detect, track, and analyze viscous fingers in fluid mixing simulations. 2) Evolution visualization visually represents the complex results of feature tracking across numerous events and time steps, requiring clear and aesthetically appealing designs. Dutta et al.~\cite{7192664} propose a distribution-based method using incremental Gaussian Mixture Models (GMMs) to track vaguely defined evolving features. Saikia et al.~\cite{https://doi.org/10.1111/cgf.13163} present a global graph-based method to track and compare features in time-dependent scalar fields. 3) Rendering large time-varying volume data requires efficient encoding techniques for acceleration and coherent transfer functions for clear temporal feature observation. Ljung et al.~\cite{https://doi.org/10.1111/cgf.12934} review transfer function techniques for direct volume rendering from the perspectives of data characteristics and user interactions. Wang et al.~\cite{8727480} introduce a ray-based proxy method using histograms and depth information to reconstruct time-varying volume data. Zhou et al.~\cite{https://doi.org/10.1111/cgf.13399} propose an information-theoretic method to automatically select representative time steps from large-scale time-varying volume datasets. For scientific time-varying volumetric datasets, the primary focus is to design efficient and effective feature-centric visualization, which motivates us to prioritize the encoding of the spatial and temporal patterns toward specific features rather than processing the entire dataset.

\subsection{Implicit Neural Representation}
Implicit Neural Representation (INR) is a technique that encodes data, such as images, 3D shapes, or continuous signals, using neural networks. For large-scale time-varying volume visualization, recent studies have introduced INR-based compression methods \cite{lu2021compressive, tang2020deep, 10371224, tang2023ecnr, https://doi.org/10.1111/cgf.14955} to reduce model complexity and mitigate I/O bottlenecks. INR-powered super-resolution frameworks have been developed to enable efficient rendering of time-varying volumetric data at high fidelity. Han et al.\cite{9852325} proposed SSR-TVD, a spatial super-resolution for time-varying data analysis and visualization, STNet, a generative framework for synthesizing spatiotemporal super-resolution volumes, and CoordNet, a coordinate-based deep learning framework that improves generalization in time-varying volumetric data visualization. Tang et al.~\cite{TANG2024103874} proposed STSR-INR, a super-resolution for multivariate time-varying volumetric data on both spatial and temporal domains.
Yariv et al.~\cite{NEURIPS2021_25e2a30f} enhance neural volume rendering by implicitly representing geometry, improving both shape reconstruction and volume density modeling. For accelerating rendering performance, Weiss et al.~\cite{weiss2022fast} propose fV-SRN, an advanced adaptation of Scene Representation Networks (SRN)~\cite{NEURIPS2019_b5dc4e5d} that dramatically speeds up volumetric reconstruction. However, the training time for INR remains significantly long when processing complex, large-scale, time-varying datasets. Therefore, developing an optimized input encoding approach becomes essential to reduce convergence time when training an INR for time-varying volumetric data.

\subsection{Network Input Encoding}
Network input or positional encoding is an emerging technique used to map raw input coordinates into a higher-dimensional space before feeding them into an INR. This helps the network better represent high-frequency details and converge faster. Input encoding can be classified into two main categories: frequency encoding and parametric encoding. 1) Frequency encoding was proposed as a positional encoding method utilizing Fourier features to map coordinates into high-frequency sine/cosine waves. Tancik et al.~\cite{NEURIPS2020_55053683} demonstrate that applying Fourier feature mappings to input coordinates enables MLPs to learn high-frequency functions effectively in low-dimensional tasks. Barron et al.\cite{Barron_2021_ICCV} replaces NeRF’s ray sampling with cone tracing and integrated positional encoding to reduce aliasing and improve multiscale rendering. Mildenhall et al.\cite{10.1145/3503250} synthesizes photorealistic novel views by optimizing a 5D neural radiance field via differentiable volume rendering from input images. 2) Parametric encoding~\cite{Sun_2022_CVPR, Fridovich-Keil_2022_CVPR} introduces supplementary trainable parameters organized in auxiliary grid or tree structures, extending beyond traditional network weights and biases. Sun et al.~\cite{Sun_2022_CVPR} optimize voxel grids directly for radiance fields, achieving NeRF-level quality with short training and sharp surface modeling. Fridovich-Keil et al. propose Plenoxels~\cite{Fridovich-Keil_2022_CVPR} that replaces neural networks with a sparse voxel grid for radiance fields, achieving NeRF-quality rendering $100\times$ faster. The recent Multi-resolution Hash Encoding (MHE)~\cite{10.1145/3528223.3530127} achieves state-of-the-art convergence speed by integrating an embedding grid architecture with multi-resolution and hash-based indexing, enabling near-instant reconstruction of diverse neural graphics primitives. Wu et al.\cite{10175377} introduce MHE to model static scientific volumetric data and provide an interactive rendering pipeline to render visualization directly from the INR. While effective for static volumes, MHE's architectural limitations hinder direct application to time-varying volumetric data.
\section{Methods}
The design of F-Hash consists of three components: 1) A feature-based coreset selection method to select feature-related training samples for Meta-learning. 2) A group of adaptive Tesseract grids with various levels of resolution to hold the mapped higher-dimensional embedding vectors. 3) Collision-free bijective hash function for each resolution level with 100\% bucket utilization. The Tesseract grids and hash functions are adaptive to the feature of interest for the best convergence speed up and encoding parameter reduction. The feature-centric design of our input encoding method makes it efficient and compatible with various time-varying features, enabling effective visualization from feature tracking to evolution visualization.

\subsection{Feature-Based Coreset Selection}\label{sec:coreset}
We adapt the idea of Meta-learning and provide a coreset selection method to collect only the feature-related subset of the entire time-varying volumetric data as the training data. The motivation is derived from two observations: 1) The analysis of time-varying volume data from various scientific domains primarily focuses on specific features of interest, such as dark matter halos in computational cosmology or flames in combustion science, as they hold key research significance. For the testing datasets we evaluated, only less than $20\%$ of the entire volumetric region is related to the informative features of interest. Thus, this inspires us to selectively model only the relevant regions rather than the entire dataset for improved efficiency. 2) Due to the large number of training samples, universal INR modeling directly from the raw time-varying data results in an inevitably large batch size for a feasible training time. A large batch size results in a lower-variance gradient, making optimization smoother (less noisy updates) and helping in avoiding bad local minima since the loss landscape is averaged over more samples. However, modern GPUs are better optimized for smaller, parallel workloads than for a single computationally heavy one. Smaller batches fit entirely in the GPU cache, allowing frequent memory reuse and faster computing. Beyond a certain batch size, scheduling overhead increases, and some cores may remain underutilized due to memory bandwidth limits. This issue becomes more obvious when training an INR with a large number of training samples, like the time-varying volumetric data. Based on the above observations, to tackle this challenge and enhance convergence speed, we must minimize the size of the training data, thereby improving the convergence speed per epoch. The x, y, and z dimensions of each time step are normalized to range $[-1, 1]$. The key frame detection methods~\cite{10.1145/3571735, 9203859, 8601376} are used to detect key frames that display critical evolutionary events in chronological order. \cref{fig:coreset} summarizes the key steps of processing the raw time-varying volumetric data with $n$ key frames to select a representative coreset for training.

\textbf{Feature Extraction:}
First, the time-varying features for each key frame, from time step $t_0$ to $t_{n-1}$, are extracted from the respective normalized input volume by specific feature extraction methods. We considered three common types of time-varying features\cite{bai2020time}: Spatial-first features, Temporal-first features, and 4D features. For time step $t_k$, the extracted feature, $f_k$, is in the form of a bag of vertices where each vertex is the corresponding sample in the volume.

\begin{figure}[t]
    \centering
    \begin{subfigure}[b]{0.32\linewidth}
        \centering
        \includegraphics[trim=0 0 0 0,clip,width=\linewidth]{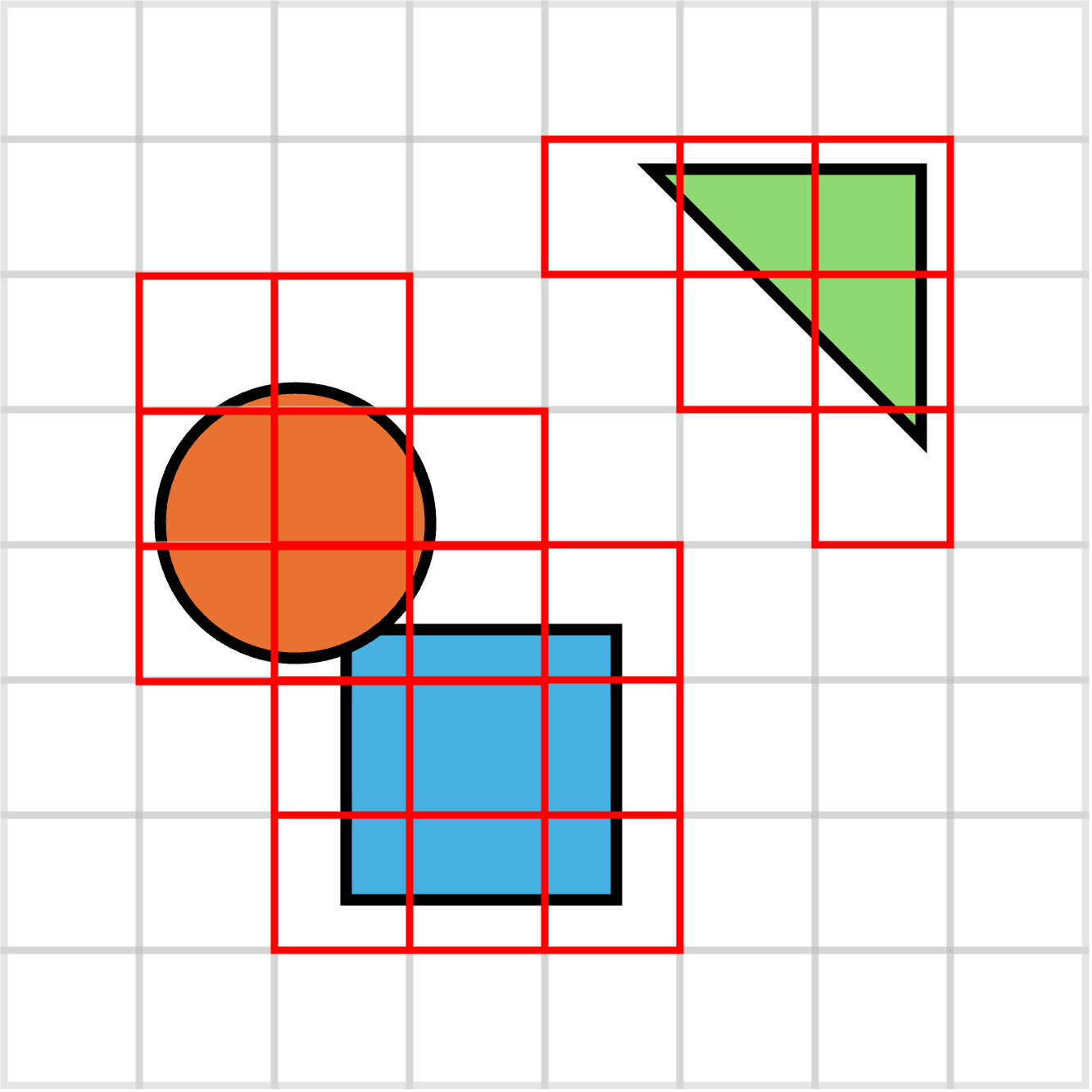}
        \caption{Interval Features}
        \label{fig:depth_iml_num_para}
    \end{subfigure}
    \hfill
    \begin{subfigure}[b]{0.32\linewidth}
        \centering 
        \includegraphics[trim=0 0 0 0,clip,width=\linewidth]{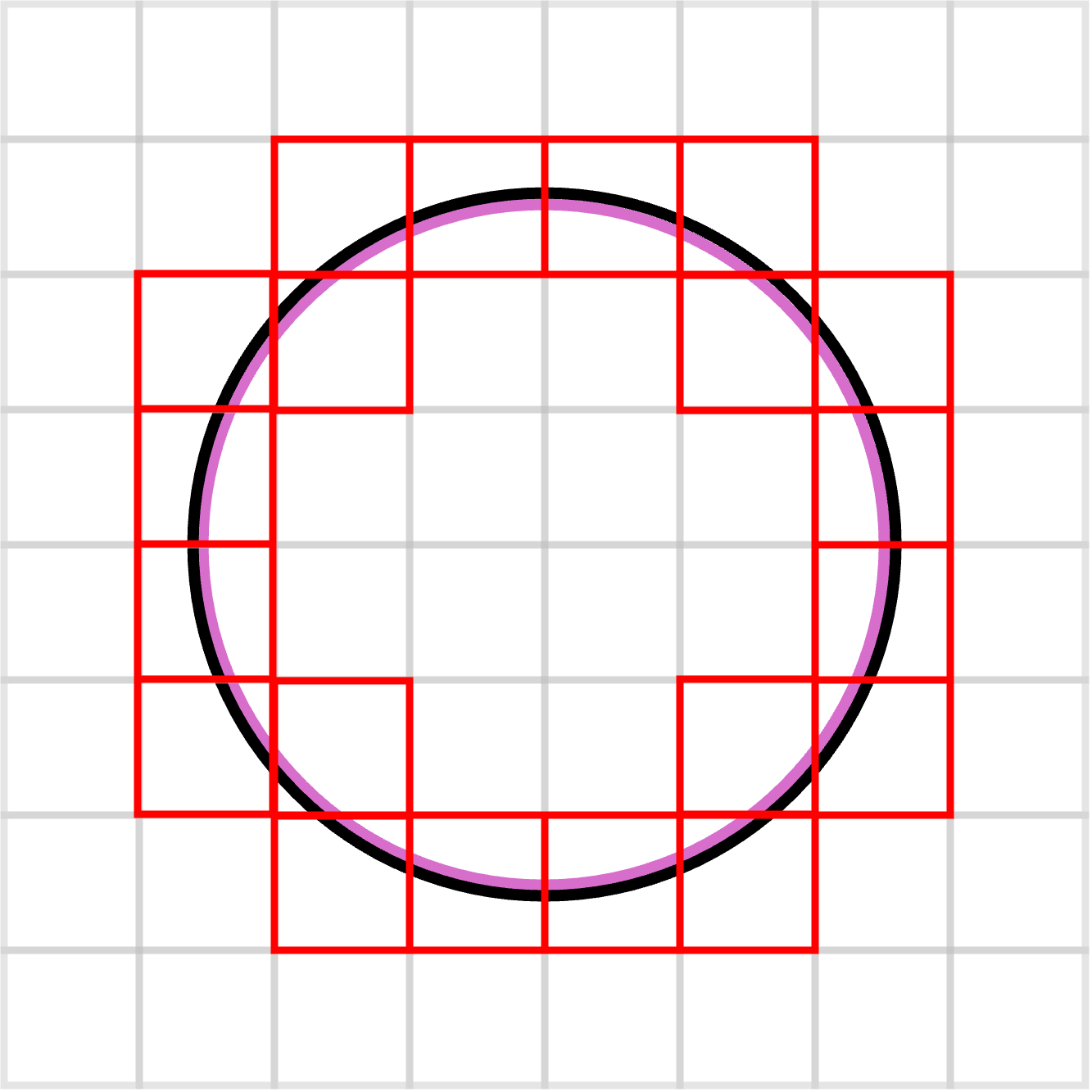}
        \caption{Isosurface Feature}
        \label{fig:depth_iml_psnr_sr}
    \end{subfigure}
    \hfill
    \begin{subfigure}[b]{0.32\linewidth}
        \centering 
        \includegraphics[trim=0 0 0 0,clip,width=\linewidth]{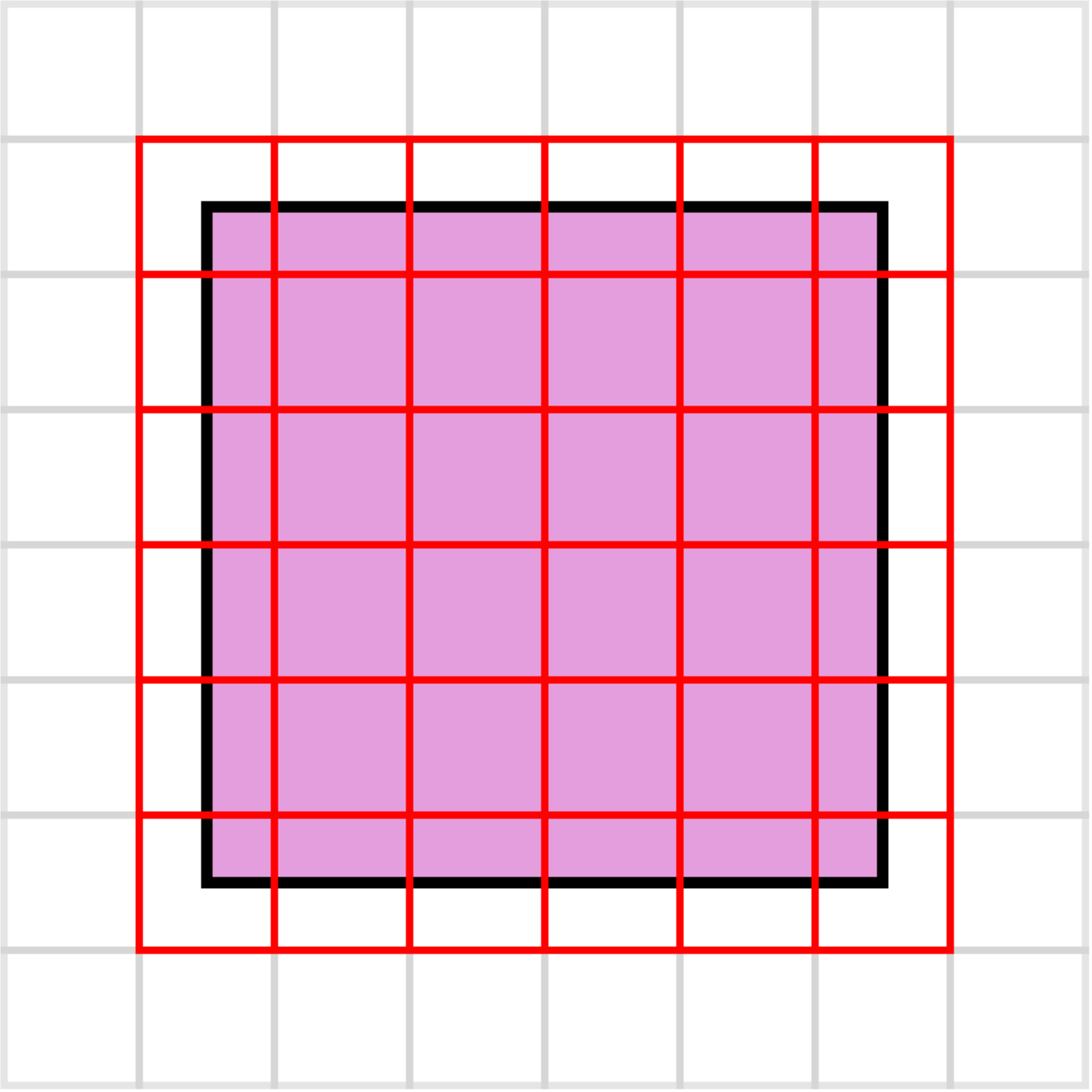}
        \caption{Segmentation Feature}
        \label{fig:depth_iml_psnr_fr}
    \end{subfigure}
    \caption{Occupancy grid (red empty squares) extraction for three types of Spatial-first time-varying features, interval, isosurface, and segmentation. The black boundary around each feature is the region of dilation.}
    \label{fig:occupancy}    
\end{figure}

\textbf{Feature Dilation:}
During the visualization stage, ray casting requires value and gradient queries of samples on the rays shooting from pixels of the visualization image. The sample on the off-grid location needs to be interpolated from the eight corners of the cell containing the sample. Therefore, it is necessary to include the neighboring vertices of $f_k$ as a comprehensive feature region $d_k$ to ensure an accurate interpolation on the boundary. The feature dilation performs an efficient search for neighboring vertices through a hash-based dictionary data structure. An occupancy grid, which will be used during rendering for acceleration, also needs to be extracted in this step. The detailed usage of the occupancy grid is discussed in \cref{sec:rendering}. \cref{fig:occupancy} demonstrates how the occupancy grid is derived for various time-varying features. The occupancy grid has the same spatial dimension as the individual volume. The value of its voxel is set to 1 if $d_k$ intersects with the voxel or 0 otherwise. The occupancy grid is stored as bits to minimize its memory footprint. Morton linearization is used to map the voxel to a 1D vector of bits for spatial locality. When rendering multiple features encoded by F-Hash, multiple occupancy grids can be merged to optimize the overall rendering performance.

\textbf{Bounding Box and Temporal Fusion:}
The range of $d_k$ in x, y, and z dimensions forms a bounding box $b_k$. Temporal fusion finds the minimal superset (coreset) of all the bounding boxes across time. We name this bounding box as Feature Bounding Box (FBB). FBB is critical in two aspects: 1) FBB contains the coreset of training samples for F-Hash. 2) The spatial multi-resolution configuration is derived from FBB. Detailed configuration policy is discussed in \cref{sec:multi_res_policy}.

\textbf{Position Translation and Scaling:}
Position translation shifts the sample coordinate in $b_k$ from the original coordinate system to the coordinate system of FBB. The new x, y, and z coordinates of each vertex in $b_k$ are calculated as:
\begin{equation}
b_k^{FBB}.x/y/z = b_k.x/y/z - Origin^{FBB}.x/y/z
\end{equation}
where $Origin^{FBB}$ is the centroid coordinate of FBB in the original coordinate system. Then we do position scaling to scale the coordinates of samples within FBB to the range of $[-1, 1]$:
\begin{equation}
b_k^{FBB}.x/y/z = b_k^{FBB}.x/y/z \times (\frac{2}{Size^{FBB}.x/y/z})
\end{equation}
where $Size^{FBB}.x/y/z$ is the size of FBB on the x, y, or z dimension.

\textbf{Time Step Concatenation:}
We concatenate the time step to corresponding samples in $d_k$ to construct 4D training samples $d_k^t$, which is composed of input $([t, x, y, z])$ and label ($v$) pair. The final coreset for training is constructed by:
\begin{equation}
Coreset = \bigcup_{k=0}^{n-1} d_k^t
\end{equation}

It is worth noticing that not all the samples in the FBB are in the coreset for training. A more aggressive method of defining the coreset is to only consider the regions where features are present. This results in irregularly shaped multi-resolution embedding grids instead of rectangular prism grids. Although this method generates an even smaller coreset and reduced model size through fewer encoding parameters, it introduces a significant computational bottleneck in looking up the hash table due to the arbitrary shape of the embedding grid and its large number of entries. We experiment with implementations using both a hash-based dictionary data structure and a parallel library of PyTorch tensor Argmax, and the results show a significant degradation in training time due to the hash lookup overhead.

\subsection{Multi-Resolution Tesseract Encoding Design}\label{sec:multi_res_policy}
In this section, we detail the design of the proposed multi-resolution Tesseract encoding, which has a grid-based parametric encoding structure. While most input encoding research papers use ``feature'' to name the high-dimensional vectors within the encoding grid, we instead use the term ``embedding'' to prevent confusion with the time-varying features. 

\subsubsection{Spatial Multi-Resolution Configuration:}
We derive the spatial-only multi-resolution configuration for x, y, and z dimensions from the FBB because it provides a compact spatial distribution and a minimal upper bound of the resolution needed. We use the native resolution of FBB ($S_x, S_y, S_z$) as the level with the highest resolution (level 1) for all dimensions.
\begin{equation}
Res_x^1 = S_x,\;\;\;Res_y^1 = S_y,\;\;\;Res_z^1 = S_z
\end{equation}
The native number of levels for each dimension is determined by:
\begin{equation}
L_x = \lceil log_f S_x \rceil,\;\;\;L_y = \lceil log_f S_y \rceil,\;\;\;L_z = \lceil log_f S_z \rceil
\end{equation}
The shared number of resolutions $L_s$ is the maximum of the three:
\begin{equation}
L_s = max(L_x, \;\; L_y, \;\; L_z)
\end{equation}

\begin{figure}[t]
    \centering 
    \includegraphics[trim=0 0 0 0,clip,width=\linewidth]{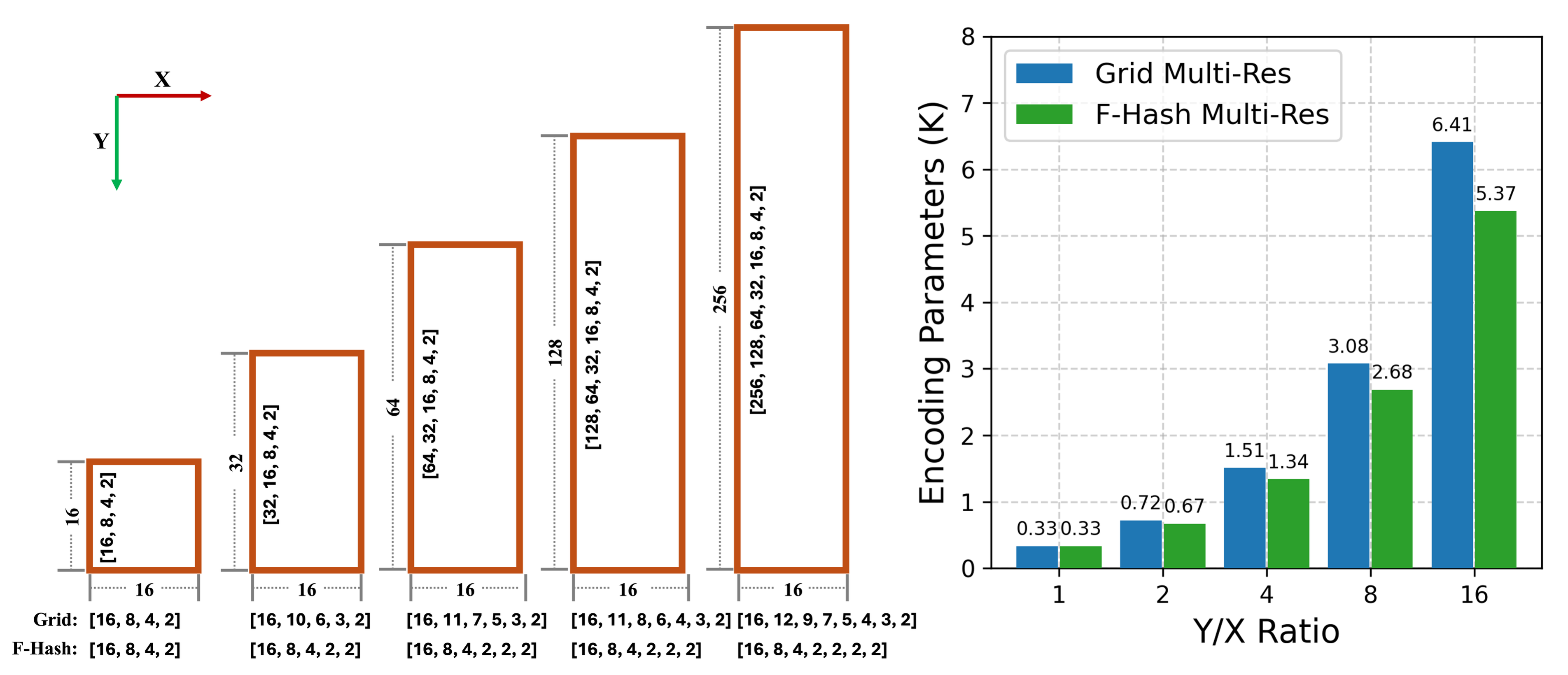}
    \caption{Multi-resolution configuration of Grid-based methods and F-Hash on grids with high aspect ratio.}
    \label{fig:yx_ratio}
\end{figure}

The resolution with a level $l$ greater than 1 is iteratively determined by the resolution of the previous level:
\begin{equation}
    Res^{l}= 
\begin{cases}
    \lceil \frac{Res^{l-1}}{f} \rceil & \text{if } Res^{l-1} > f\\
    Tail(l)                           & \text{otherwise}
\end{cases}
,\;\;\;\;l \in \{2, 3, \dots, L_s-1\}
\end{equation}
\begin{equation}
    Tail(l)= 
\begin{cases}
    2    & \text{if } l \leq L_s \\
    stop & \text{otherwise}
\end{cases}
\end{equation}
where $f$ is the fold parameter with an integer value starting from 2. A higher fold gives a larger resolution difference between neighboring levels, resulting in a smaller number of resolution levels. Since the size of FBB on x, y, or z dimensions can be different, we pad the smaller dimension with the minimal resolution (2) at the tail to save encoding parameters while still aligning the number of resolution levels as $l_s$. Compared with existing MHE-based methods using evenly partitioned resolutions across levels, our method has several advantages: 
\begin{itemize}[leftmargin=*]
  \item Our method divides the resolution of the current level by a fold parameter to calculate the resolution of the next level. Our policy of configuring resolutions requires fewer encoding parameters when having the same number of resolution levels as MHE.
  \item For a specific resolution level, resolution remains identical across the dimensions in grid-based input encoding methods like MHE. F-Hash determines the resolution of each dimension independently, resulting in a smaller number of encoding parameters when handling rectangular cuboid volumetric frames with unequal edge lengths. \cref{fig:yx_ratio} shows an example comparing the number of encoding parameters when handling 2D rectangle grids with a high aspect ratio.
  \item For various resolution levels, the MHE hash tables for different resolution levels share the same size, even though the embedding grid becomes smaller for lower resolution levels, resulting in bucket waste in those levels. The hash tables of F-Hash are resolution-dependent, scaling their sizes proportionally with the embedding grid size of the current resolution level.
  \item In MHE, the number of resolution levels and their respective resolutions are predefined hyperparameters that require manual tuning. In contrast, our approach automatically optimizes those parameters from the spatial information of the time-varying feature of interest.
\end{itemize}

\subsubsection{Temporal Multi-resolution Configuration}
In order to expand the input encoding to higher dimensional data than 3D spatial volume, we propose multi-resolution Tesseract encoding with temporal dimension to directly model time-varying volumetric data. Due to the dynamic changes over temporal dimension and arbitrary length of time sequences, we need to treat temporal dimension independently to 3D spatial dimensions. In order to minimize the total number of encoding parameters, as mentioned in \cref{sec:coreset}, we utilize key frame extraction methods to select the time steps of the informative frames where critical evolutionary events happen. We then apply a similar spatial resolution policy to configure the resolution on the temporal domain using fold. In real-world scenarios, compared to the size of spatial dimensions, the range of the temporal dimension of the key frames is much smaller, and the padding of 2 is normally needed at the tail of temporal resolution levels. 

\subsubsection{Hash Function Design}\label{sec:hash_function}
The proposed feature-based hash function (F-Hash) is to map each corner position of Tesseract embedding grid into a bucket in the hash table. The main goals of the design include no hash collision, no bucket waste, and fast looking up. Compared with existing multi-resolution hash encoding methods, the proposed F-Hash meets all the requirements, making it a more capable input encoding solution. Our F-Hash utilizes simple linearization to map 4D coordinates to the hash table index. This enables a single look-up to retrieve the mapping for all 16 corners of the Tesseract through simple shifting. The bucket index of an embedding grid corner $(t, x, y, z)$ of resolution level $l$ can be calculated by:

{\small
\begin{equation}
F\text{-}Hash^{l}(t, x, y, z) = t\times Res_x^{l}\times Res_y^{l}\times Res_z^{l} + z\times Res_x^{l}\times Res_y^{l} + y\times Res_x^{l} + x
\end{equation}
}

\begin{figure}[t]
    \centering 
    \includegraphics[trim=0 2 0 0,clip,width=\linewidth]{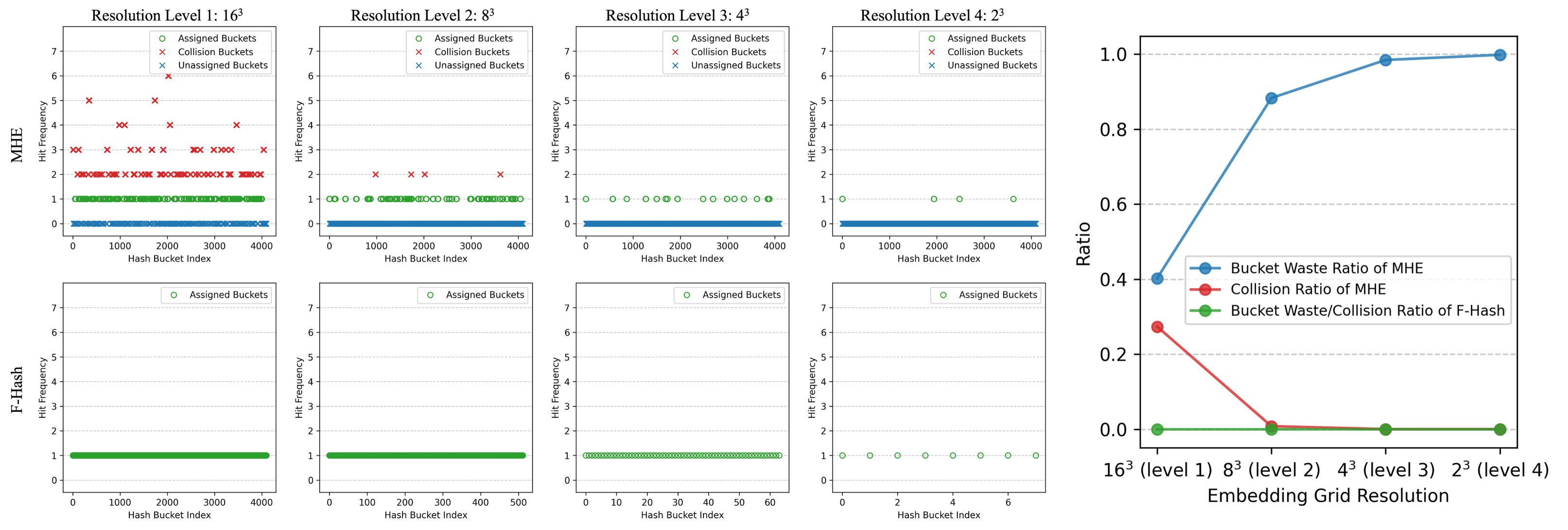}
    \caption{Comparison between MHE and F-Hash on collision and bucket waste performance in an example with 4 resolution levels.}
    \label{fig:collision_waste}
\end{figure}

\cref{{fig:hash_function}} demonstrate the F-Hash mapping from the neighboring embedding cells to the hash table entries for each resolution level. MHE uses a simple spatial hash function~\cite{teschner2003optimized}, which presents inevitable collision on lower resolution levels and bucket waste on higher resolution levels as shown \cref{fig:collision_waste}. F-Hash is a bijective minimal perfect hash function (MPHF) without collision and bucket waste, enabling more accurate modeling with more compact encoding parameters. Other linearizations like Morton and Z-order can also be used to construct F-Hash with preserved locality.

\subsubsection{Input Encoding}
\begin{figure}[t]
    \centering 
    \includegraphics[trim=0 0 0 0,clip,width=0.95\linewidth]{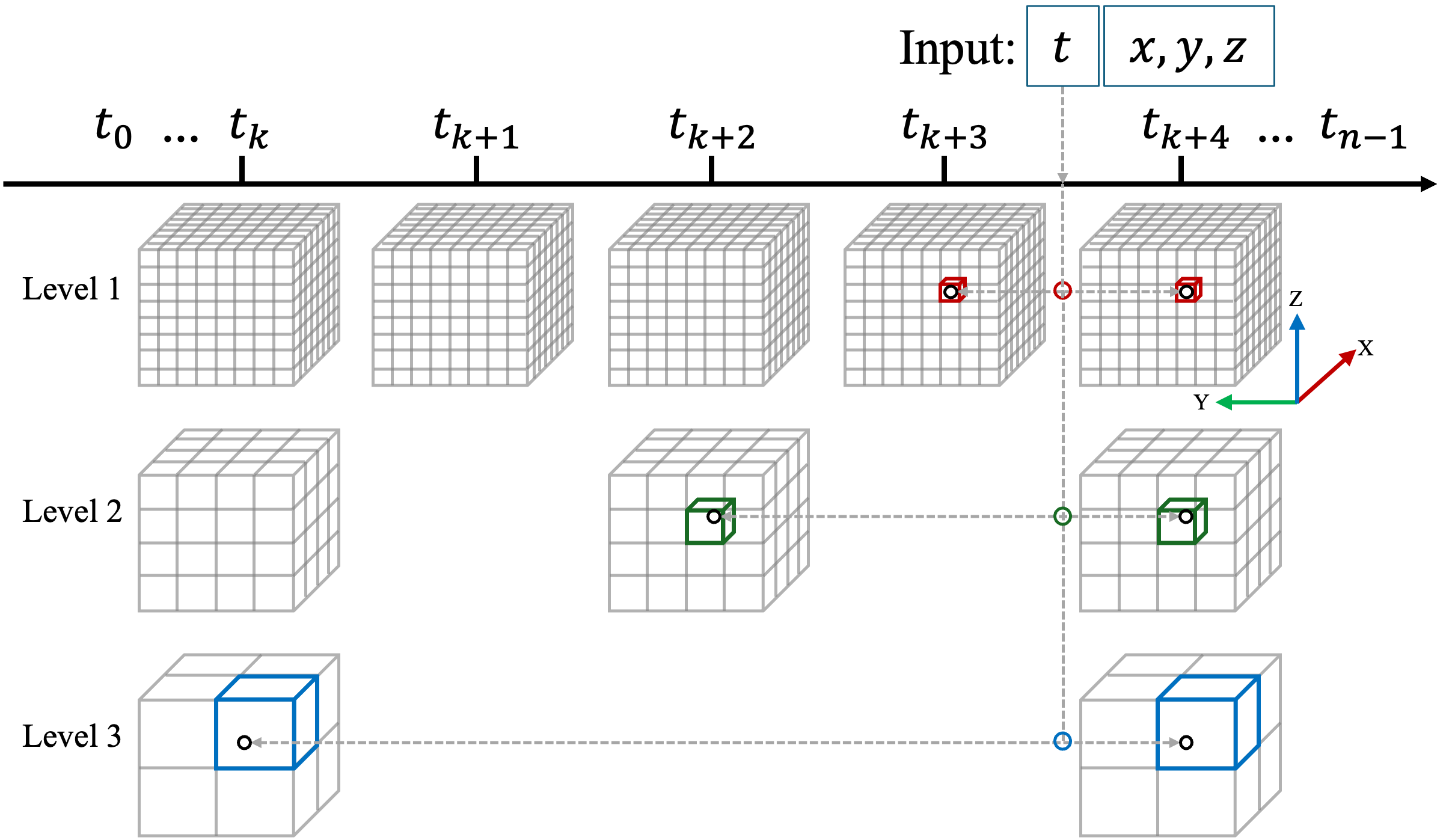}
    \caption{Multi-Resolution Tesseract Encoding. For the given input $(t, x, y, z)$, the neighboring time steps from level 1 to level 3 are, $[t_{k+3}, t_{k+4}]$, $[t_{k+2}, t_{k+4}]$, and $[t_{k}, t_{k+4}]$. Embedding cells are color-coded for each resolution level.}
    \label{fig:find_grid}
\end{figure}

The encoding steps for the proposed multi-resolution Tesseract encoding consist of the following steps: 

\textbf{1. Locate neighboring time steps:} The input, $(t, x, y, z)$, needs to find its neighboring time steps across the temporal dimension for each resolution level according to $t$. We name the time step before and after $t$ as $t_{Previous}$ and $t_{Next}$. \cref{fig:find_grid} shows the embedding grids of the proposed Tesseract encoding with three spatial resolution levels.

\begin{figure}[t]
    \centering 
    \includegraphics[trim=0 0 0 0,clip,width=0.95\linewidth]{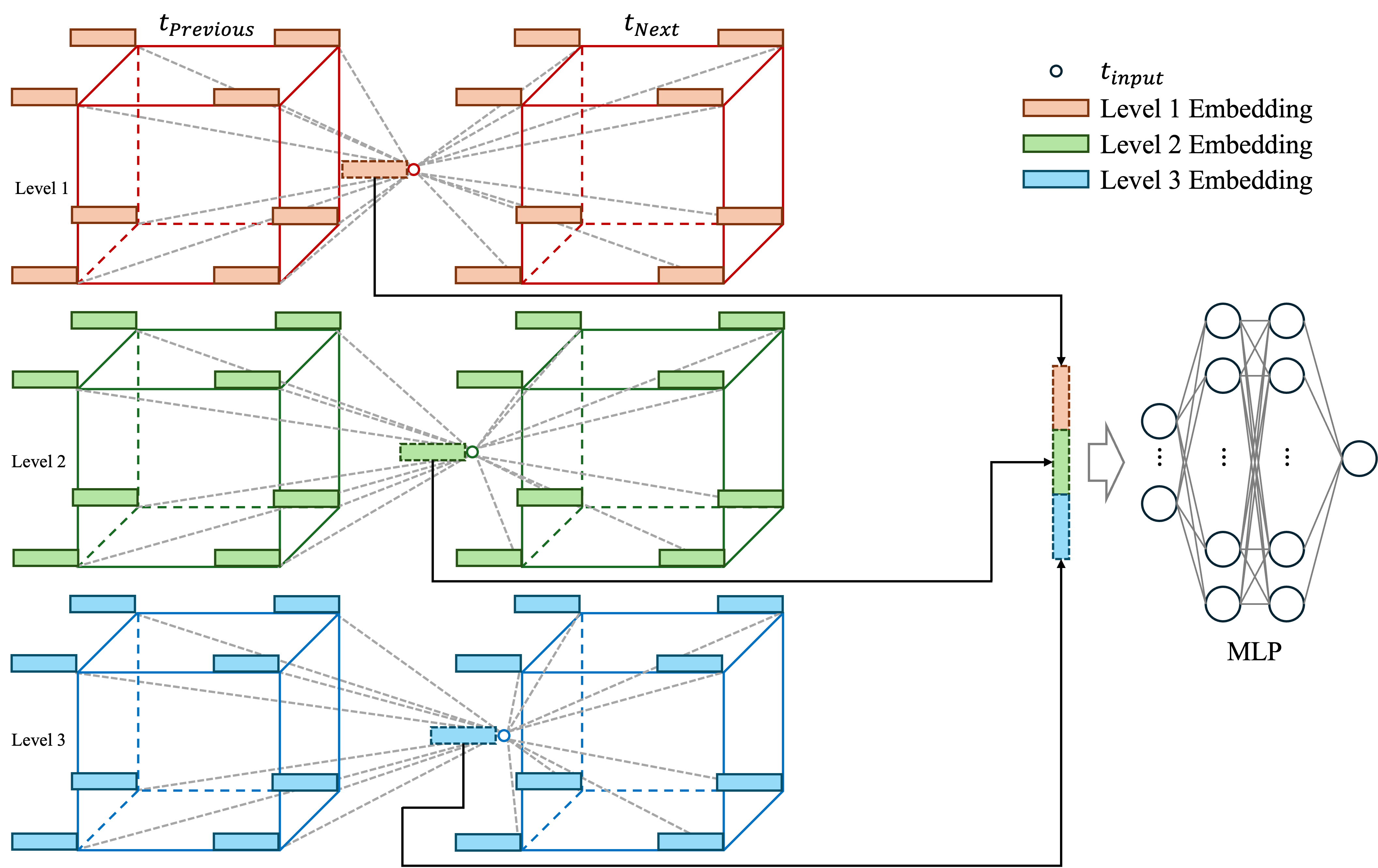}
    \caption{Quadrilinear interpolation.}
    \label{fig:interpolation}
\end{figure}

\begin{figure}[t]
    \centering
        \includegraphics[trim=0 0 0 0,clip,width=\linewidth]{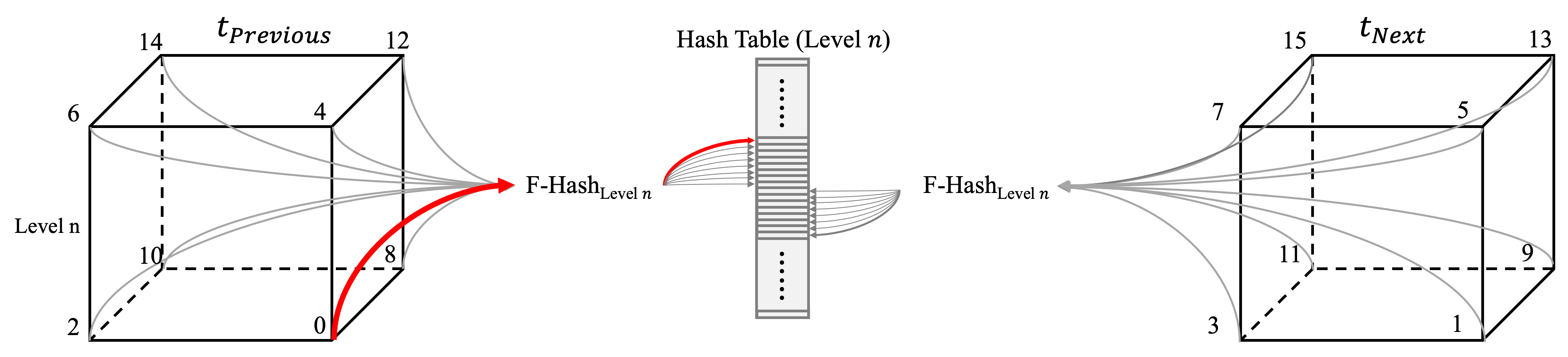}
    \caption{Hash function design for mapping the multi-resolution Tesseract embedding grid to hash buckets of resolution level $n$.}
    \label{fig:hash_function}
\end{figure}

\textbf{2. Locate embedding cells:} Once the neighboring time steps are found, the embedding cells need to be located from the embedding grid at $t_{Previous}$ and $t_{Next}$ for each resolution level. The embedding grid of each resolution level is a cube covering a range of $[-1, 1]$ for each dimension. The embedding cell is the smallest unit for a specific resolution level containing the input location $(x, y, z)$.

\textbf{3. Look up embedding vectors:} Each of the eight corners of the embedding cell is mapped to an embedding vector through the proposed F-hash. There are 16 embedding vector lookups in total for both the embedding cells at $t_{Previous}$ and $t_{Next}$. 

\textbf{4. Interpolate embedding vectors:} A quadrilinear interpolation across both spatial and temporal is performed for an aggregated embedding vector $V$ for each resolution level. The spatial trilinear interpolation is performed first within the embedding cells, followed by the linear interpolation across the temporal dimension: 
{\small
\begin{equation}
V^{Previous}(x, y, z) = \sum_{i=0}^{1} \sum_{j=0}^{1} \sum_{k=0}^{1} V_{ijk}^{Previous} \cdot x^i (1-x)^{1-i} \cdot y^j (1-y)^{1-j} \cdot z^k (1-z)^{1-k}\\
\end{equation}
}
{\small
\begin{equation}
V^{Next}(x, y, z) = \sum_{i=0}^{1} \sum_{j=0}^{1} \sum_{k=0}^{1} V_{ijk}^{Next} \cdot x^i (1-x)^{1-i} \cdot y^j (1-y)^{1-j} \cdot z^k (1-z)^{1-k}\\
\end{equation}
}
{\small
\begin{equation}
V(t) = \sum_{l=t_{Previous}}^{t_{Next}} V^{l} \cdot t^l (1-t)^{t_{Next}-l}.
\end{equation}
}

All aggregated embedding vectors $V$ from all resolution levels are concatenated to form the high-dimensional input for the downstream MLP. \cref{fig:interpolation} demonstrates the interpolating process. To accelerate rendering in volumetric neural representations, a shallow network (with only one or minimal hidden layers) is used instead of a deep architecture. This reduces training and inferencing time significantly because the embedding grid encoding offloads the geometric complexity from the network, allowing a lightweight model to represent complex data. All the embedding vectors of Tesseract grids are initialized by uniform initialization before training.

\subsection{Rendering}\label{sec:rendering}
We implement a time-varying volume visualization algorithm to directly render the INR modeled through F-Hash. We design the rendering pipeline using a similar idea of sample streaming method~\cite{10175377} for visualizing volumetric neural representation. We further improve the algorithm by adding an adaptive ray marching (ARM) algorithm to optimize thread usage on GPU for reduced rendering latency. The idea of ARM is to efficiently utilize the GPU threads by increasing the ray marching pace over time as more rays get finished. \cref{alg:framework} details the rendering algorithm. Our renderer supports popular visualizations for time-varying volumetric datasets, including feature tracking and evolution visualization. The occupancy grid derived from the coreset selection step is used to accelerate the rendering by skipping the empty space. Super-resolution in the temporal domain is also supported. For a given unseen time step, a linearly interpolated occupancy grid for the specific time step is first calculated. Then, the trained INR can be directly inferenced for rendering. Since the INR is trained on the key frames, which cover all the critical evolutionary changes, the interpolated occupancy grid remains consistent with adjacent key frames. Although training INR with input encoding is fast, the convergence time of online training on time-varying volumetric data is still relatively long for interactive visualization. Therefore, our rendering pipeline pretrain INR before visualization.

\begin{algorithm}[t]
\caption{Time-varying volumetric neural representation rendering pipeline using sample streaming with Adaptive Ray Marching (ARM)}\label{alg:framework}
\textbf{Input:} Time Step (t), View, Transfer Functions(TF), and F-Hash trained INR \\
\textbf{Output:} Visualization image
\begin{algorithmic}[1]
\State $N_f = 0$ \Comment{Finished samples}
\State $R = getRays(view)$
\If {$N_f <= R_{max}$}
    \State $N_a = getAliveRayNum()$
    \If{$N_a == 0$} \Comment{When all rays finished}
        \State Break
    \Else
        \State $N_s = max(min(N_{r}/N_a, 64), 1)$ \Comment Adjust Marching Pace
        \State $N_f += N_s$
        \State $S = getSamples(N_{s}, Occupancy Grid)$ \Comment{Sample streaming}
        \State $S = t \| S$ \Comment{Concatenate time step}
        \State $V = model(S)$ \Comment{Inference F-Hash INR}
        \State $RGBA = applyTF(V, TF)$
        \State $compositing(RGBA)$
        \State $updateAlive()$
    \EndIf
\EndIf
\end{algorithmic}
\end{algorithm}

\begin{figure}[t]
    \centering
        \begin{subfigure}[b]{0.288\linewidth}
            \centering
            \includegraphics[trim=0 0 0 0,clip,width=\linewidth]{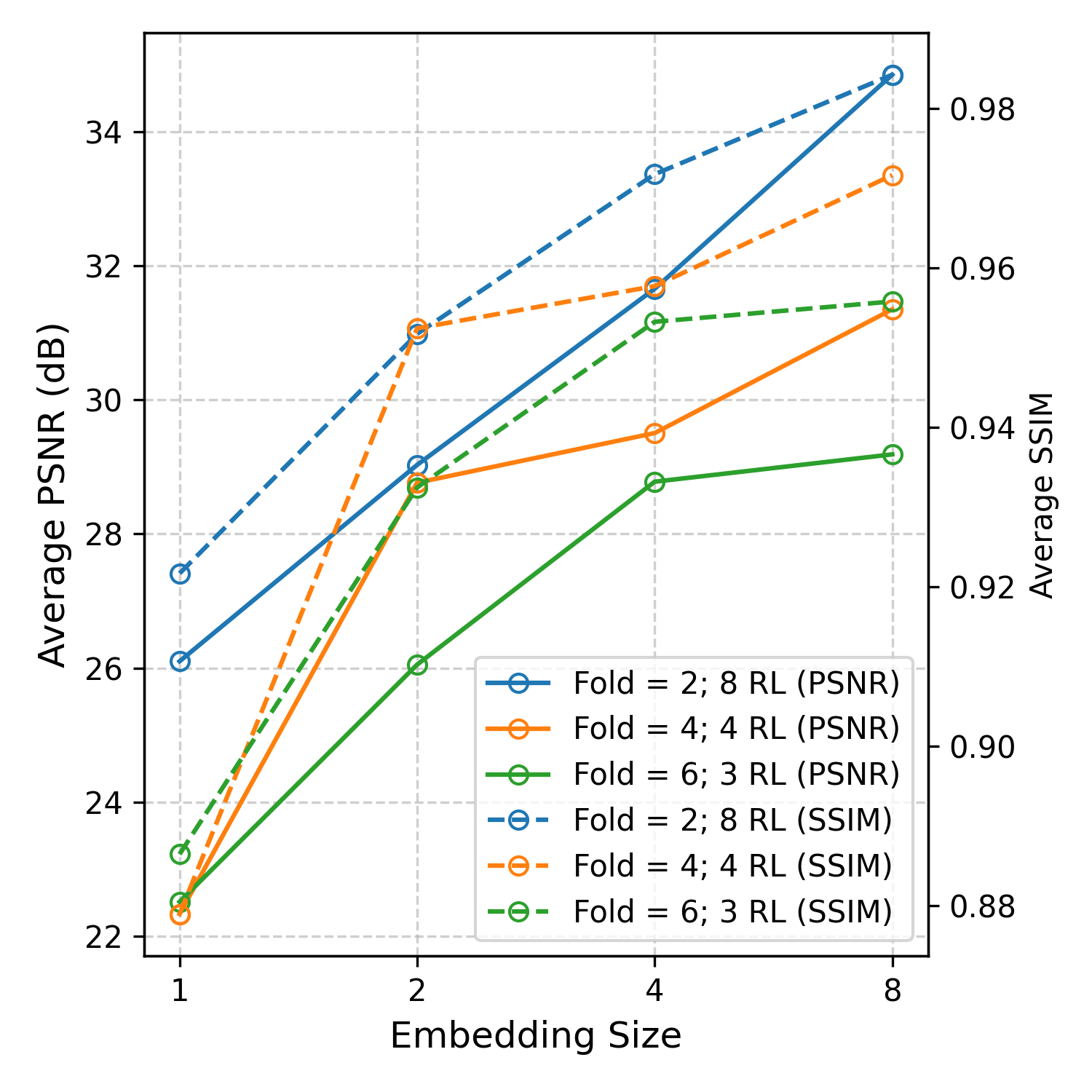}
            \caption{Prediction accuracy (IE)}
            \label{fig:ablation_psnrAndssim_ie}
        \end{subfigure}
        \begin{subfigure}[b]{0.7\linewidth}
            \centering
            \includegraphics[trim=0 0 0 0,clip,width=\linewidth]{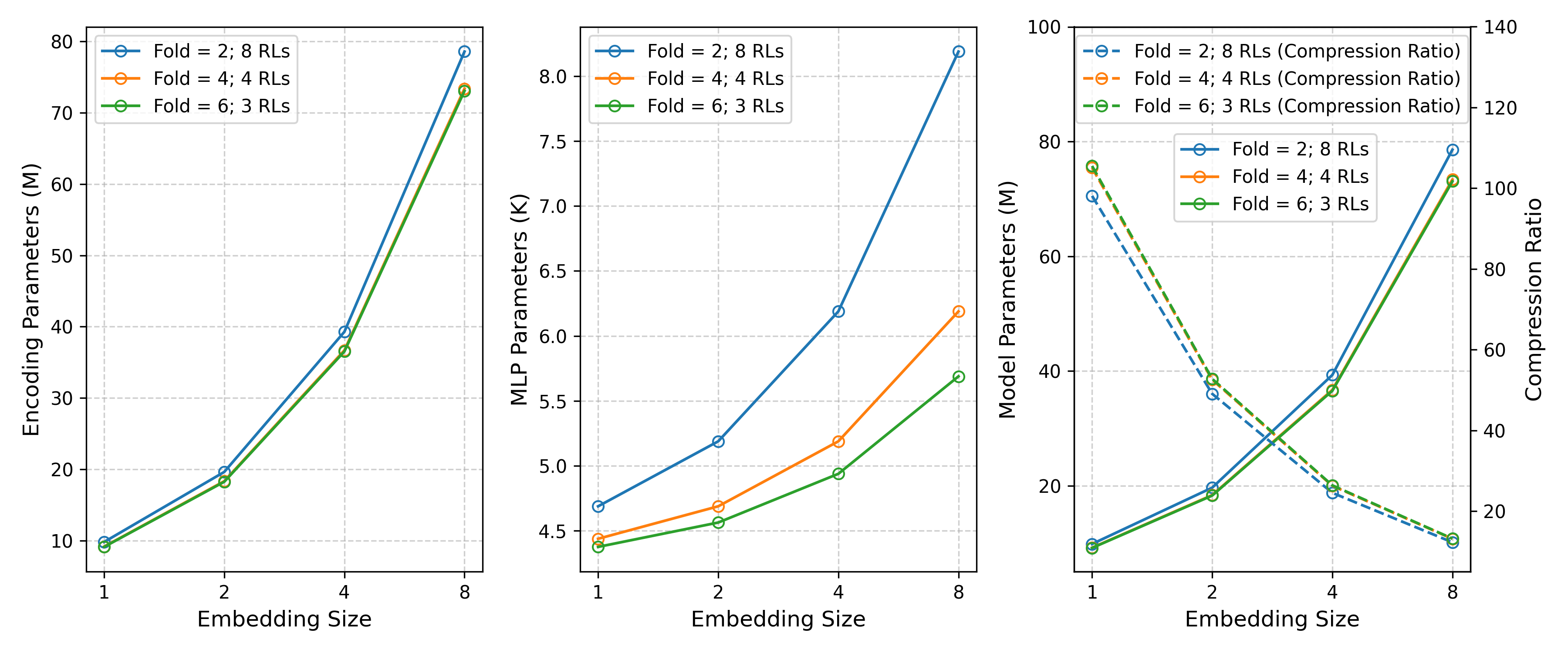}
            \caption{Model parameter size (IE)}
            \label{fig:ablation_size_ie}
        \end{subfigure}
        \begin{subfigure}[b]{0.288\linewidth}   
            \centering 
            \includegraphics[trim=0 0 0 0,clip,width=\linewidth]{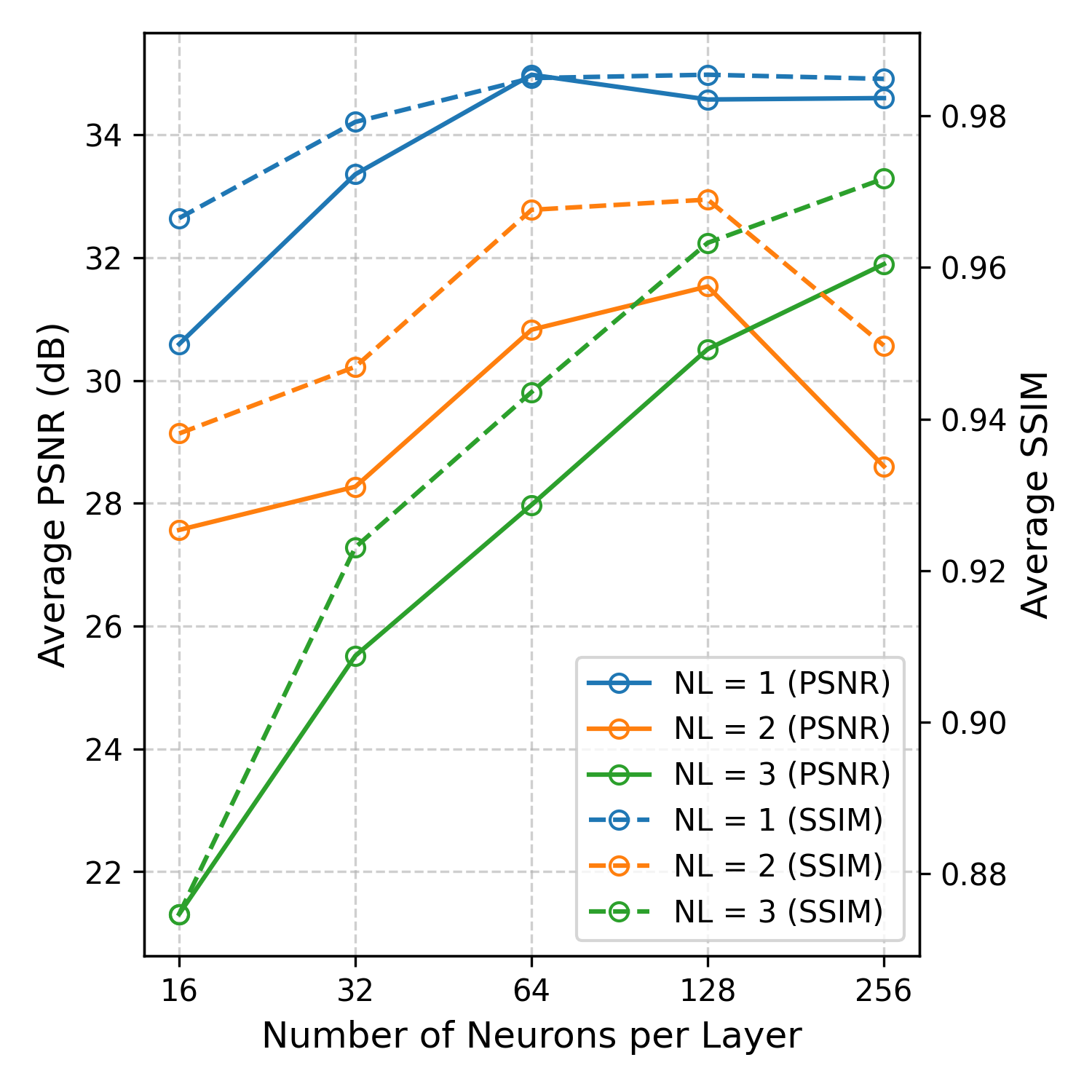}
            \caption{Prediction accuracy (HP)}
            \label{fig:ablation_psnrAndssim_hp}
        \end{subfigure}
        \begin{subfigure}[b]{0.7\linewidth}
            \centering
            \includegraphics[trim=0 0 0 0,clip,width=\linewidth]{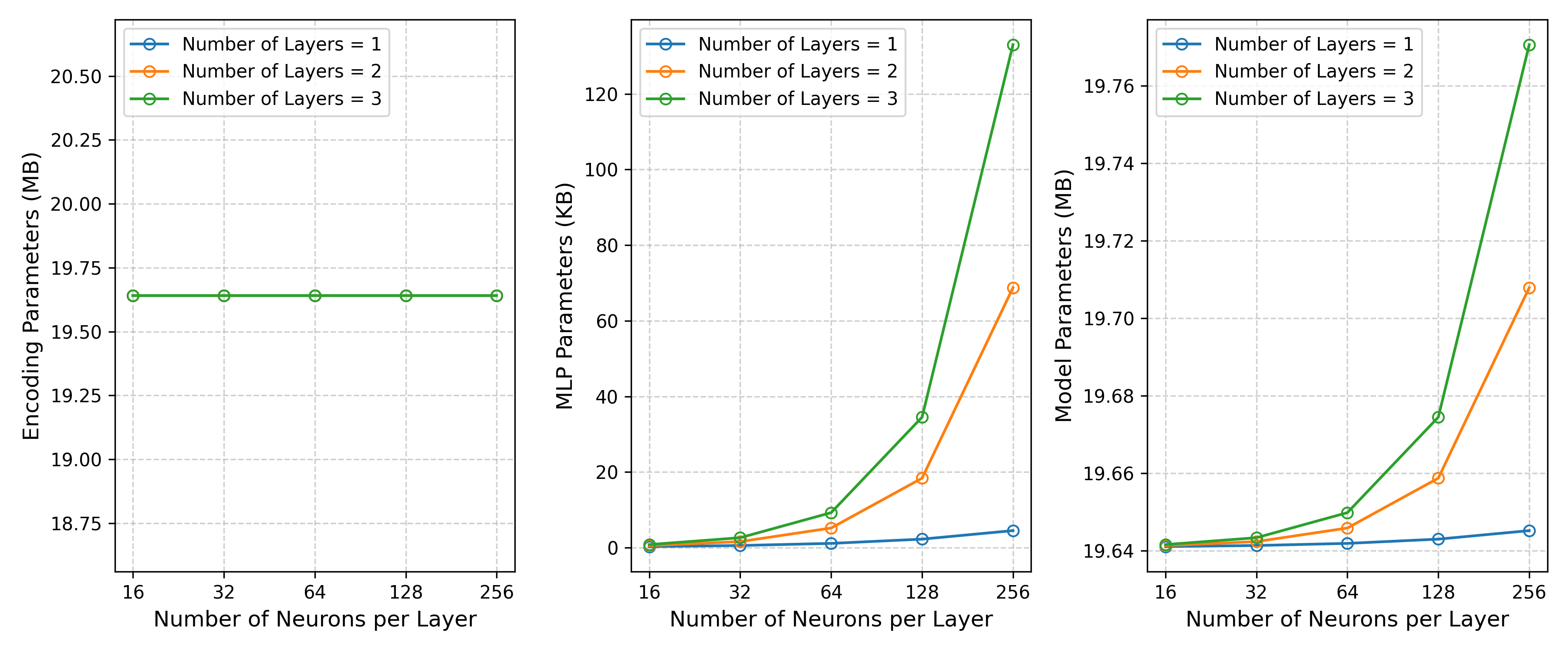}
            \caption{Model parameter size (HP)}
            \label{fig:ablation_size_hp}
        \end{subfigure}
    \caption{Prediction accuracy (the 10th iteration, results of more iterations can be found in Appendix) and model parameter size evaluation. (a) and (b) are results of input encoding (IE) configuration with different folds/resolution levels (RL) and embedding size. (c) and (d) are results of Hyperparameter (HP) configuration with different numbers of layers and number of neurons per layer.}
    \label{fig:end2end_vs_standalone}
\end{figure}




\begin{table}[t]
  \caption{Training time under different input encoding configurations.}
   \vspace{-2mm}
  \label{tab:ablation_training_time}
  \scriptsize%
	\centering%
  \begin{adjustbox}{width=0.5\textwidth}
      \begin{tabu}{ c | c | c | c | c   }
      \toprule
      Training Time (Minutes) $\downarrow$ & Embedding = 1 & Embedding = 2  & Embedding = 4  & Embedding = 8\\
      \midrule
      Fold = 2 (Res lvl = 8) & 8.4 & 16.8 & 33.7 & 67.3\\
      \midrule
      Fold = 4 (Res lvl = 4) & 7.9 & 15.7 & 31.4 & 62.9\\
      \midrule
      Fold = 6 (Res lvl = 3) & 7.8 & 15.6 & 31.3 & 62.6\\
      \bottomrule
      \end{tabu}
 \end{adjustbox}
\end{table}

\begin{table}[t]
  \caption{Time-varying datasets information of variable, resolution, and size. Intensity, Angular, and MixFrac variables are selected as the volumetric scalar fields of each dataset for the experiment.}
  \label{tab:datasets}
  \scriptsize%
	\centering%
  \begin{adjustbox}{width=0.48\textwidth}
      \begin{tabu}{ c c c c c }
          \toprule
          Dataset & Variable & Resolution ($X\times Y\times Z)\times T$ & Data Type & Size \\
          \midrule
          Combustion   & CHI/HO2/OH/\textbf{MixFrac} & $(200\times172\times54)\times136$ & float32 & 0.94 GB\\
          \midrule
          Argon Bubble & \textbf{Intensity}        & $(128^2\times 256)\times241$ & float32 & 3.77 GB\\
          \midrule
          Supernova    & \textbf{Angular}/Entropy  & $(216^3)\times136$ & float32 &  5.11 GB \\
          \bottomrule
      \end{tabu}
  \end{adjustbox}
\end{table}
\section{Ablation Study}
\subsection{Input Encoding Configuration}
The proposed embedding grids learn both the geometric and temporal dynamics of the time-varying data. We evaluate the following two key configurations of the input encoding: 1) \textbf{Fold Parameter:}
The fold parameter determines the multi-resolution configuration. The default fold is set to 2, providing the most number of levels and the highest possible resolution on each level. Larger fold values give a smaller number of levels with lower resolution for subsequent levels, except for level 1. 2) \textbf{Embedding Size:} The embedding size determines the number of embedding weights assigned for each corner of the Tesseract grid. A larger embedding size results in a higher dimension of the input encoding for the downstream MLP. We fix the MLP parameters (number of MLP layers = 2, number of neurons per layer = 64) while exploring the input encoding parameter space. 
As \cref{fig:ablation_psnrAndssim_ie} shows, using a smaller fold or a larger embedding size gives better training accuracy per training iteration and improves convergence speed.
\cref{fig:ablation_size_ie} shows that changing the embedding size will change the size of both the encoding parameter and the MLP parameters more dramatically than changing the fold parameter, resulting in a larger INR. A similar correlation can be observed in \cref{tab:ablation_training_time} on training time, where increasing the embedding size significantly prolongs the training duration. This is because the embedding size will linearly increase the total number of trainable parameters of the INR, requiring more gradient computation and updates during backpropagation and therefore increasing the per-iteration training time. The last subfigure of \cref{fig:ablation_size_ie} also shows the respective compression ratio using various folds and embedding sizes when encoding the Argon Bubble dataset with 9 key frames. Based on our observations, we set both the fold parameter and embedding size to 2 to achieve high accuracy while maintaining a moderate model size and reasonable training time.

\subsection{Hyperparameter Tuning}
Together with the embedding grids, the downstream neural network, as shown in \cref{fig:interpolation}, will predict the value at the query spatial-temporal location from the high-dimensional input encoding. We evaluate the following two key hyperparameters of the network: 1) \textbf{Number of MLP Layers:} The number of MLP layers determines how deep the neural network is. A deeper network is better for hierarchical feature learning, but harder to train. 2) \textbf{Number of Neurons Per Layer:} The number of neurons per layer determines the learning capability of the network. A wider network (more neurons per layer) can approximate simple functions efficiently. We fix the input encoding parameters (fold = 2, embedding size = 2) while exploring the MLP parameter space for modeling the Argon Bubble dataset. 
As shown in \cref{fig:ablation_psnrAndssim_hp}, adding more layers to the MLP slightly improves training accuracy per iteration. While increasing the number of neurons in the lower range improves training accuracy, it diminishes in the higher range.
As shown in \cref{fig:ablation_size_hp}, although adding more neurons causes the MLP to grow exponentially, this doesn't substantially affect the total INR size since the input encoding accounts for the most parameters. Adjusting the MLP's depth and width has minimal impact on training time, with all configurations completed in approximately 39 minutes. The proposed Tesseract embedding grid effectively captures both spatial and temporal characteristics through a high-dimensional encoding, enabling a shallow MLP to accurately model the entire time-varying volumetric data. We select the number of MLP layers as 2 and the number of neurons per layer as 64 for our experiments.

\section{Experiments and Evaluation}
\subsection{Experimental Setup}
\subsubsection{Datasets}
For quality and performance evaluation, we select 3 large-scale time-varying volumetric datasets with distinct spatial and temporal features collected from diverse domains. Detailed information is listed in \cref{tab:datasets}. The variables we used in our experiment are MixFrac from the Combustion dataset, Intensity from the Argon Bubble dataset, and Angular from the Supernova dataset. The spatial ranges of x, y, and z dimensions of each time step volume of all datasets are normalized within $[-1, 1]$. The values of each dataset are normalized across the 4D domain within the range $[0, 1]$. There are 10, 9, and 10 key frames extracted for the respective Combustion, Argon Bubble, and Supernova datasets.

\subsubsection{Evaluation}
The experiments are designed to investigate input encoding methods from two aspects: convergence performance during training and model performance during rendering. PSNR and SSIM are used to evaluate the reconstruction accuracy. For convergence evaluation, we use convergence time and training time. We select representative time-varying visualization tasks from the literature reviews~\cite{6185547, 10.1007/s12650-020-00654-x} for a comprehensive evaluation of the practicality and viability of our method. We select the mainstream time-varying volume visualization techniques including Spatial-first Feature Tracking (Interval, Isosurface, and Segmentation features) and Evolution Visualization (Animation). The interval feature is extracted by selecting samples with values within a range of interest. The isosurface features represent the samples of constant value. The segmentation features represent foreground elements extracted from a 3D background space. Evolution visualization is to provide a meaningful overview through the visual dynamics over time.
We compare our method with recent parametric-based input encoding works, which are Dense Grid Single-Resolution Encoding (DG Single-Res)~\cite{Chan_2022_CVPR}, Dense Grid Multi-Resolution Encoding (DG Multi-Res)~\cite{10.1145/3478513.3480569}, a Pytorch implementation of Multi-resolution Hash Encoding (MHE)~\cite{10.1145/3528223.3530127}, and Multi-resolution Hash Encoding accelerated by fully fused MLP~\cite{10.1145/3450626.3459812} through Tinycudann framework (MHE Tinycudann). We exclude frequency-based encoding method for its poor performance. 

\begin{figure}[t]
    \centering
        \begin{subfigure}[b]{0.32\linewidth}
            \centering
                \includegraphics[trim=0 290 0 0,clip,width=\linewidth]{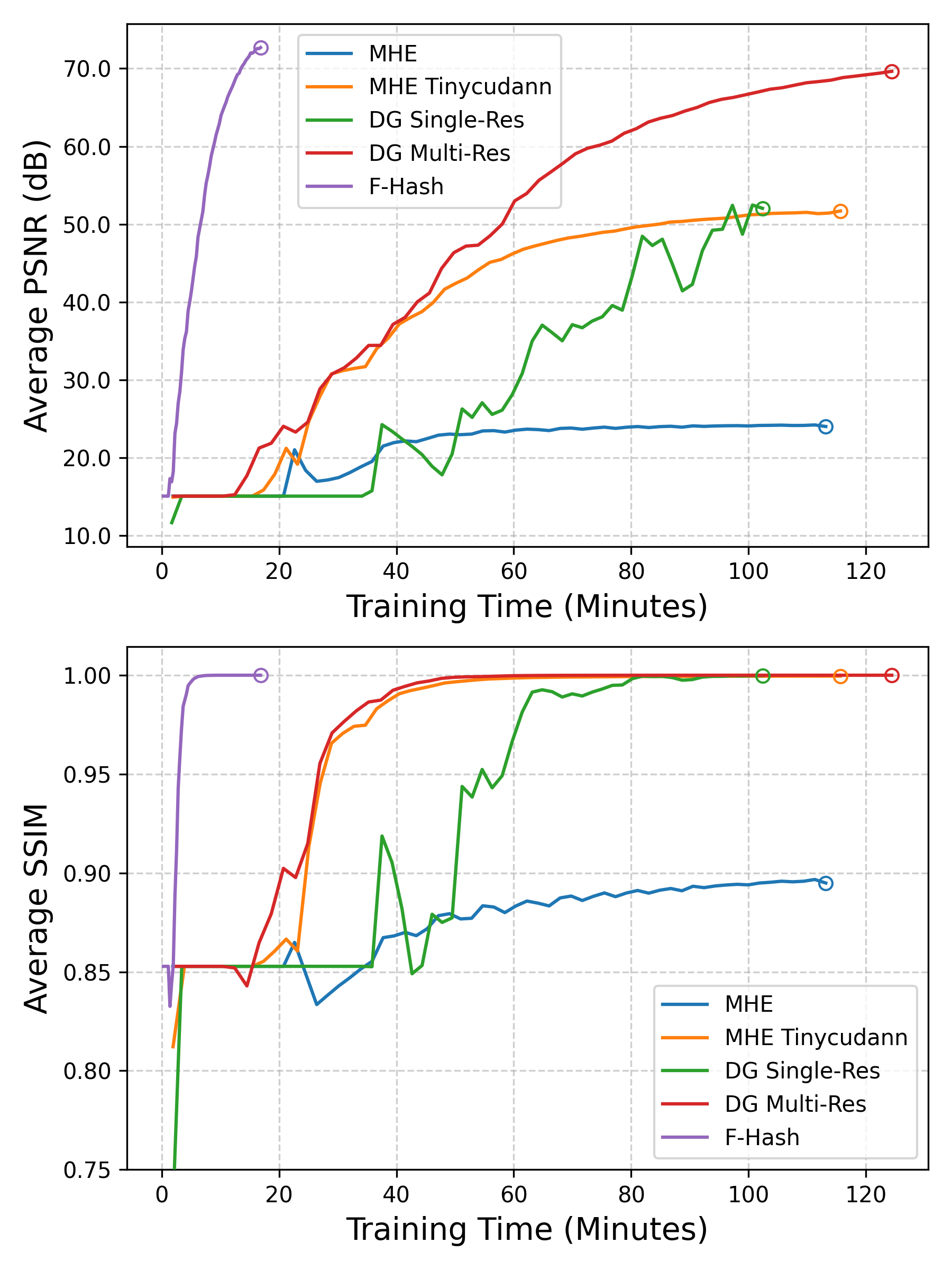}
            \caption{Interval}
            \label{fig:argon_convergence_curves_a}
        \end{subfigure}
        \hfill
        \begin{subfigure}[b]{0.32\linewidth}
            \centering 
            \includegraphics[trim=0 290 0 0,clip,width=\linewidth]{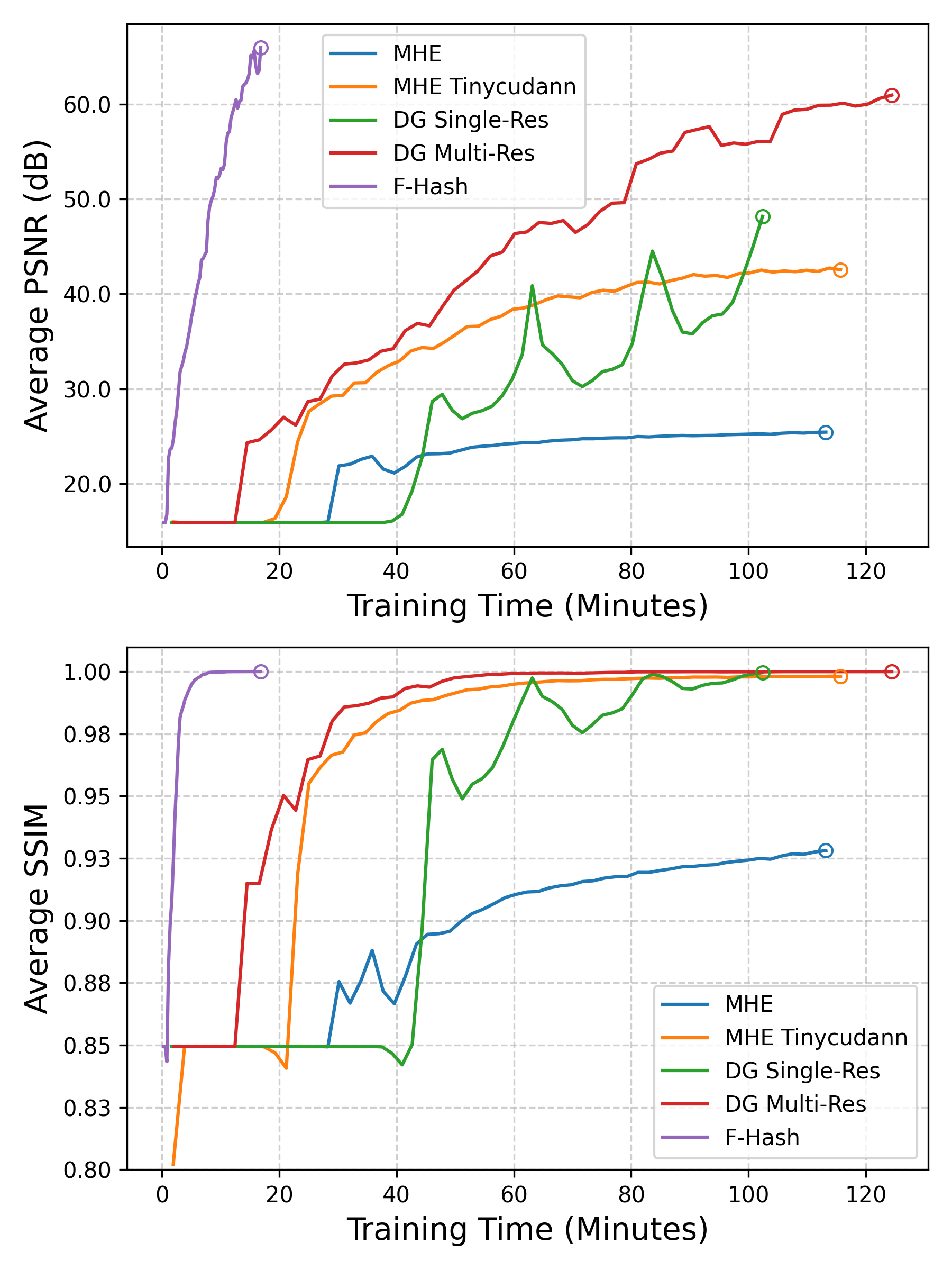}
            \caption{Isosurface}
            \label{fig:argon_convergence_curves_b}
        \end{subfigure}
        \hfill
        \begin{subfigure}[b]{0.32\linewidth}
            \centering 
            \includegraphics[trim=0 290 0 0,clip,width=\linewidth]{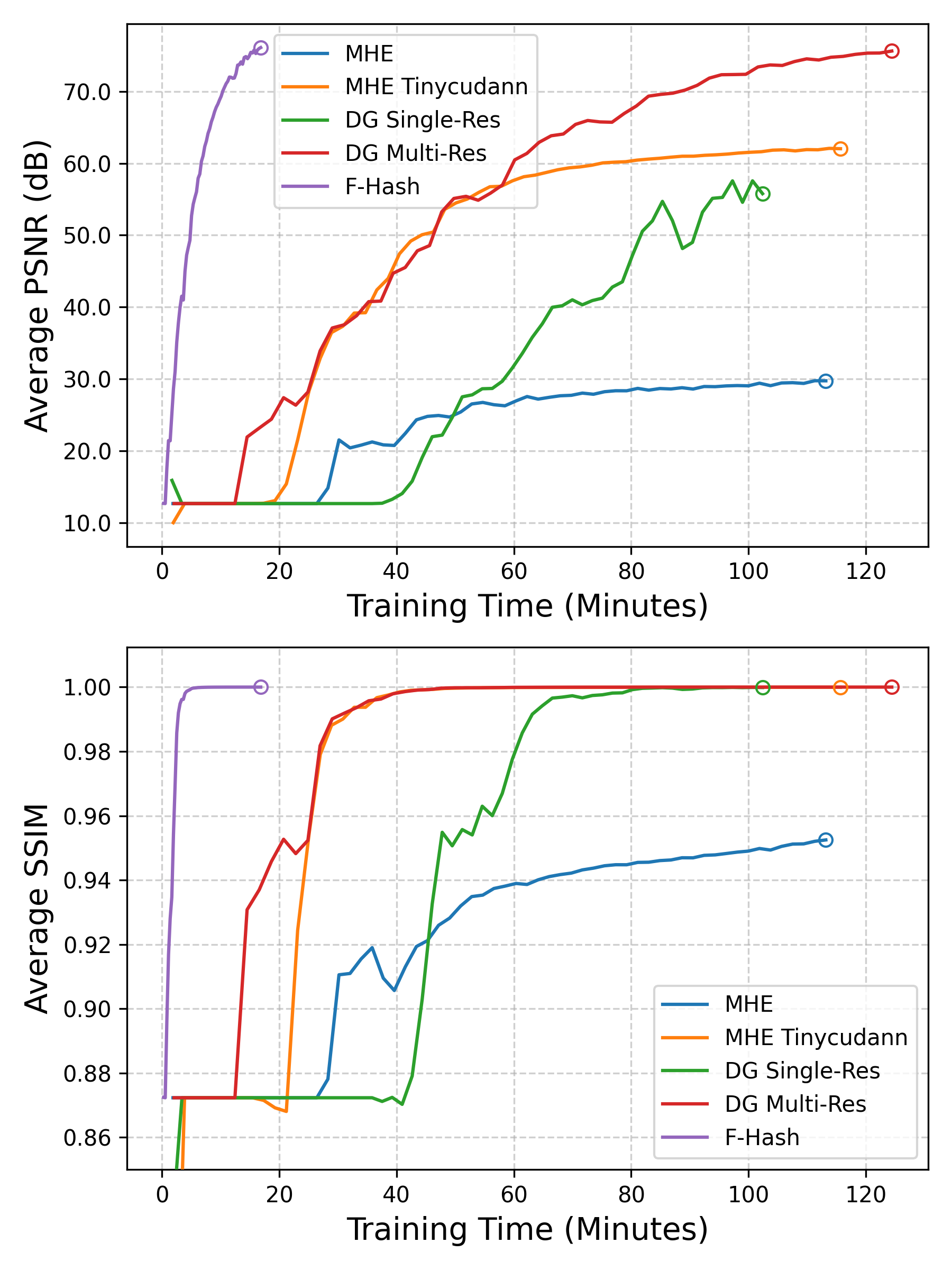}
            \caption{Segmentation}
            \label{fig:argon_convergence_curves_c}
        \end{subfigure}
    \vspace{-3mm}
    \caption{Convergence speed for training the INR using different input encoding methods for various features of the Argon Bubble dataset.}
    \label{fig:argon_convergence_curves}    
\end{figure}

\begin{figure}[t]
    \centering
        \begin{subfigure}[b]{0.32\linewidth}
            \centering
            \includegraphics[trim=0 0 0 0,clip,width=\linewidth]{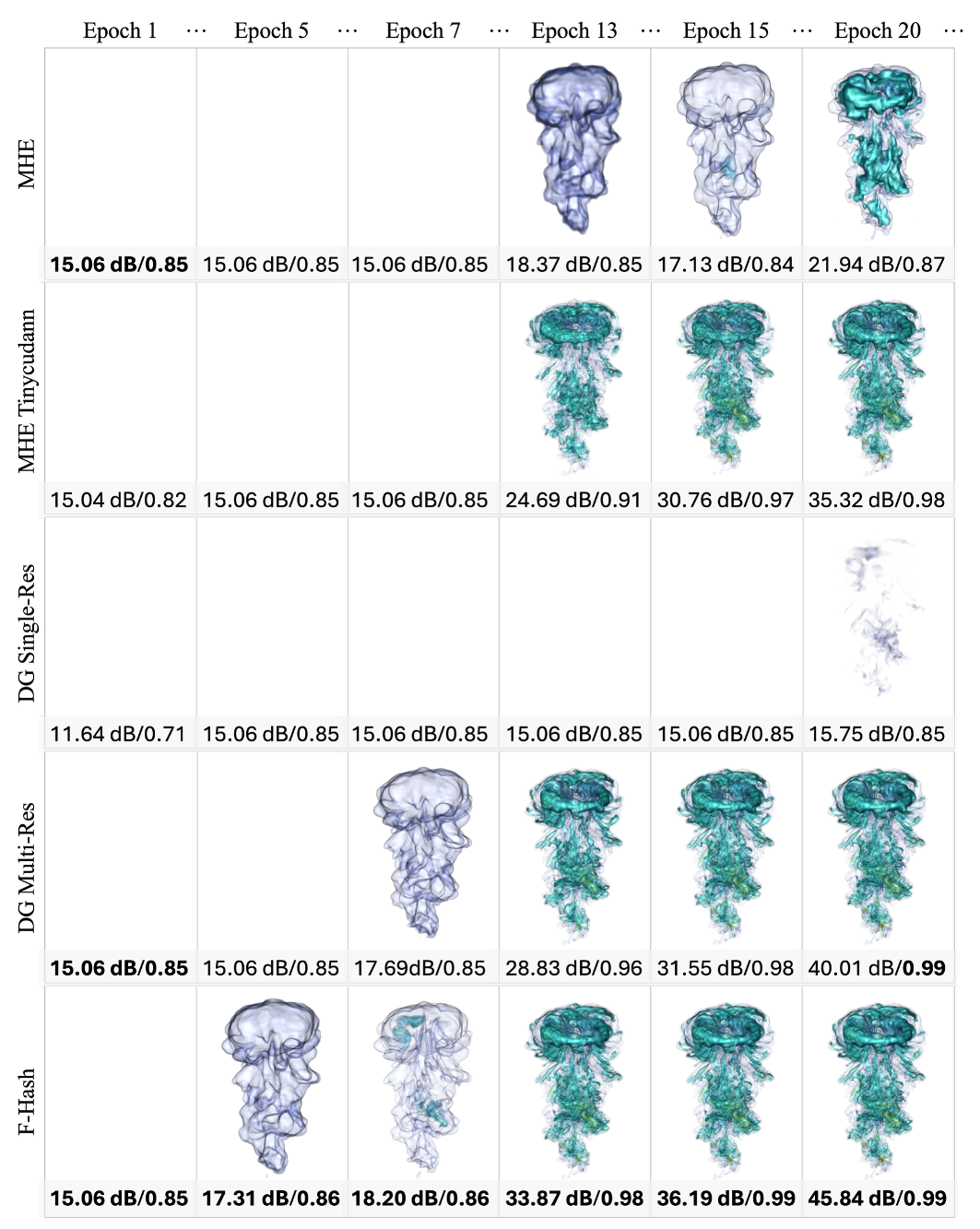}
            \caption{Interval}
            \label{fig:compare_argon_interval}
        \end{subfigure}
        \hfill
        \begin{subfigure}[b]{0.32\linewidth}
            \centering 
            \includegraphics[trim=0 0 0 0,clip,width=\linewidth]{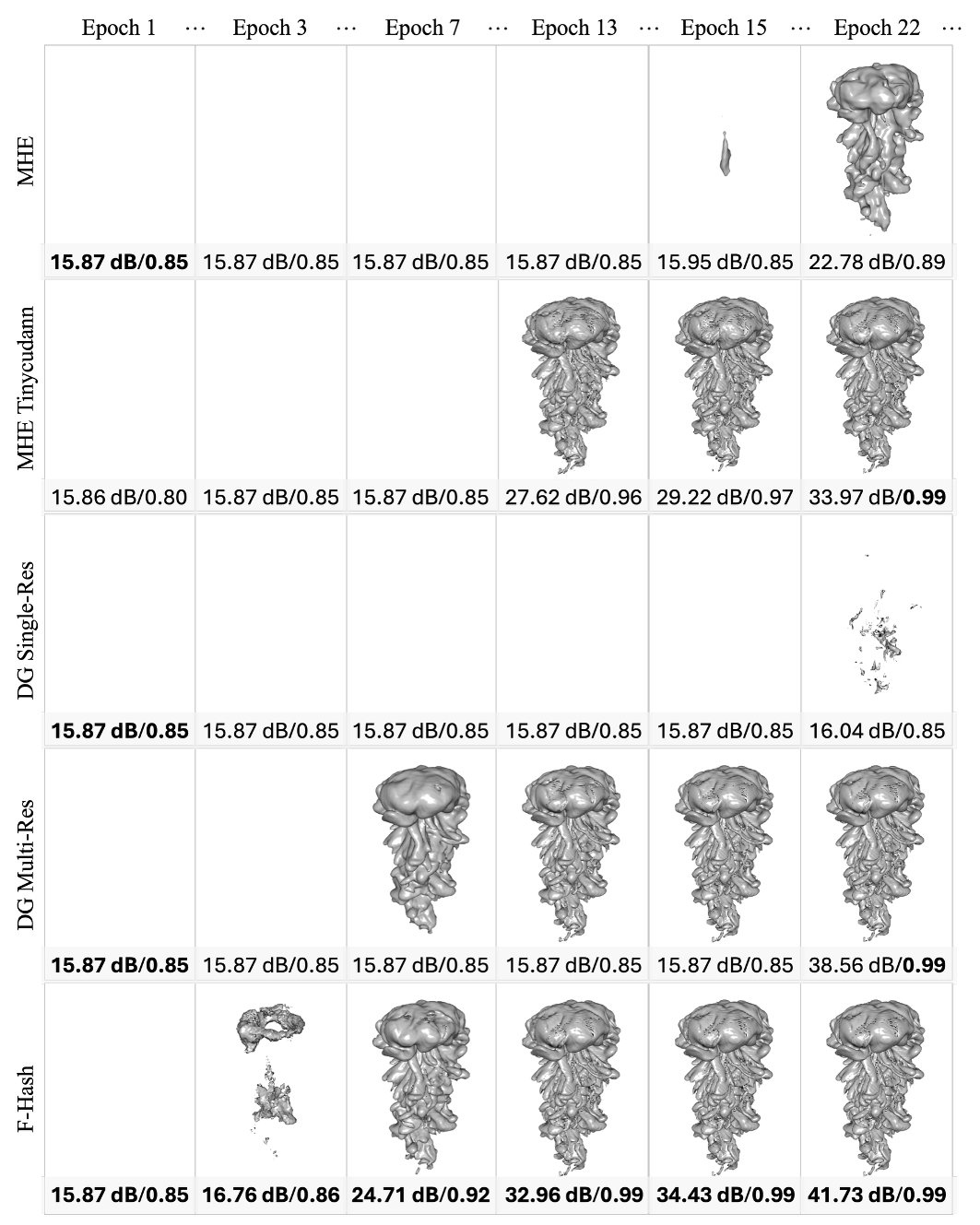}
            \caption{Isosurface}
            \label{fig:compare_argon_iso}
        \end{subfigure}
        \hfill
        \begin{subfigure}[b]{0.32\linewidth}
            \centering 
            \includegraphics[trim=0 0 0 0,clip,width=\linewidth]{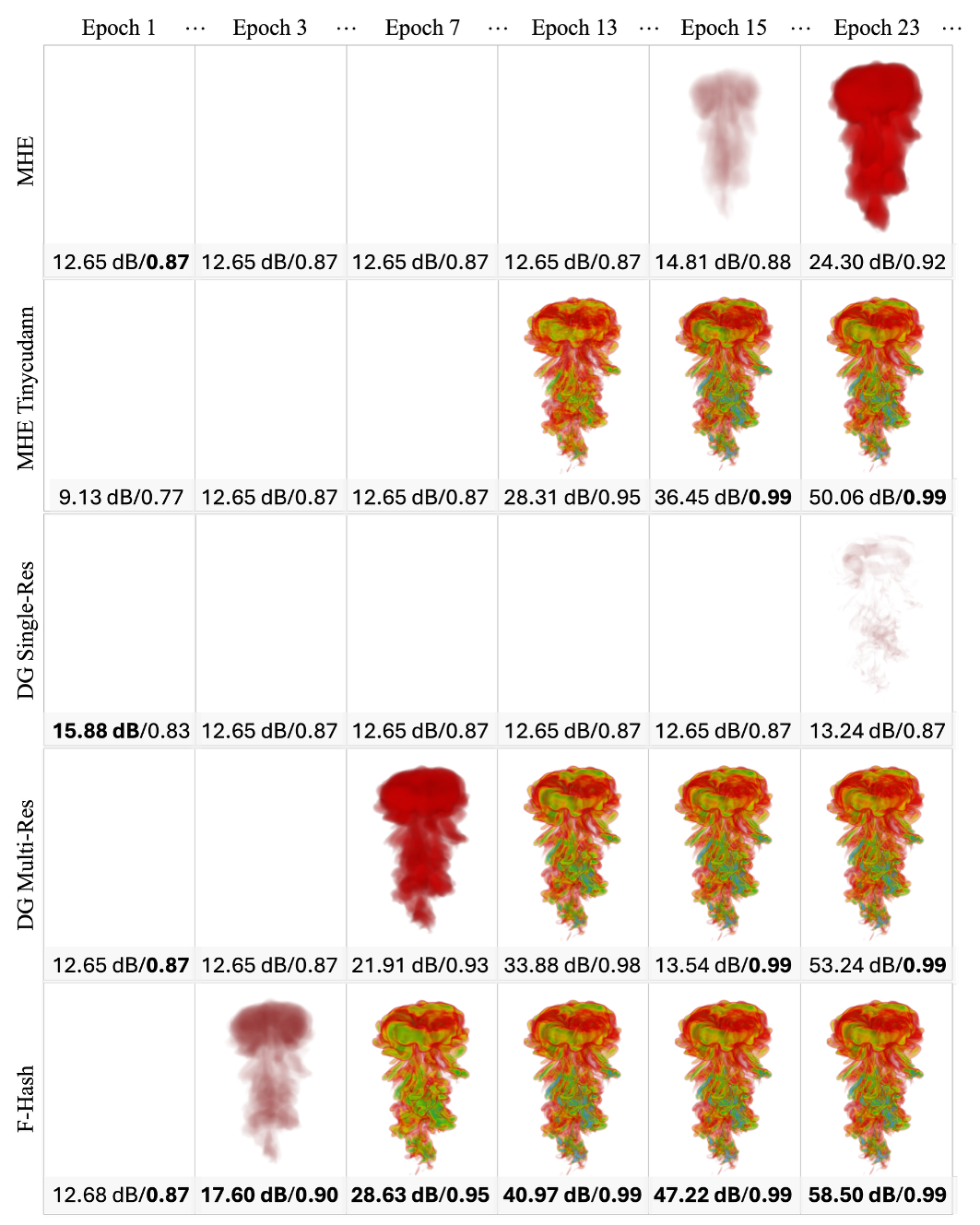}
            \caption{Segmentation}
            \label{fig:compare_argon_segment}
        \end{subfigure}
    \vspace{-3mm}
    \caption{Convergence visualization for the Argon Bubble dataset.}
    \label{fig:argon_convergence_images}    
\end{figure}

\begin{figure}[t]
    \centering
        \begin{subfigure}[b]{0.32\linewidth}
            \centering
            \includegraphics[trim=0 290 0 0,clip,width=\linewidth]{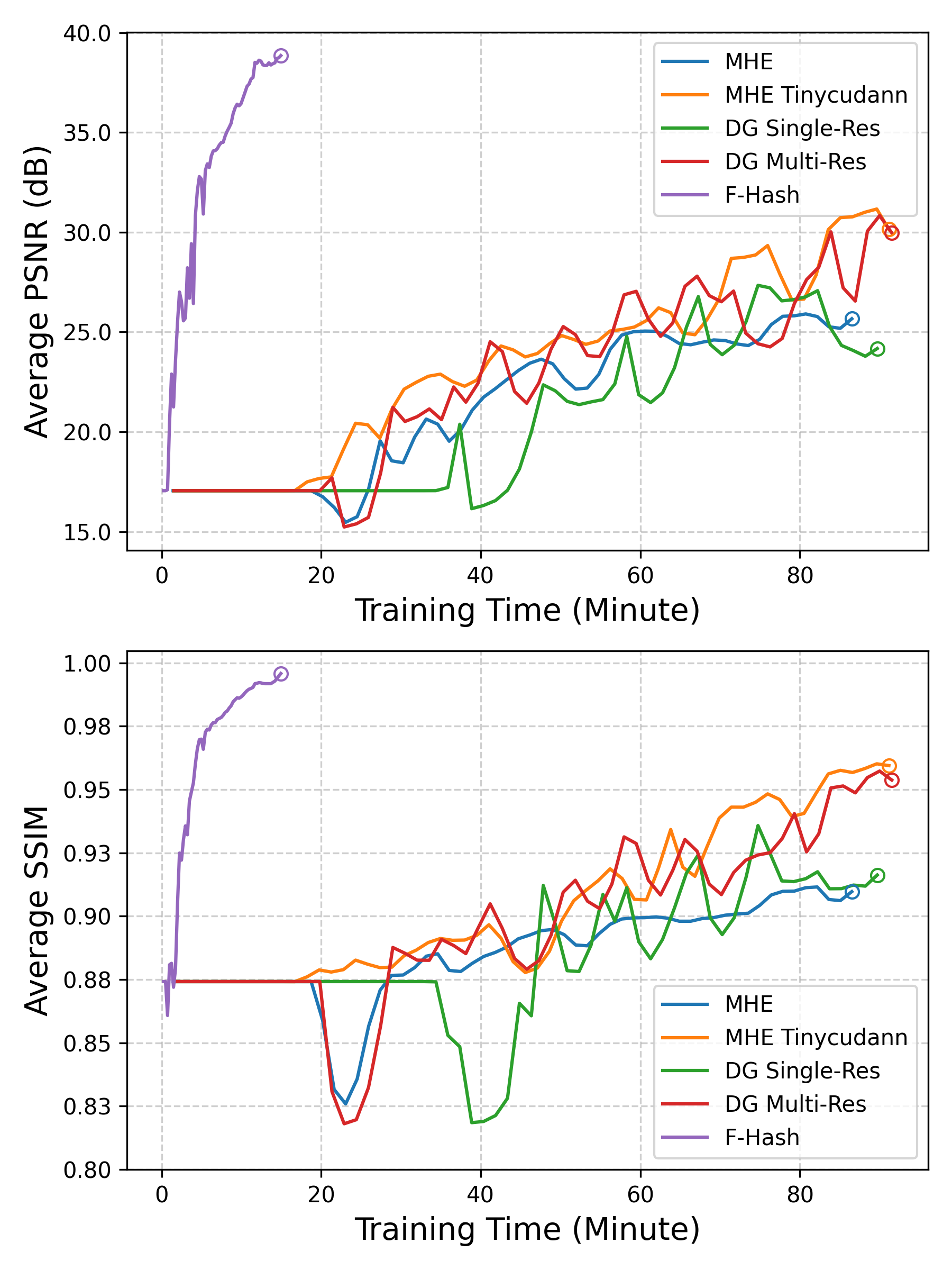}
            \caption{Interval}
            \label{fig:combustion_convergence_curves_a}
        \end{subfigure}
        \hfill
        \begin{subfigure}[b]{0.32\linewidth}
            \centering 
            \includegraphics[trim=0 290 0 0,clip,width=\linewidth]{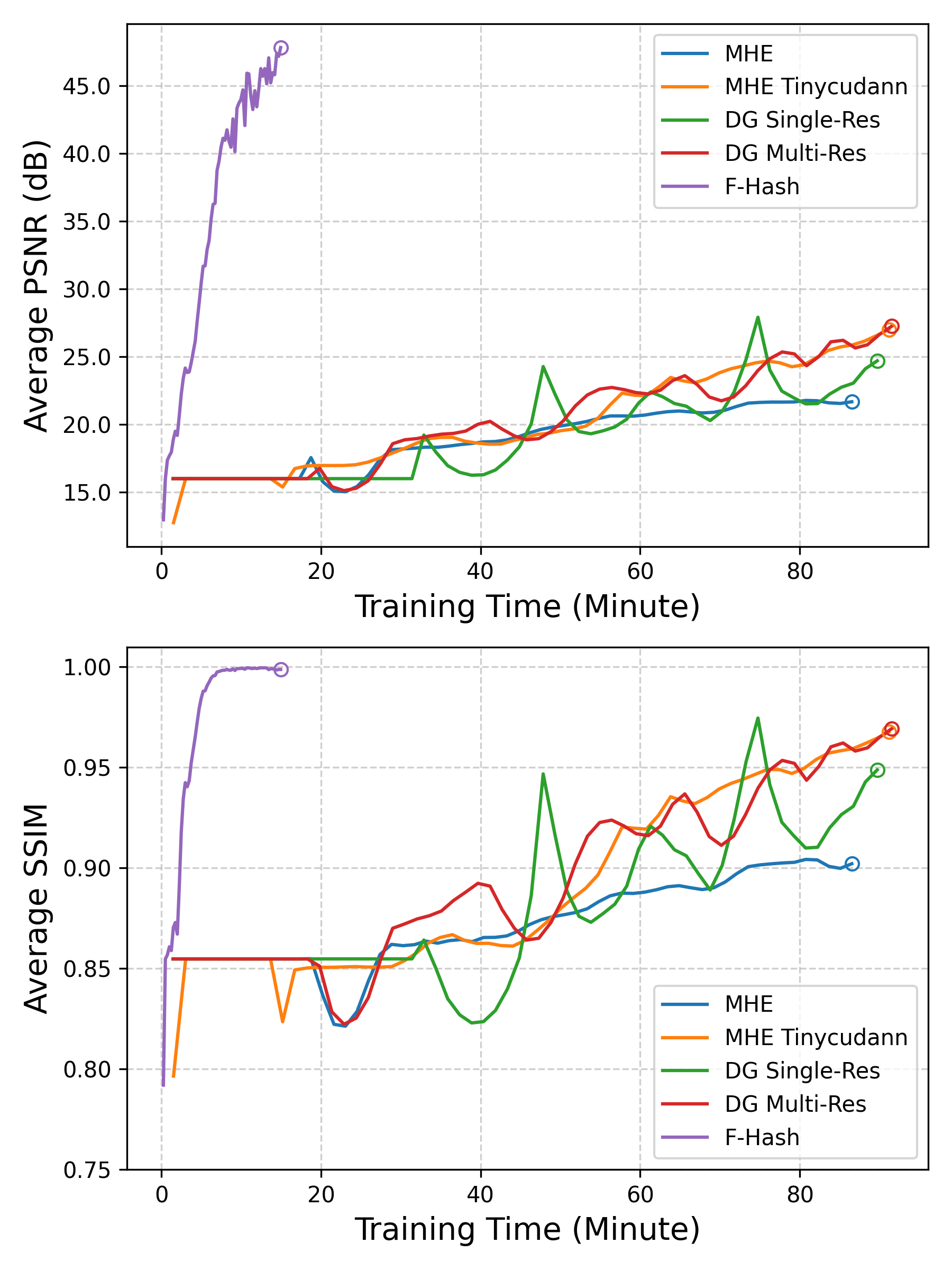}
            \caption{Isosurface}
            \label{fig:combustion_convergence_curves_b}
        \end{subfigure}
        \hfill
        \begin{subfigure}[b]{0.32\linewidth}
            \centering 
            \includegraphics[trim=0 290 0 0,clip,width=\linewidth]{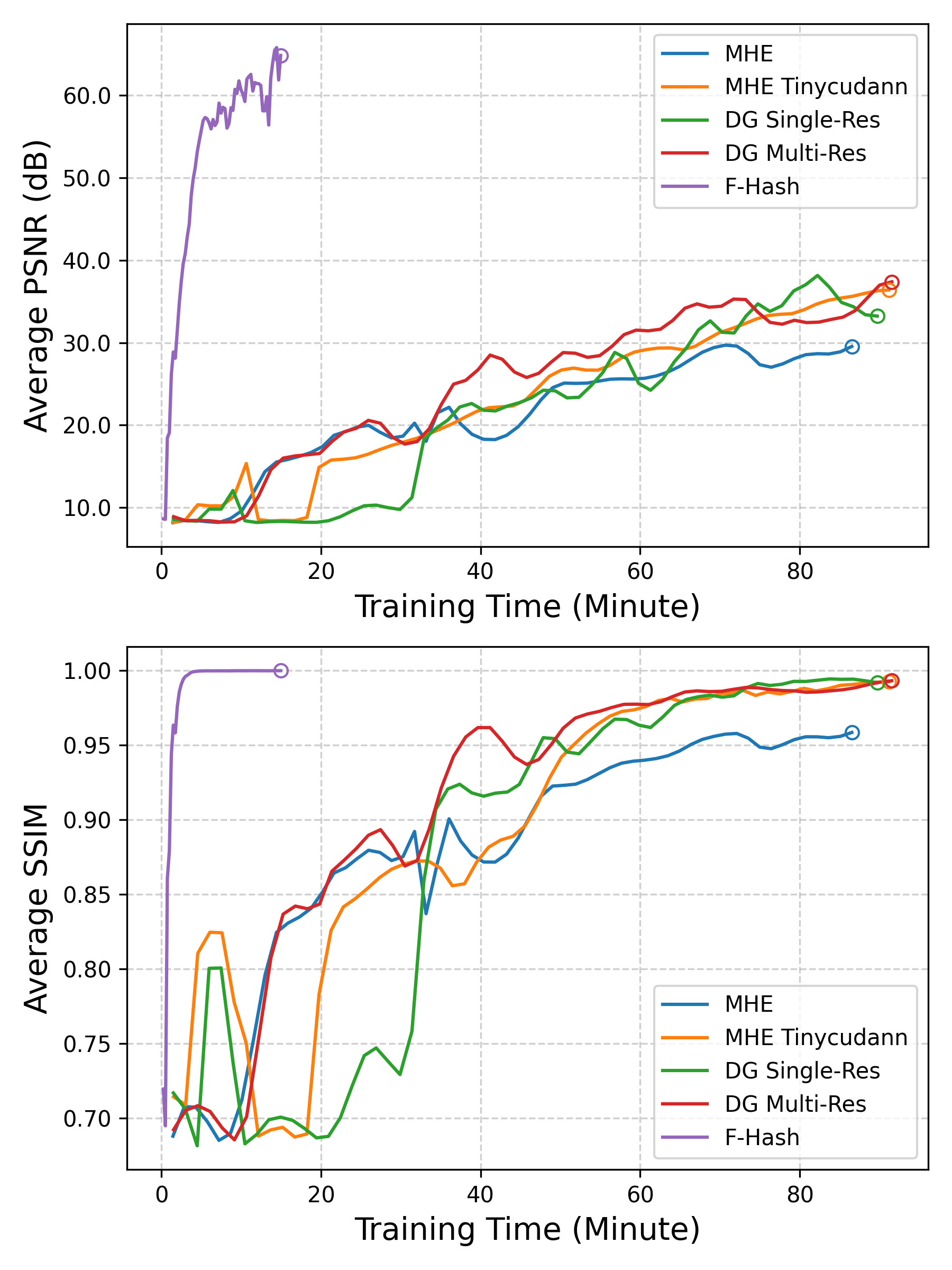}
            \caption{Segmentation}
            \label{fig:combustion_convergence_curves_c}
        \end{subfigure}
    \vspace{-3mm}
    \caption{Convergence speed for training the INR using different input encoding methods for various features of the Combustion dataset.}
    \label{fig:combustion_convergence_curves}    
\end{figure}

\begin{figure}[t]
    \centering
        \begin{subfigure}[b]{0.32\linewidth}
            \centering
            \includegraphics[trim=0 0 0 0,clip,width=\linewidth]{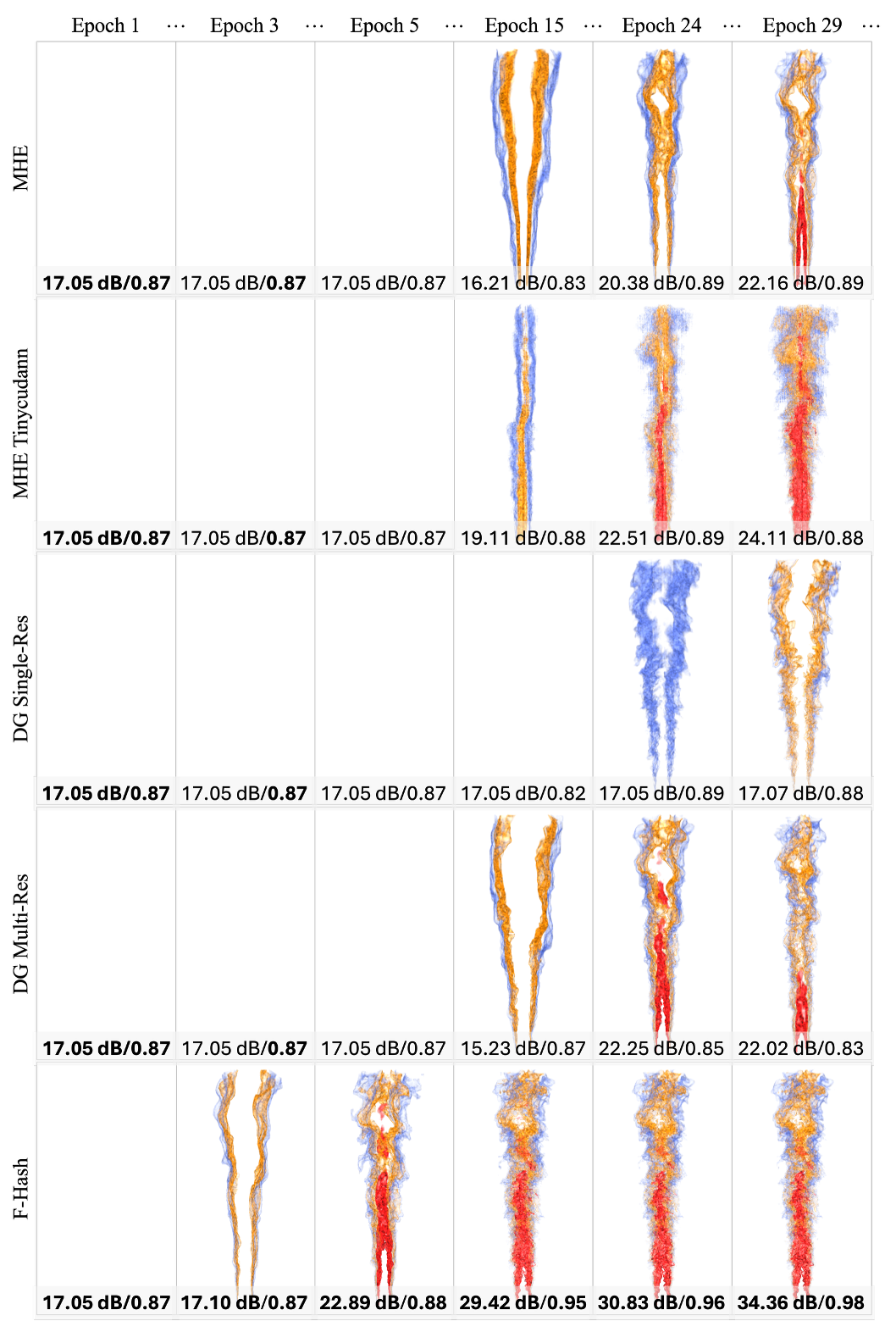}
            \caption{Interval}
            \label{fig:compare_combustion_interval}
        \end{subfigure}
        \hfill
        \begin{subfigure}[b]{0.32\linewidth}
            \centering 
            \includegraphics[trim=0 0 0 0,clip,width=\linewidth]{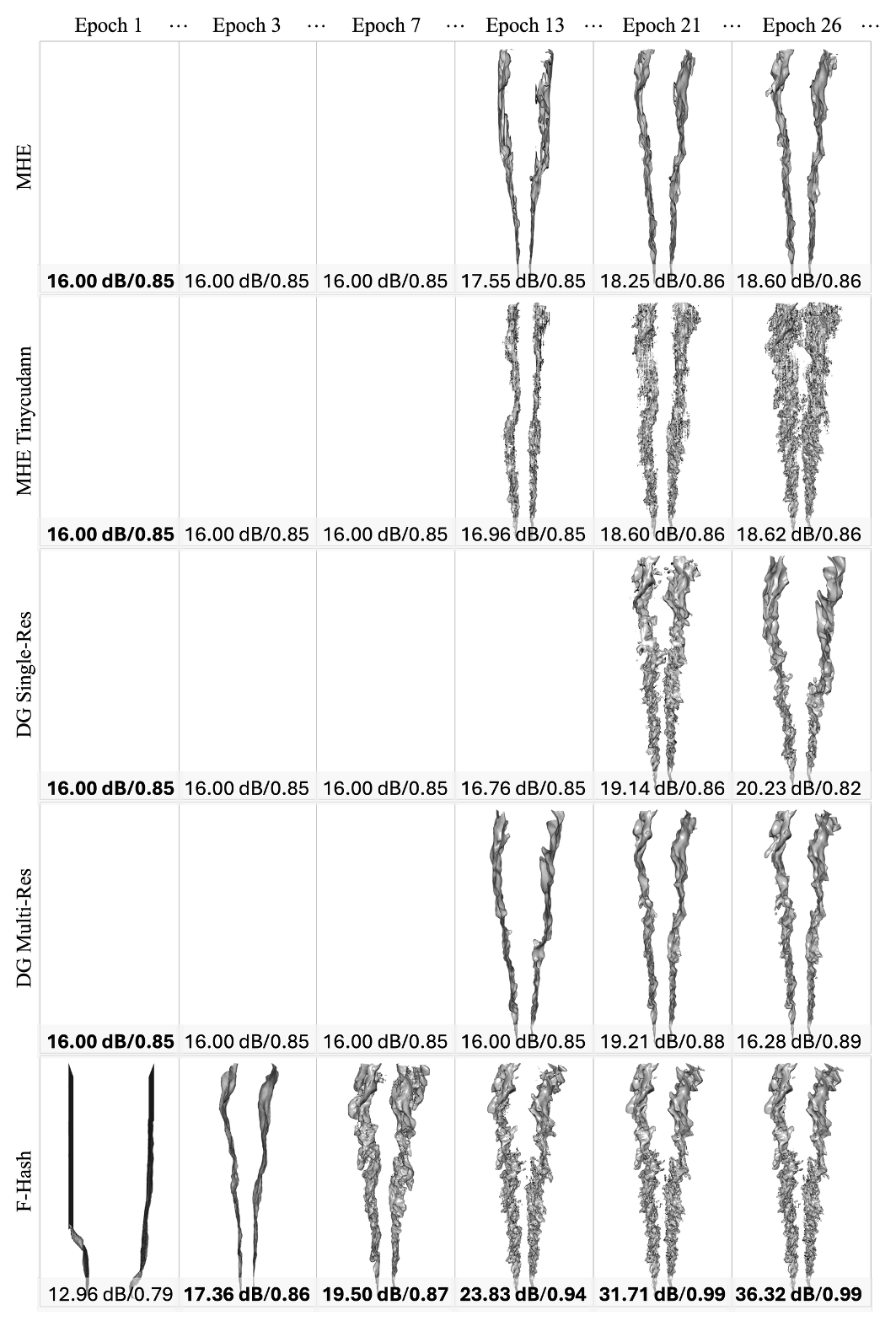}
            \caption{Isosurface}
            \label{fig:compare_combustion_iso}
        \end{subfigure}
        \hfill
        \begin{subfigure}[b]{0.32\linewidth}
            \centering 
            \includegraphics[trim=0 0 0 0,clip,width=\linewidth]{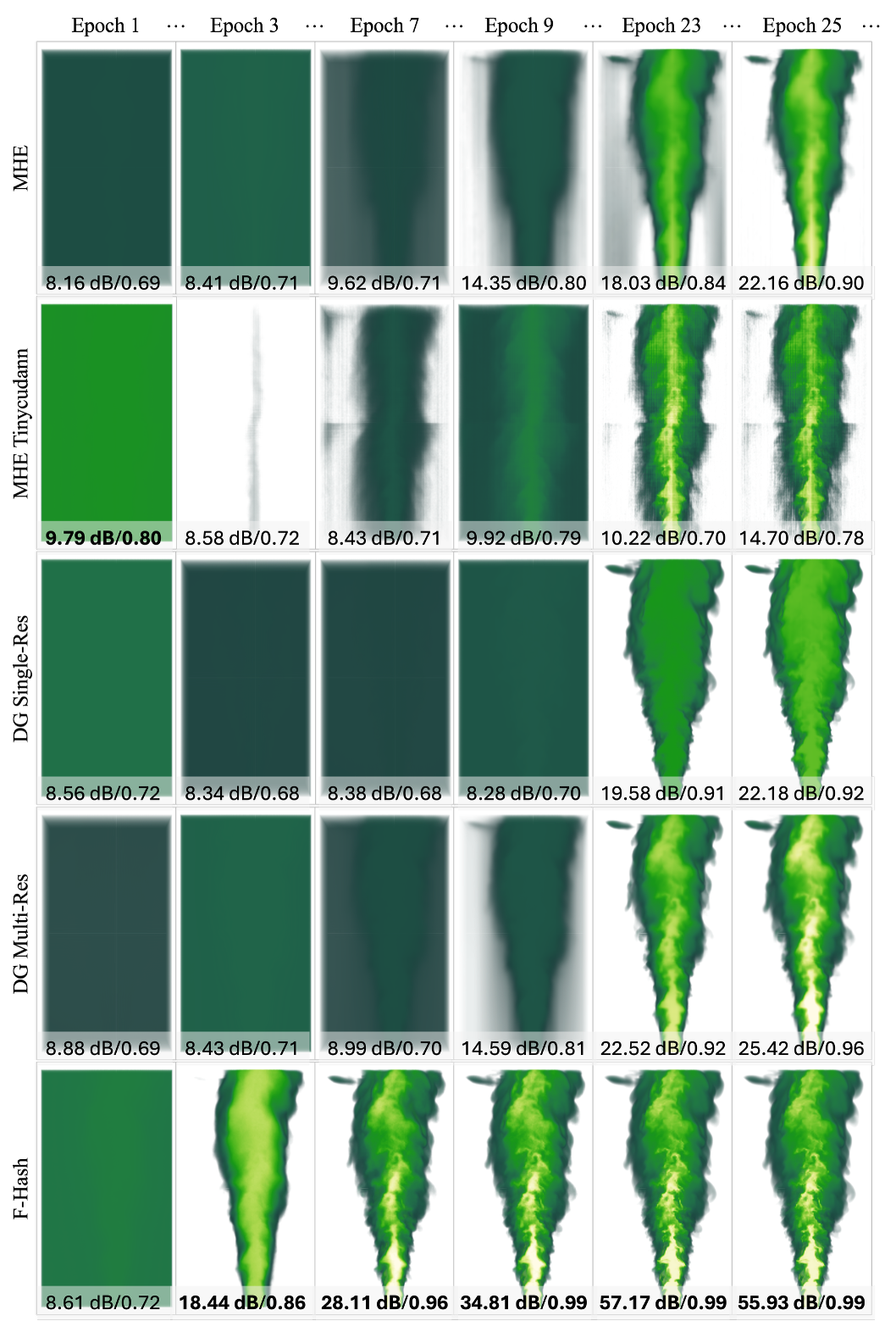}
            \caption{Segmentation}
            \label{fig:compare_combustion_segment}
        \end{subfigure}
    \vspace{-3mm}
    \caption{Convergence visualization for the Combustion dataset.}
    \label{fig:combustion_convergence_images}    
\end{figure}

\subsubsection{Training}
The INRs with input encoding are trained using the PyTorch software stack to accelerate the training and inferencing performance on a single NVIDIA RTX A6000 GPU. The Adaptive Moment Estimation (Adam) optimizer is used with an initial learning rate of 0.01 for all the experiments. The training stops at the 60th epoch/iteration. Although using a relatively small batch size makes the training faster, the loss curve will dramatically oscillate during the training due to the frequent updates of local gradient descent, making the convergence comparison through accuracy very random. However, a very large batch size slows down the training dramatically. We use the batch size of $2^{21}$ for all the experiments to balance the training time and training accuracy.

\begin{figure}[t]
    \centering
        \begin{subfigure}[b]{0.32\linewidth}
            \centering
            \includegraphics[trim=0 290 0 0,clip,width=\linewidth]{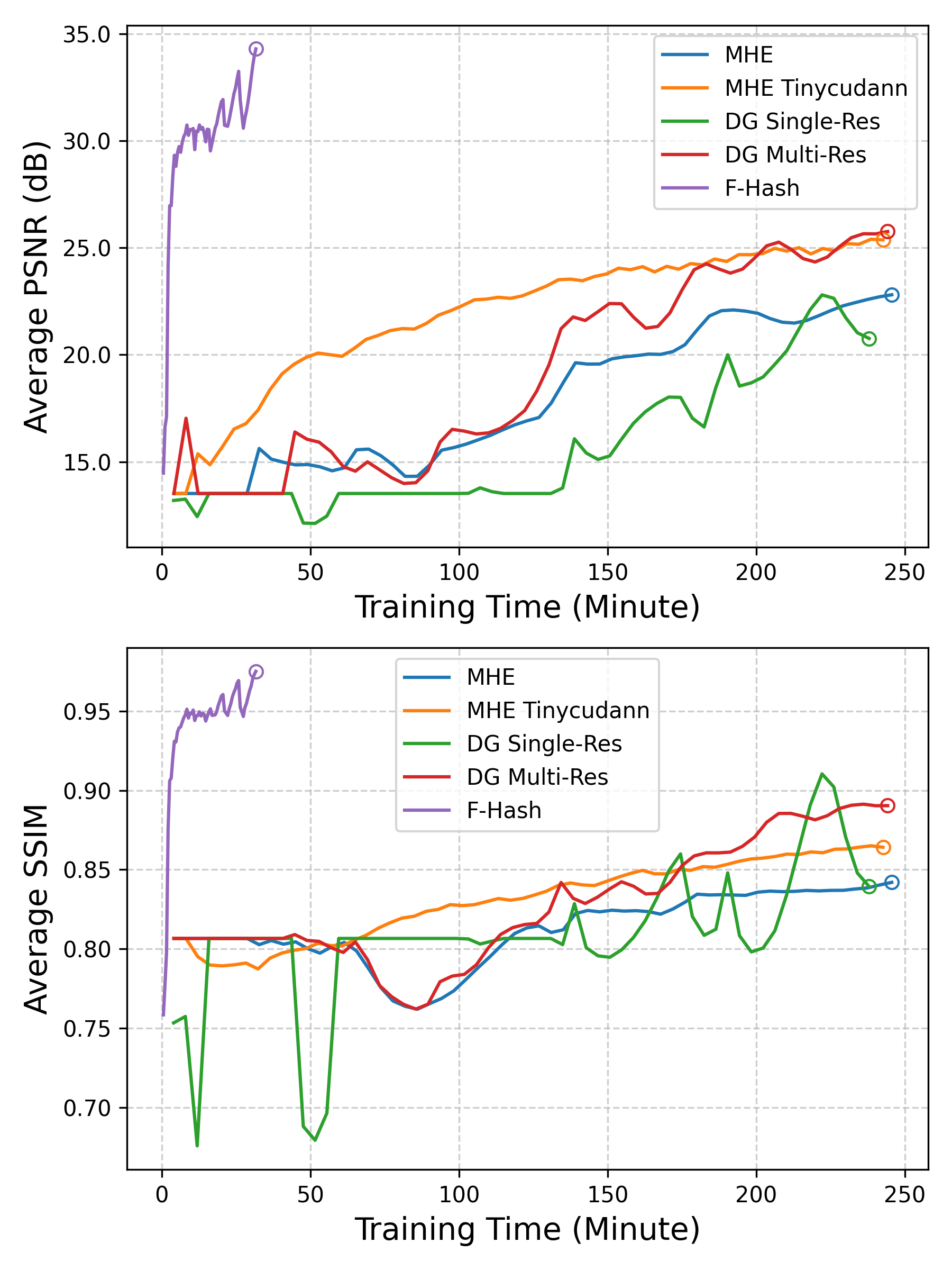}
            \caption{Interval}
            \label{fig:supernova_convergence_curves_a}
        \end{subfigure}
        \hfill
        \begin{subfigure}[b]{0.32\linewidth}
            \centering 
            \includegraphics[trim=0 290 0 0,clip,width=\linewidth]{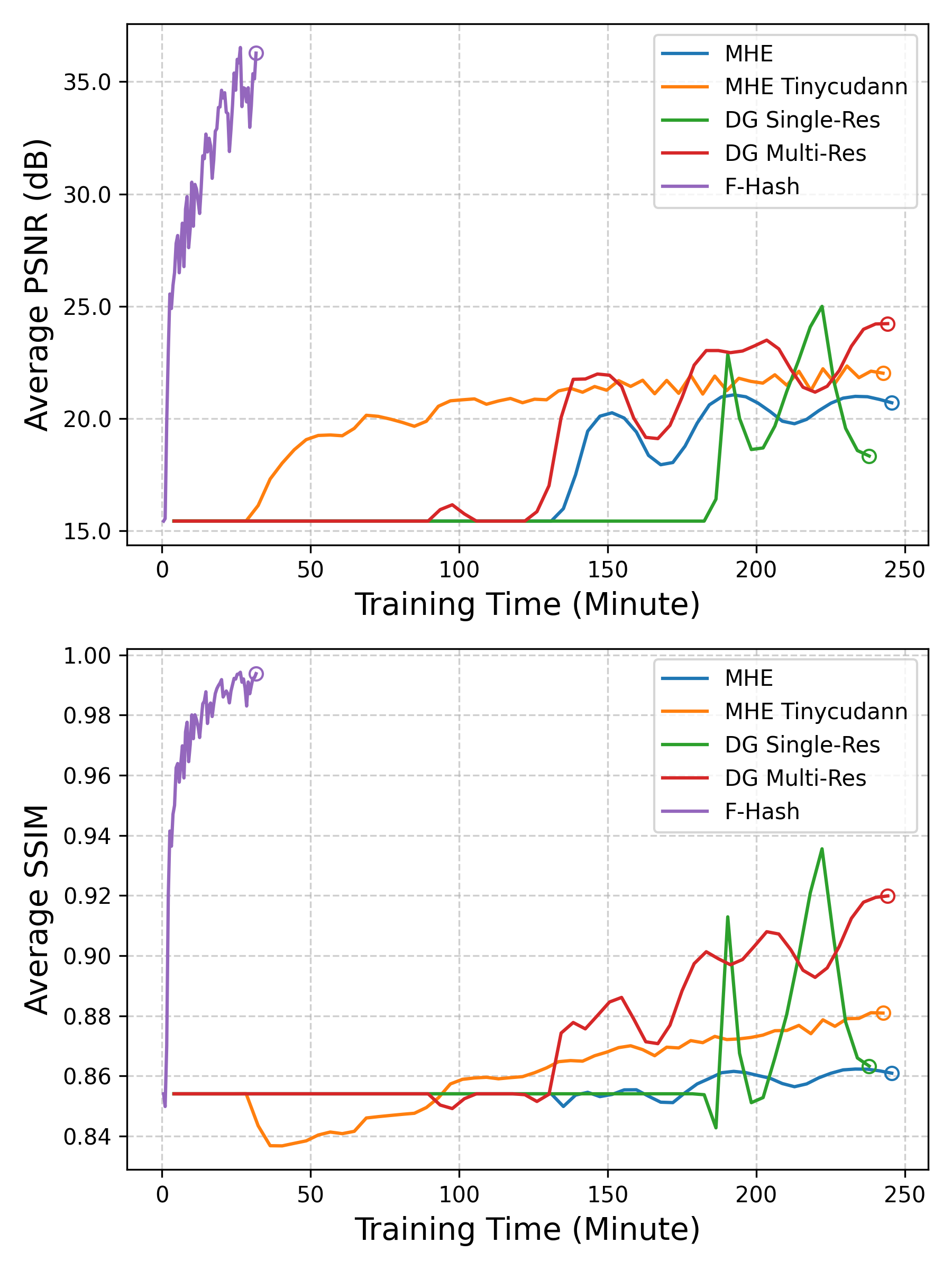}
            \caption{Isosurface}
            \label{fig:supernova_convergence_curves_b}
        \end{subfigure}
        \hfill
        \begin{subfigure}[b]{0.32\linewidth}
            \centering 
            \includegraphics[trim=0 290 0 0,clip,width=\linewidth]{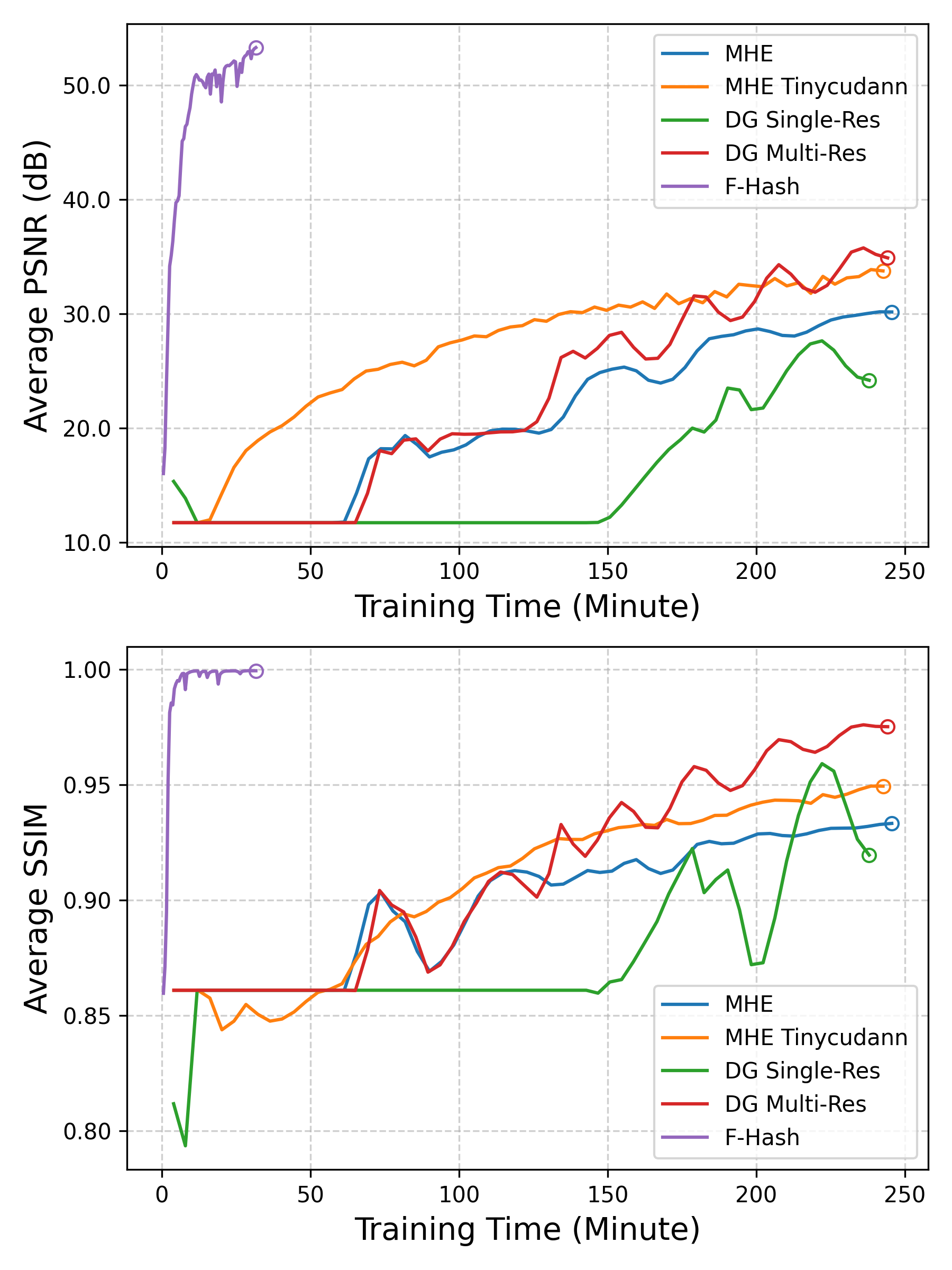}
            \caption{Segmentation}
            \label{fig:supernova_convergence_curves_c}
        \end{subfigure}
    \vspace{-3mm}
    \caption{Convergence speed for training the INR using different input encoding methods for various features of the Supernova dataset.}
    \label{fig:supernova_convergence_curves}    
\end{figure}

\begin{figure}[t]
    \centering
        \begin{subfigure}[b]{0.32\linewidth}
            \centering
            \includegraphics[trim=0 0 0 0,clip,width=\linewidth]{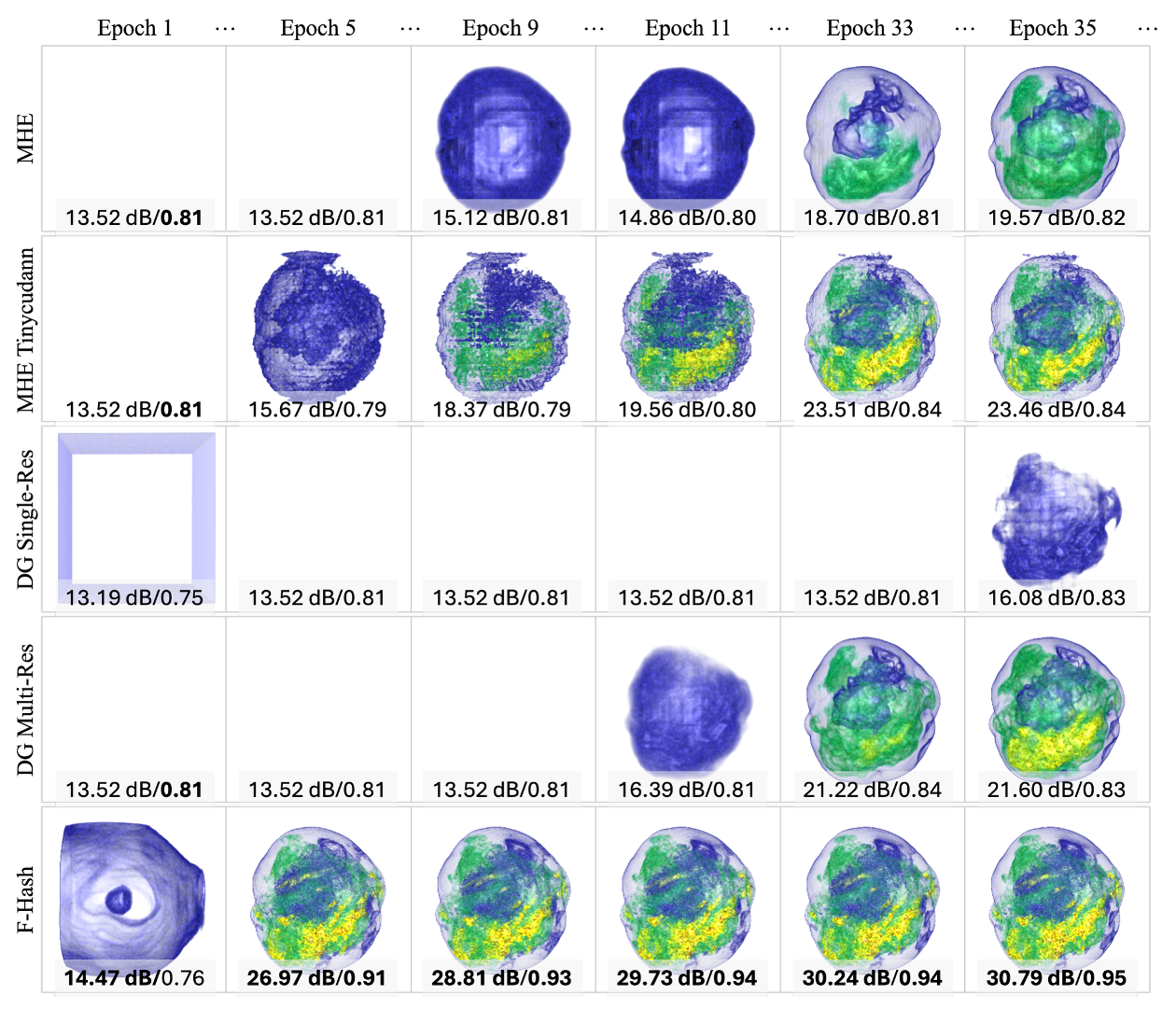}
            \caption{Interval}
            \label{fig:compare_supernova_interval}
        \end{subfigure}
        \hfill
        \begin{subfigure}[b]{0.32\linewidth}
            \centering 
            \includegraphics[trim=0 0 0 0,clip,width=\linewidth]{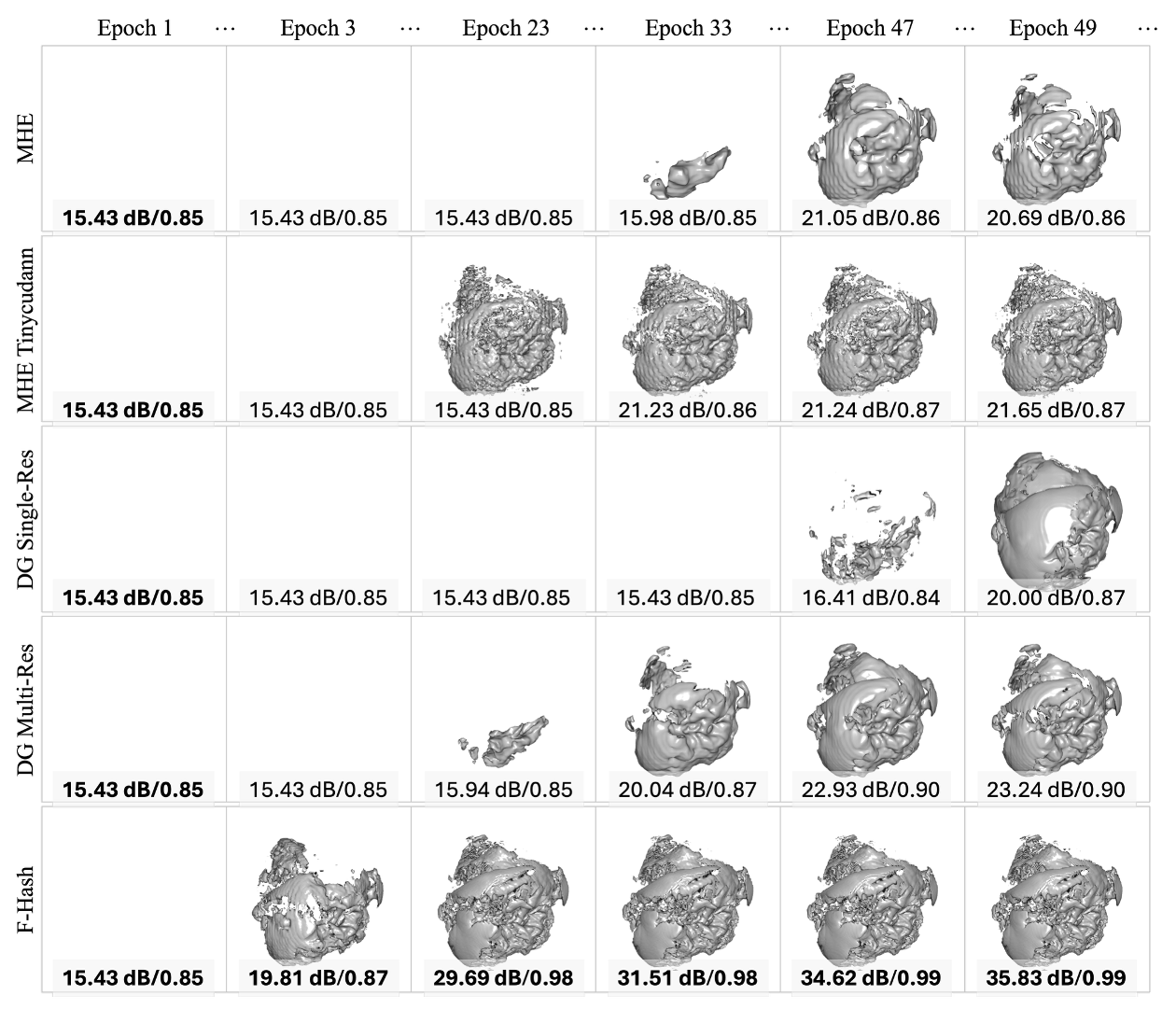}
            \caption{Isosurface}
            \label{fig:compare_supernova_iso}
        \end{subfigure}
        \hfill
        \begin{subfigure}[b]{0.32\linewidth}
            \centering 
            \includegraphics[trim=0 0 0 0,clip,width=\linewidth]{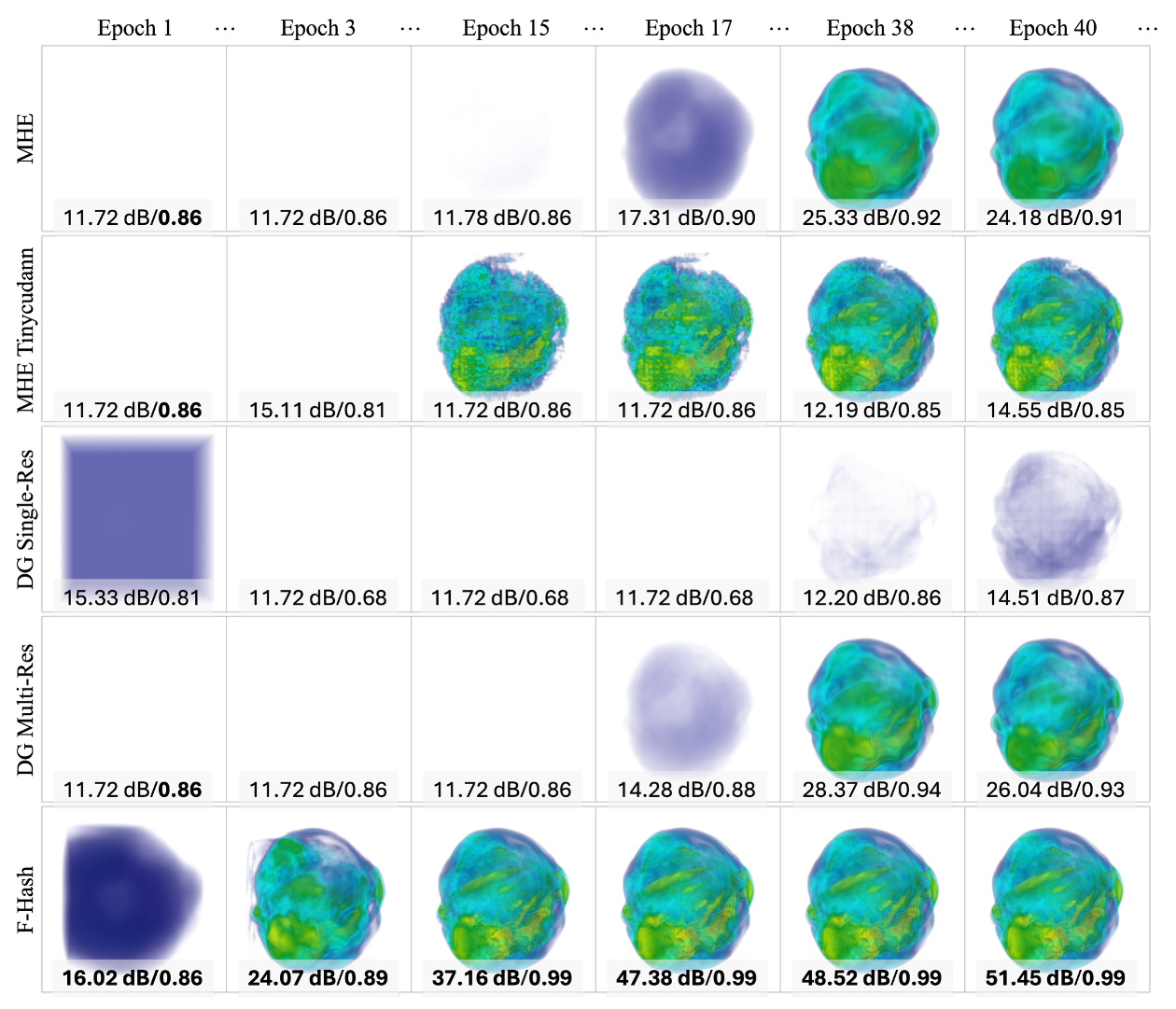}
            \caption{Segmentation}
            \label{fig:compare_supernova_segment}
        \end{subfigure}
    \vspace{-3mm}
    \caption{Convergence visualization for the Supernova dataset.}
    \label{fig:supernova_convergence_images}    
\end{figure}

\subsection{Input Encoding Evaluation}
\subsubsection{Convergence Time}
We train the INR using different input encoding methods and evaluate convergence speed for visualizing three types of features: interval, isosurface, and segmentation. The embedded Tesseract grid design enables F-Hash to represent time-varying volumetric data as a unified neural representation, unlike other input encoding methods that require separate modeling of each key frame. We track the training time until the 60th iteration for each method and evaluate the convergence progress based on the average reconstruction accuracy of the visualization image generated from all key frames for each feature. \cref{fig:argon_convergence_curves_a} shows how fast each method converges in terms of PSNR and SSIM as the training goes on for the interval feature of the Argon Bubble dataset. F-Hash is significantly faster (reaching high PSNR/SSIM quickly) than other methods that finish the training around the same time. Similar results can be observed in the isosurface (\cref{fig:argon_convergence_curves_b}) and segmentation (\cref{fig:argon_convergence_curves_c}) features. For the smaller Combustion dataset, all the methods can finish the training faster, as shown in \cref{fig:combustion_convergence_curves}. Due to the more complicated spatiotemporal structure of the dataset, other methods struggle to converge under 60 epochs, while F-Hash demonstrates superior efficiency in learning dynamic features. While the Supernova dataset, being both the largest and most dynamic, presents challenges for all methods in achieving high accuracy compared to the other two datasets, F-Hash still outperforms the other input encoding methods as shown in \cref{fig:supernova_convergence_curves}. The reasons for the observation are: 1) F-Hash's coreset selection reduces the total number of training samples, resulting in a less number of batches for each training iteration. 2) The proposed multi-resolution Tesseract embedding grid is capable of learning complex spatiotemporal features with a smaller number of trainable encoding parameters, resulting in faster backpropagation during training. 3) Our design of Tesseract embedding grid jointly considers both spatial and temporal information, providing a more informative high-dimensional representation for the MLP to learn. \cref{fig:argon_convergence_images}, \cref{fig:combustion_convergence_images}, and \cref{fig:supernova_convergence_images} provide visualizations of convergence through the generated rendering image for each feature of the three datasets, which provides an informative insight about how each input encoding method prioritizes which region of data to converge first. The two MHE methods converge from the regions with a lower value range, and the MHE with Tinycudann is more efficient in capturing high-frequency details. The DG Multi-Res tends to converge a more global region, while the DG Single-Res starts by converging regions with more random locations. F-Hash can quickly converge for the entire feature from low- to high-value regions. The qualitative evaluation of the convergence time for a given reconstruction accuracy is detailed in the left-hand side \cref{tab:convergence_time_and_encoding_parameter}. For all selected levels of PSNR accuracy (20, 30, and 40 dB), F-Hash is the fastest to converge. 

\subsubsection{Number of Parameter}
The configuration of each input encoding method while modeling different datasets is listed on the right-hand side of \cref{tab:convergence_time_and_encoding_parameter}. The fold parameter of F-Hash is set to 2 for all the experiments for the highest modeling capacity. MHE-based methods adjust the hash table size to match the volume size of the frame. The number of resolution levels also increases when handling larger time-varying volumetric data. Since the FBB is adaptive to the specific feature, enabling efficient encoding through dynamic adjustment of the number of input encoding parameters when handling interval, isosurface, or segmentation features. As a result, F-Hash has the least number of input encoding parameters and achieves the shortest convergence time for all testing datasets.

\begin{table*}[t]
  \caption{Left is the convergence time measurement. The budget of the maximal training iteration allowed is set to 60. If a method can't reach a specific PSNR within the training budget, the time is marked as Not Reached (NR). On the right, the first two rows are the encoding configuration, while the last row is the number of parameters in mebi-unit (M).}
  \label{tab:convergence_time_and_encoding_parameter}
  \scriptsize%
	\centering%
  \begin{adjustbox}{width=\textwidth}
      \begin{tabu}{ c | c | c c c c c c c c c c c c c c c | c c c c c c c }
      \toprule
      \multirow{3}{*}{Dataset} & \multirow{3}{*}{PSNR} & \multicolumn{15}{|c|}{Convergence Time (min) $\downarrow$} & \multicolumn{7}{|c}{Encoding Configuration / Parameter Size (M) $\downarrow$} \\
        & &
      \multicolumn{3}{|c|}{MHE} &
      \multicolumn{3}{|c|}{MHE Tinycudann} &
      \multicolumn{3}{|c|}{DG Single-Res} &
      \multicolumn{3}{|c|}{DG Multi-Res} &
      \multicolumn{3}{|c|}{F-Hash} &
      \multicolumn{1}{|c|}{MHE} &
      \multicolumn{1}{|c|}{MHE Tinycudann} &
      \multicolumn{1}{|c|}{DG Single-Res} &
      \multicolumn{1}{|c|}{DG Multi-Res} &
      \multicolumn{3}{|c}{F-Hash} \\
        & &
      \multicolumn{3}{|c|}{Int  \ \ \ \ \ \ \ \ \ Iso \ \ \ \ \ \ \ \ \ Seg} &
      \multicolumn{3}{|c|}{Int  \ \ \ \ \ \ \ \ \ Iso \ \ \ \ \ \ \ \ \ Seg} &
      \multicolumn{3}{|c|}{Int  \ \ \ \ \ \ \ \ \ Iso \ \ \ \ \ \ \ \ \ Seg} &
      \multicolumn{3}{|c|}{Int  \ \ \ \ \ \ \ \ \ Iso \ \ \ \ \ \ \ \ \ Seg} &
      \multicolumn{3}{|c|}{Int  \ \ \ \ \ \ \ \ \ Iso \ \ \ \ \ \ \ \ \ Seg} &
      \multicolumn{1}{|c|}{} &
      \multicolumn{1}{|c|}{} &
      \multicolumn{1}{|c|}{} &
      \multicolumn{1}{|c|}{} &
      \multicolumn{3}{|c}{Int  \ \ \ \ \ \ \ \ \ \ \ \ \ \ \ \ \ \ \ \ \ \ Iso \ \ \ \ \ \ \ \ \ \ \ \ \ \ \ \ \ \ \ \ \ \ Seg} \\
      \midrule
       \multirow{3}{*}{Combustion} & 20 dB & 
      33.2 & 51.9 & 31.7 & 
      24.3 & 54.7 & 45.6 & 
      37.4 & 46.3 & 35.9 & 
      29.0 & 39.7 & 25.9 & 
      \textbf{1.0}  & \textbf{2.2} & \textbf{1.2} & 
      \multirow{1}{*}{Hash table size = $2^{20}$} & \multirow{1}{*}{Hash table size = $2^{20}$} & \multirow{1}{*}{} & \multirow{1}{*}{} & \multirow{1}{*}{Fold = 2} & \multirow{1}{*}{Fold = 2} & \multirow{1}{*}{Fold = 2} \\
      \cmidrule(lr){2-17}
       & 30 dB &
       NR & NR & NR &
       83.6 & NR & 86.6 &
       NR & NR & 67.2 &
       83.9 & NR & 58.0 &
       \textbf{4.2} & \textbf{5.0} & \textbf{2.0} 
       & Res level = 6  & Res level = 6  & Res level = 6  & Res level = 6 & Res level = 6  & Res level = 6  & Res level = 6  \\
      \cmidrule(lr){2-17}
       & 40 dB &
       NR & NR & NR &
       NR & NR & NR &
       NR & NR & NR &
       NR & NR & NR &
       NR & \textbf{7.4} & \textbf{2.9} 
       & 120.0 M & 45.8 M & 45.0 M & 51.5 M & \textbf{16.1} M &  \textbf{17.4} M & \textbf{24.2} M  \\
       \midrule
       \multirow{3}{*}{Argon Bubble} & 20 dB & 
      22.6 & 30.2 & 30.2 & 
      21.2 & 23.2 & 23.1 & 
      37.5 & 44.4 & 46.1 & 
      16.6 & 14.5 & 14.5 & 
      \textbf{2.2}  & \textbf{1.1}  & \textbf{1.1} & 
      \multirow{1}{*}{Hash table size = $2^{21}$} & \multirow{1}{*}{Hash table size = $2^{21}$} & \multirow{1}{*}{} & \multirow{1}{*}{} & \multirow{1}{*}{Fold = 2} & \multirow{1}{*}{Fold = 2} & \multirow{1}{*}{Fold = 2} \\
      \cmidrule(lr){2-17}
       & 30 dB &
       NR   & NR   & NR &
       28.9 & 32.8 & 26.9 &
       61.4 & 59.7 & 59.7 &
       29.0 & 29.0 & 27.0 &
       \textbf{3.4}  & \textbf{3.1}  & \textbf{2.2} 
       & Res level = 7 & Res level = 7  & Res level = 7  & Res level = 7 & Res level = 6  & Res level = 7  & Res level = 7  \\
      \cmidrule(lr){2-17}
       & 40 dB &
       NR   &  NR  & NR &
       48.2 & 73.2 & 36.6 &
       80.2 & 63.1 & 68.3 &
       43.6 & 49.8 & 35.3 &
       \textbf{4.8}  & \textbf{5.9}  & \textbf{3.4} 
       & 252.0 M & 77.2 M & 72.0 M & 82.3 M & \textbf{16.7 M}  & \textbf{19.5 M}  & \textbf{19.7 M}  \\
       \midrule
       \multirow{3}{*}{Supernova} & 20 dB & 
      163.7 & 147.3 & 135.1 & 
      52.6 & 68.7 & 40.4 & 
      190.4 & 190.3 & 186.4 & 
      134.3 & 134.3 & 126.2 & 
      \textbf{2.1}   & \textbf{2.1}   & \textbf{1.6} & 
      \multirow{1}{*}{Hash table size = $2^{23}$} & \multirow{1}{*}{Hash table size = $2^{23}$} & \multirow{1}{*}{} & \multirow{1}{*}{} & \multirow{1}{*}{Fold = 2} & \multirow{1}{*}{Fold = 2} & \multirow{1}{*}{Fold = 2} \\
      \cmidrule(lr){2-17}
       & 30 dB &
       NR  & NR   & 237.4 &
       NR  & NR   & 137.5 &
       NR  & NR   & NR &
       NR  & NR   & 179.1 &
       \textbf{7.4} & \textbf{10.0} & \textbf{2.6} 
       & Res level = 7  & Res level = 7  & Res level = 7  & Res level = 7 & Res level = 8  & Res level = 8  & Res level = 8  \\
      \cmidrule(lr){2-17}
       & 40 dB &
       NR & NR & NR &
       NR & NR & NR &
       NR & NR & NR &
       NR & NR & NR &
       NR & NR & \textbf{5.8}
       & 1120.0 M  & 205.8 M & 135.0 M & 154.3 M & \textbf{108.2 M}  & \textbf{86.9 M}  & \textbf{107.7 M}  \\
      \bottomrule
      \end{tabu}
 \end{adjustbox}
\end{table*}

\subsection{Performance Evaluation}
\subsubsection{Reconstruction Accuracy}
\begin{figure*}[t]
    \centering
    \begin{subfigure}[b]{0.337\linewidth}
        \centering
        \includegraphics[trim=2 2 2 2,clip,width=\linewidth]{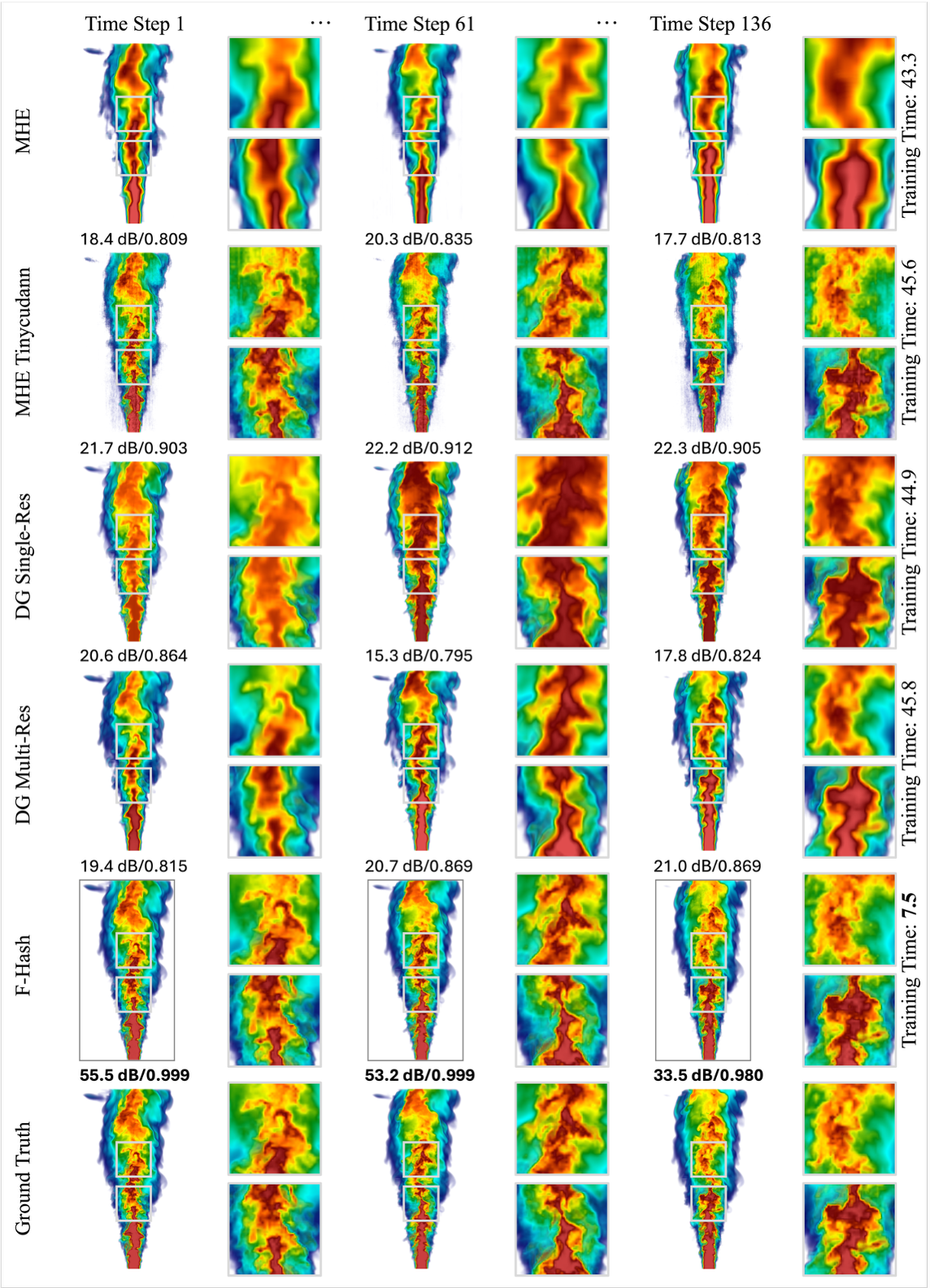}
        \caption{Combustion}
        \label{fig:animation_combustion}
    \end{subfigure}
    \begin{subfigure}[b]{0.32\linewidth}
        \centering 
        \includegraphics[trim=9 2 2 2,clip,width=\linewidth]{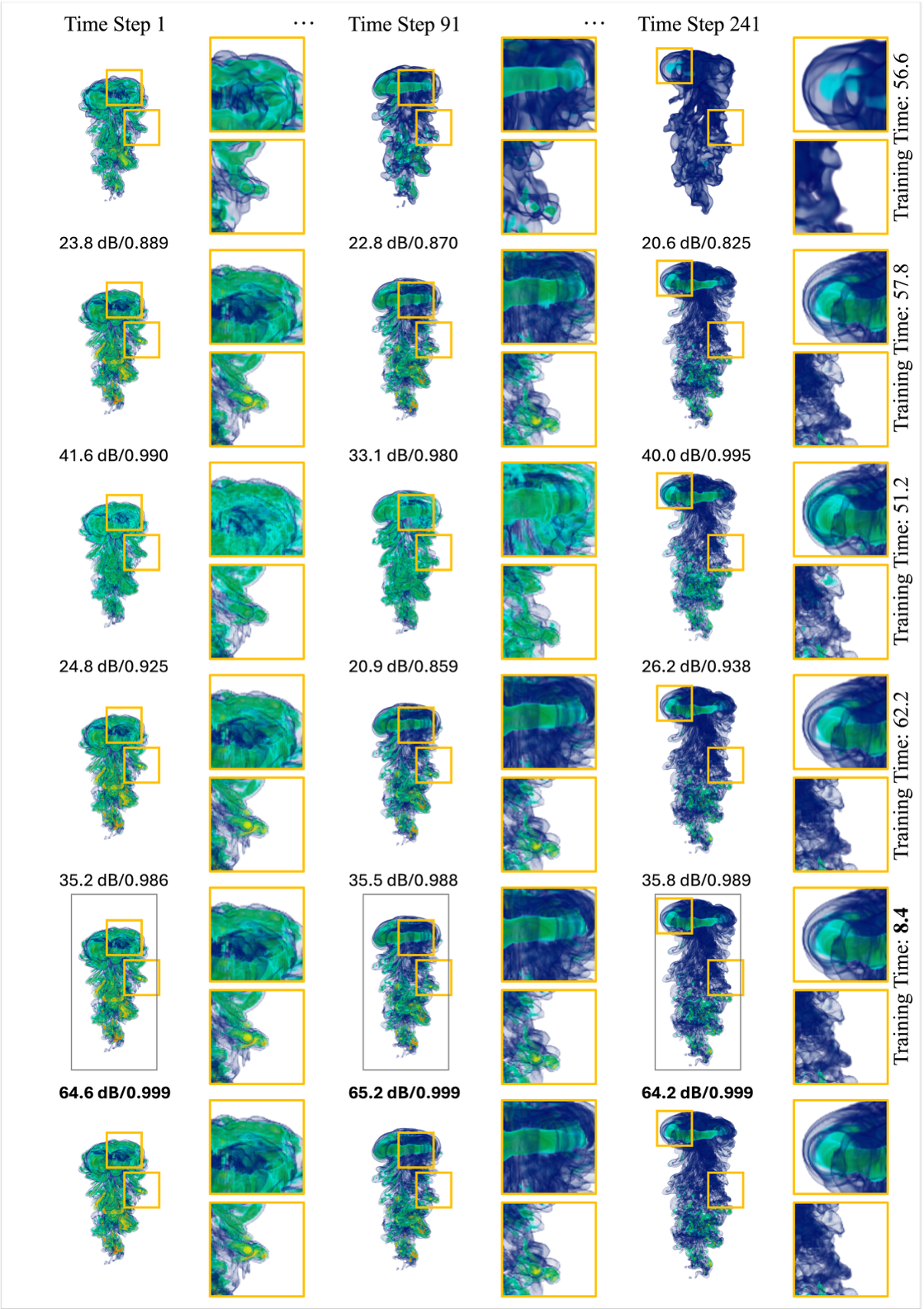}
        \caption{Argon Bubble}
        \label{fig:animation_argon}
    \end{subfigure}
    \begin{subfigure}[b]{0.32\linewidth}
        \centering 
        \includegraphics[trim=9 2 2 2,clip,width=\linewidth]{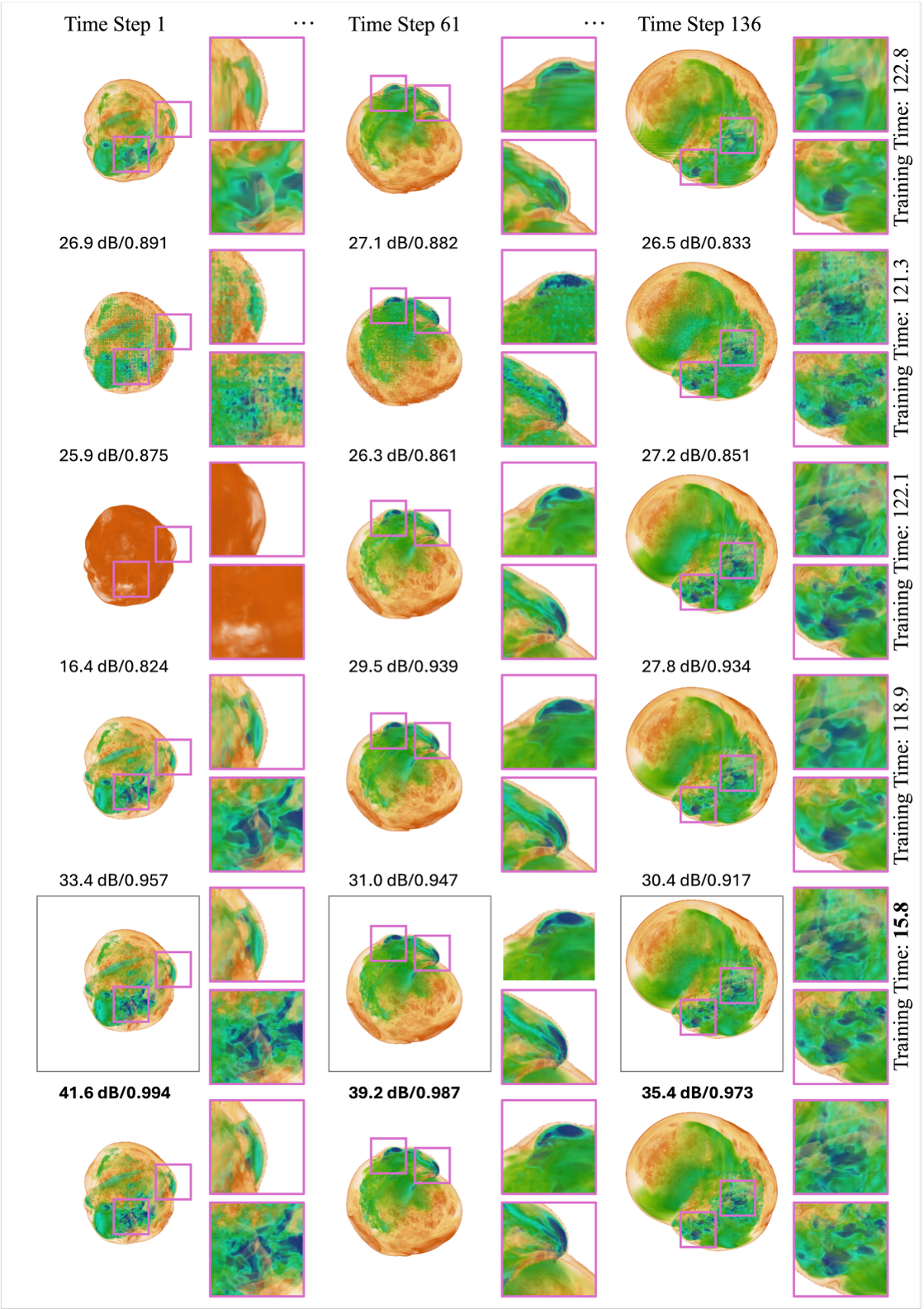}
        \caption{Supernova}
        \label{fig:animation_supernova}
    \end{subfigure}
    \caption{Evolution visualization using different input encoding methods. Reconstruction accuracy is labeled in the format of PSNR/SSIM for each rendering. For the F-Hash, 
    3D feature region (FBB) for each datasets are labeled as the gray bounding boxes with dimension of $80\times 83\times 204$, $200\times 110\times 54$, and $156\times 183\times 185$, which takes $63.95\%$, $32.30\%$, $52.41\%$ of the Combustion, Argon Bubble, and Supernova frame volume.}
    \label{fig:animation}    
\end{figure*}

In this evaluation, we let all input encoding methods finish for the same number of iterations (30th) and compare their performances on reconstruction accuracy and training time. We use the popular evolution visualization as the time-varying visualization task on the segmentation feature. \cref{fig:animation} demonstrates a visual comparison of reconstruction accuracy of frames over time for the three testing datasets. More frames are shown in \cref{fig:teaser} with convergence speed up for 30 dB PSNR and encoding parameters reduction. F-Hash achieves the highest reconstruction accuracy compared to other input encoding methods while maintaining a short training time. Quantitative measurements of average PSNR/SSIM across frames and training time are listed in \cref{tab:animation}. We can observe that, together with the model size, the size of the F-Hash feature region (coreset) plays a key role in determining the training time. A larger feature region (coreset) tends to have a longer training time.

\subsubsection{Compression}
Since only a subset (key frames) of the total frames are considered to construct the INR. The trained INR also performs a compression to the original large-scale time-varying volumetric data. The compression ratio of F-Hash is the highest among all the input encoding methods as listed in the left-hand side of \cref{tab:compression}. However, compared with specialized volume compressors like the state-of-the-art learning-based compression methods, NeurComp~\cite{lu2021compressive}\footnote{https://github.com/matthewberger/neurcomp}, and lossy compression methods, SZ3~\cite{9866018}\footnote{https://github.com/szcompressor/SZ3} and TTHRESH~\cite{8663447}\footnote{https://github.com/rballester/tthresh}, F-Hash falls short of these approaches in compressing large-scale volumetric data.


\begin{table}[t]
  \caption{Measurements of average PSNR/SSIM across frames and training time.}
  \label{tab:animation}
  \scriptsize%
	\centering%
  \begin{adjustbox}{width=0.48\textwidth}
      \begin{tabu}{ c | c c c c c  }
      \toprule
      \multirow{2}{*}{Dataset} & \multicolumn{5}{c}{Avg PSNR (dB) $\uparrow$ / Avg SSIM $\uparrow$/ Training Time (min) $\downarrow$}  \\
        &
      MHE &
      MHE Tinycudann &
      DG Single-Res &
      DG Multi-Res &
      F-Hash\\
      \midrule
      Combustion & 18.9 / 0.818 / 43.4  & 23.7 / 0.906 / 45.7 & 20.2 / 0.829 / 44.8  & 20.5 / 0.850 / 45.9 & \textbf{47.1} / \textbf{0.994} / \textbf{7.4} \\
      \midrule
      Argon Bubble & 22.3 / 0.862 / 56.5 & 28.3 / 0.987 / 57.9 & 23.9 / 0.908 / 51.1 & 35.6 / 0.987 / 62.3 & \textbf{64.6} / \textbf{0.999} / \textbf{8.5} \\
      \midrule
      Supernova & 26.7 / 0.870 / 122.7 & 26.6 / 0.861 / 121.4 & 24.5 / 0.899 / 122.2 & 31.7 / 0.941 / 118.8 & \textbf{38.7} / \textbf{0.986} / \textbf{15.7} \\
      \bottomrule
      \end{tabu}
 \end{adjustbox}
\end{table}

\begin{table}[t]
  \caption{Compression ratio of different input encoding methods, and specialized compression methods with $\geq 45dB$ average frame PSNR.}
  \label{tab:compression}
  \scriptsize%
	\centering%
  \begin{adjustbox}{width=0.48\textwidth}
      \begin{tabu}{ c | c | c | c | c | c c c | c | c | c}
      \toprule
      \multirow{2}{*}{Dataset} & \multirow{2}{*}{MHE} & \multicolumn{1}{c|}{MHE}  & \multicolumn{1}{c|}{DG} & \multicolumn{1}{c|}{DG} & \multicolumn{3}{c|}{F-Hash} & \multirow{2}{*}{NeurComp} & \multirow{2}{*}{SZ3} & \multirow{2}{*}{TTHRESH} \\
        & & Tinycudann & Single-Res  & Multi-Res & Int \ & Iso & Seg & &\\
      \midrule
      Combustion & $1.9\times$  & $3.5\times$ & $5.1\times$  & $4.5\times$ & $14.3\times$ & $13.2\times$ & $9.5\times$ & $1407.3\times$ & $51.4\times$ & $1023.3 \times$\\
      \midrule
      Argon Bubble & $3.7\times$ & $6.2\times$ & $12.8\times$ & $11.2\times$ & $55.2\times$ & $47.1\times$ & $46.8\times$ & $1223.8\times$ & $54.6\times$ & $1107.4\times$\\
      \midrule
      Supernova & $1.1\times$ & $2.4\times$ & $9.2\times$ & $8.1\times$ & $11.5\times$ & $14.3\times$ & $11.6\times$ & $1371.2\times$ & $61.7\times$ & $1240.3\times$\\
      \bottomrule
      \end{tabu}
 \end{adjustbox}
\end{table}

\subsubsection{Rendering}
We also evaluate the rendering latency using our proposed ARM methods. We measure the average rendering time of generating a visualization image from a random view on a specific frame during an evolution visualization task. Since only the F-Hash encoded INR supports super-resolution, for a fair comparison, only the key frame is rendered. \cref{tab:render} shows the rendering latency across all input encoding methods. We can see that inferencing F-Hash encoded INR is faster with the help of an occupancy grid and a smaller model size. Using ARM can further reduce the rendering latency of the sample streaming algorithm for more interactive visualization.

\begin{table}[t]
  \caption{Average rendering latency.}
  \label{tab:render}
  \scriptsize%
	\centering%
  \begin{adjustbox}{width=0.48\textwidth}
      \begin{tabu}{ c | c c c c c  }
      \toprule
      \multirow{2}{*}{Dataset} & \multicolumn{5}{c}{Sample Streaming/ Sample Streaming + ARM (ms)}  \\
        &
      MHE &
      MHE Tinycudann &
      DG Single-Res &
      DG Multi-Res &
      F-Hash\\
      \midrule
      Combustion & 385.4 / 253.5  & 211.1 / 136.6  & 144.5 / 97.1 & 165.2 / 110.9  & \textbf{77.6 / 53.8} \\
      \midrule
      Argon Bubble & 898.5 / 592.4  & 531.5 / 351.9 & 256.9 / 168.8  & 293.6 / 192.4 & \textbf{63.01 / 43.4} \\
      \midrule
      Supernova & 3592.5 / 2367.5 & 1686.1 / 1112.6  & 432.8 / 284.6 & 494.8 / 326.6 & \textbf{345.3 / 228.1} \\
      \bottomrule
      \end{tabu}
 \end{adjustbox}
\end{table}




\section{Limitations and Future Work}
In this work, we propose an efficient encoding architecture that greatly enhances the convergence speed of training implicit neural networks for time-varying volumetric data. Our experiments demonstrate that F-Hash achieves the fastest convergence speed with more efficient use of encoding parameters compared with existing input encoding methods. The proposed input encoding design is easy to adapt for applications utilizing INR.
Recent INR-based research in compression, visualization, and super-resolution can benefit from our multi-resolution Tesseract encoding to improve convergence speed. Our method can be easily extended to other spatiotemporal data types, including video data, weather/climate data, and geospatial time series.
The limitations of the work include: 1) While F-Hash achieves state-of-the-art convergence speed, its training time remains relatively slow to realize online training due to the large size of the time-varying data. 2) The current compression performance of F-Hash is not as good as existing learning-based compressors or other lossy compressors. 3) The dimension of the coreset (FBB) becomes closer to the input volumetric when the feature of interest are spatially distant across frames, although this is rare in real-world scenario, leading to longer training time. Future work can explore the various aspects of the architecture. The existing fold parameter is predefined, a data-adaptive approach could optimize these parameters, thereby further reducing the required encoding parameters. We also want to further optimize the coreset selection step to construct a more compact feature region with a fast look-up from the embedding entry to the hash bucket. 

\section{Conclusion}
In this work, we present F-Hash, a novel feature-based multi-resolution Tesseract encoding achieving state-of-the-art convergence speed for training an INR for time-varying volumetric data. Our encoding method is a unified encoding solution for various time-varying features. We validate the proposed architecture through comprehensive qualitative and quantitative experiments on three large-scale time-varying volumetric datasets. The proposed adaptive ray marching algorithm improves the rendering latency of visualizing time-varying volumetric neural representation. Our input encoding method can empower a wide range of applications leveraging implicit neural representation for efficient training from complex and multi-dimensional data.



\bibliographystyle{abbrv-doi-hyperref}

\bibliography{template}

\begin{thebibliography}{10}

\bibitem{bai2020time}
Z.~Bai, Y.~Tao, and H.~Lin.
\newblock Time-varying volume visualization: A survey.
\newblock {\em Journal of Visualization}, 23:745--761, 2020.

\bibitem{10.1007/s12650-020-00654-x}
Z.~Bai, Y.~Tao, and H.~Lin.
\newblock Time-varying volume visualization: a survey.
\newblock {\em J. Vis.}, 23(5):745–761,  17 pages, Oct. 2020. \href{https://doi.org/10.1007/s12650-020-00654-x}
{doi: {{%
10\hspace{.1pt}\discretionary{.}{%
}{.}\hspace{.4pt}1007\discretionary{/}{%
}{/}s12650\discretionary{%
}{-}{-}020\discretionary{%
}{-}{-}00654\discretionary{%
}{-}{-}x}}}


\bibitem{8663447}
R.~Ballester-Ripoll, P.~Lindstrom, and R.~Pajarola.
\newblock Tthresh: Tensor compression for multidimensional visual data.
\newblock {\em IEEE Transactions on Visualization and Computer Graphics}, 26(9):2891--2903, 2020. \href{https://doi.org/10.1109/TVCG.2019.2904063}
{doi: {{%
10\hspace{.1pt}\discretionary{.}{%
}{.}\hspace{.4pt}1109\discretionary{/}{%
}{/}TVCG\hspace{.1pt}\discretionary{.}{%
}{.}\hspace{.4pt}2019\hspace{.1pt}\discretionary{.}{%
}{.}\hspace{.4pt}2904063}}}


\bibitem{Barron_2021_ICCV}
J.~T. Barron, B.~Mildenhall, M.~Tancik, P.~Hedman, R.~Martin-Brualla, and P.~P. Srinivasan.
\newblock Mip-nerf: A multiscale representation for anti-aliasing neural radiance fields.
\newblock In {\em Proceedings of the IEEE/CVF International Conference on Computer Vision (ICCV)}, pp. 5855--5864, October 2021.

\bibitem{9903564}
D.~Bauer, Q.~Wu, and K.-L. Ma.
\newblock Fovolnet: Fast volume rendering using foveated deep neural networks.
\newblock {\em IEEE Transactions on Visualization and Computer Graphics}, 29(1):515--525, 2023. \href{https://doi.org/10.1109/TVCG.2022.3209498}
{doi: {{%
10\hspace{.1pt}\discretionary{.}{%
}{.}\hspace{.4pt}1109\discretionary{/}{%
}{/}TVCG\hspace{.1pt}\discretionary{.}{%
}{.}\hspace{.4pt}2022\hspace{.1pt}\discretionary{.}{%
}{.}\hspace{.4pt}3209498}}}


\bibitem{Chan_2022_CVPR}
E.~R. Chan, C.~Z. Lin, M.~A. Chan, K.~Nagano, B.~Pan, S.~De~Mello, O.~Gallo, L.~J. Guibas, J.~Tremblay, S.~Khamis, T.~Karras, and G.~Wetzstein.
\newblock Efficient geometry-aware 3d generative adversarial networks.
\newblock In {\em Proceedings of the IEEE/CVF Conference on Computer Vision and Pattern Recognition (CVPR)}, pp. 16123--16133, June 2022.

\bibitem{https://doi.org/10.1111/cgf.14955}
S.~Devkota and S.~Pattanaik.
\newblock Efficient neural representation of volumetric data using coordinate-based networks.
\newblock {\em Computer Graphics Forum}, 42(7):e14955, 2023. \href{https://doi.org/10.1111/cgf.14955}
{doi: {{%
10\hspace{.1pt}\discretionary{.}{%
}{.}\hspace{.4pt}1111\discretionary{/}{%
}{/}cgf\hspace{.1pt}\discretionary{.}{%
}{.}\hspace{.4pt}14955}}}


\bibitem{7192664}
S.~Dutta and H.-W. Shen.
\newblock Distribution driven extraction and tracking of features for time-varying data analysis.
\newblock {\em IEEE Transactions on Visualization and Computer Graphics}, 22(1):837--846, 2016. \href{https://doi.org/10.1109/TVCG.2015.2467436}
{doi: {{%
10\hspace{.1pt}\discretionary{.}{%
}{.}\hspace{.4pt}1109\discretionary{/}{%
}{/}TVCG\hspace{.1pt}\discretionary{.}{%
}{.}\hspace{.4pt}2015\hspace{.1pt}\discretionary{.}{%
}{.}\hspace{.4pt}2467436}}}


\bibitem{Fridovich-Keil_2022_CVPR}
S.~Fridovich-Keil, A.~Yu, M.~Tancik, Q.~Chen, B.~Recht, and A.~Kanazawa.
\newblock Plenoxels: Radiance fields without neural networks.
\newblock In {\em Proceedings of the IEEE/CVF Conference on Computer Vision and Pattern Recognition (CVPR)}, pp. 5501--5510, June 2022.

\bibitem{10.1145/3478513.3480569}
S.~Hadadan, S.~Chen, and M.~Zwicker.
\newblock Neural radiosity.
\newblock {\em ACM Trans. Graph.}, 40(6),  article no. 236,  11 pages, Dec. 2021. \href{https://doi.org/10.1145/3478513.3480569}
{doi: {{%
10\hspace{.1pt}\discretionary{.}{%
}{.}\hspace{.4pt}1145\discretionary{/}{%
}{/}3478513\hspace{.1pt}\discretionary{.}{%
}{.}\hspace{.4pt}3480569}}}


\bibitem{9229162}
J.~Han and C.~Wang.
\newblock Ssr-tvd: Spatial super-resolution for time-varying data analysis and visualization.
\newblock {\em IEEE Transactions on Visualization and Computer Graphics}, 28(6):2445--2456, 2022. \href{https://doi.org/10.1109/TVCG.2020.3032123}
{doi: {{%
10\hspace{.1pt}\discretionary{.}{%
}{.}\hspace{.4pt}1109\discretionary{/}{%
}{/}TVCG\hspace{.1pt}\discretionary{.}{%
}{.}\hspace{.4pt}2020\hspace{.1pt}\discretionary{.}{%
}{.}\hspace{.4pt}3032123}}}


\bibitem{HAN2022168}
J.~Han and C.~Wang.
\newblock Tsr-vfd: Generating temporal super-resolution for unsteady vector field data.
\newblock {\em Computers \& Graphics}, 103:168--179, 2022. \href{https://doi.org/10.1016/j.cag.2022.02.001}
{doi: {{%
10\hspace{.1pt}\discretionary{.}{%
}{.}\hspace{.4pt}1016\discretionary{/}{%
}{/}j\hspace{.1pt}\discretionary{.}{%
}{.}\hspace{.4pt}cag\hspace{.1pt}\discretionary{.}{%
}{.}\hspace{.4pt}2022\hspace{.1pt}\discretionary{.}{%
}{.}\hspace{.4pt}02\hspace{.1pt}\discretionary{.}{%
}{.}\hspace{.4pt}001}}}


\bibitem{9852325}
J.~Han and C.~Wang.
\newblock Coordnet: Data generation and visualization generation for time-varying volumes via a coordinate-based neural network.
\newblock {\em IEEE Transactions on Visualization and Computer Graphics}, 29(12):4951--4963, 2023. \href{https://doi.org/10.1109/TVCG.2022.3197203}
{doi: {{%
10\hspace{.1pt}\discretionary{.}{%
}{.}\hspace{.4pt}1109\discretionary{/}{%
}{/}TVCG\hspace{.1pt}\discretionary{.}{%
}{.}\hspace{.4pt}2022\hspace{.1pt}\discretionary{.}{%
}{.}\hspace{.4pt}3197203}}}


\bibitem{10371224}
J.~Han, H.~Zheng, and C.~Bi.
\newblock Kd-inr: Time-varying volumetric data compression via knowledge distillation-based implicit neural representation.
\newblock {\em IEEE Transactions on Visualization and Computer Graphics}, 30(10):6826--6838, 2024. \href{https://doi.org/10.1109/TVCG.2023.3345373}
{doi: {{%
10\hspace{.1pt}\discretionary{.}{%
}{.}\hspace{.4pt}1109\discretionary{/}{%
}{/}TVCG\hspace{.1pt}\discretionary{.}{%
}{.}\hspace{.4pt}2023\hspace{.1pt}\discretionary{.}{%
}{.}\hspace{.4pt}3345373}}}


\bibitem{9552857}
J.~Han, H.~Zheng, D.~Z. Chen, and C.~Wang.
\newblock Stnet: An end-to-end generative framework for synthesizing spatiotemporal super-resolution volumes.
\newblock {\em IEEE Transactions on Visualization and Computer Graphics}, 28(1):270--280, 2022. \href{https://doi.org/10.1109/TVCG.2021.3114815}
{doi: {{%
10\hspace{.1pt}\discretionary{.}{%
}{.}\hspace{.4pt}1109\discretionary{/}{%
}{/}TVCG\hspace{.1pt}\discretionary{.}{%
}{.}\hspace{.4pt}2021\hspace{.1pt}\discretionary{.}{%
}{.}\hspace{.4pt}3114815}}}


\bibitem{9428530}
T.~Hospedales, A.~Antoniou, P.~Micaelli, and A.~Storkey.
\newblock Meta-learning in neural networks: A survey.
\newblock {\em IEEE Transactions on Pattern Analysis and Machine Intelligence}, 44(9):5149--5169, 2022. \href{https://doi.org/10.1109/TPAMI.2021.3079209}
{doi: {{%
10\hspace{.1pt}\discretionary{.}{%
}{.}\hspace{.4pt}1109\discretionary{/}{%
}{/}TPAMI\hspace{.1pt}\discretionary{.}{%
}{.}\hspace{.4pt}2021\hspace{.1pt}\discretionary{.}{%
}{.}\hspace{.4pt}3079209}}}


\bibitem{8601376}
C.~Huang and H.~Wang.
\newblock A novel key-frames selection framework for comprehensive video summarization.
\newblock {\em IEEE Transactions on Circuits and Systems for Video Technology}, 30(2):577--589, 2020. \href{https://doi.org/10.1109/TCSVT.2019.2890899}
{doi: {{%
10\hspace{.1pt}\discretionary{.}{%
}{.}\hspace{.4pt}1109\discretionary{/}{%
}{/}TCSVT\hspace{.1pt}\discretionary{.}{%
}{.}\hspace{.4pt}2019\hspace{.1pt}\discretionary{.}{%
}{.}\hspace{.4pt}2890899}}}


\bibitem{6185547}
J.~Kehrer and H.~Hauser.
\newblock Visualization and visual analysis of multifaceted scientific data: A survey.
\newblock {\em IEEE Transactions on Visualization and Computer Graphics}, 19(3):495--513, 2013. \href{https://doi.org/10.1109/TVCG.2012.110}
{doi: {{%
10\hspace{.1pt}\discretionary{.}{%
}{.}\hspace{.4pt}1109\discretionary{/}{%
}{/}TVCG\hspace{.1pt}\discretionary{.}{%
}{.}\hspace{.4pt}2012\hspace{.1pt}\discretionary{.}{%
}{.}\hspace{.4pt}110}}}


\bibitem{8440839}
A.~Kumpf, M.~Rautenhaus, M.~Riemer, and R.~Westermann.
\newblock Visual analysis of the temporal evolution of ensemble forecast sensitivities.
\newblock {\em IEEE Transactions on Visualization and Computer Graphics}, 25(1):98--108, 2019. \href{https://doi.org/10.1109/TVCG.2018.2864901}
{doi: {{%
10\hspace{.1pt}\discretionary{.}{%
}{.}\hspace{.4pt}1109\discretionary{/}{%
}{/}TVCG\hspace{.1pt}\discretionary{.}{%
}{.}\hspace{.4pt}2018\hspace{.1pt}\discretionary{.}{%
}{.}\hspace{.4pt}2864901}}}


\bibitem{9866018}
X.~Liang, K.~Zhao, S.~Di, S.~Li, R.~Underwood, A.~M. Gok, J.~Tian, J.~Deng, J.~C. Calhoun, D.~Tao, Z.~Chen, and F.~Cappello.
\newblock Sz3: A modular framework for composing prediction-based error-bounded lossy compressors.
\newblock {\em IEEE Transactions on Big Data}, 9(2):485--498, 2023. \href{https://doi.org/10.1109/TBDATA.2022.3201176}
{doi: {{%
10\hspace{.1pt}\discretionary{.}{%
}{.}\hspace{.4pt}1109\discretionary{/}{%
}{/}TBDATA\hspace{.1pt}\discretionary{.}{%
}{.}\hspace{.4pt}2022\hspace{.1pt}\discretionary{.}{%
}{.}\hspace{.4pt}3201176}}}


\bibitem{https://doi.org/10.1111/cgf.12934}
P.~Ljung, J.~Krüger, E.~Groller, M.~Hadwiger, C.~D. Hansen, and A.~Ynnerman.
\newblock State of the art in transfer functions for direct volume rendering.
\newblock {\em Computer Graphics Forum}, 35(3):669--691, 2016. \href{https://doi.org/10.1111/cgf.12934}
{doi: {{%
10\hspace{.1pt}\discretionary{.}{%
}{.}\hspace{.4pt}1111\discretionary{/}{%
}{/}cgf\hspace{.1pt}\discretionary{.}{%
}{.}\hspace{.4pt}12934}}}


\bibitem{lu2021compressive}
Y.~Lu, K.~Jiang, J.~A. Levine, and M.~Berger.
\newblock Compressive neural representations of volumetric scalar fields.
\newblock {\em Computer Graphics Forum}, 40(3):135--146, 2021. \href{https://doi.org/10.1111/cgf.14295}
{doi: {{%
10\hspace{.1pt}\discretionary{.}{%
}{.}\hspace{.4pt}1111\discretionary{/}{%
}{/}cgf\hspace{.1pt}\discretionary{.}{%
}{.}\hspace{.4pt}14295}}}


\bibitem{lukasczyk2017}
J.~Lukasczyk, G.~Aldrich, M.~Steptoe, G.~Favelier, C.~Gueunet, J.~Tierny, R.~Maciejewski, B.~Hamann, and H.~Leitte.
\newblock Viscous fingering: A topological visual analytic approach.
\newblock In {\em Physical Modeling for Virtual Manufacturing Systems and Processes}, vol. 869 of {\em Applied Mechanics and Materials}, pp. 9--19. Trans Tech Publications Ltd, 9 2017. \href{https://doi.org/10.4028/www.scientific.net/AMM.869.9}
{doi: {{%
10\hspace{.1pt}\discretionary{.}{%
}{.}\hspace{.4pt}4028\discretionary{/}{%
}{/}www\hspace{.1pt}\discretionary{.}{%
}{.}\hspace{.4pt}scientific\hspace{.1pt}\discretionary{.}{%
}{.}\hspace{.4pt}net\discretionary{/}{%
}{/}AMM\hspace{.1pt}\discretionary{.}{%
}{.}\hspace{.4pt}869\hspace{.1pt}\discretionary{.}{%
}{.}\hspace{.4pt}9}}}


\bibitem{1182960}
K.-L. Ma.
\newblock Visualizing time-varying volume data.
\newblock {\em Computing in Science \& Engineering}, 5(2):34--42, 2003. \href{https://doi.org/10.1109/MCISE.2003.1182960}
{doi: {{%
10\hspace{.1pt}\discretionary{.}{%
}{.}\hspace{.4pt}1109\discretionary{/}{%
}{/}MCISE\hspace{.1pt}\discretionary{.}{%
}{.}\hspace{.4pt}2003\hspace{.1pt}\discretionary{.}{%
}{.}\hspace{.4pt}1182960}}}


\bibitem{Mehta_2021_ICCV}
I.~Mehta, M.~Gharbi, C.~Barnes, E.~Shechtman, R.~Ramamoorthi, and M.~Chandraker.
\newblock Modulated periodic activations for generalizable local functional representations.
\newblock In {\em Proceedings of the IEEE/CVF International Conference on Computer Vision (ICCV)}, pp. 14214--14223, October 2021.

\bibitem{10.1145/3503250}
B.~Mildenhall, P.~P. Srinivasan, M.~Tancik, J.~T. Barron, R.~Ramamoorthi, and R.~Ng.
\newblock Nerf: representing scenes as neural radiance fields for view synthesis.
\newblock {\em Commun. ACM}, 65(1):99–106,  8 pages, Dec. 2021. \href{https://doi.org/10.1145/3503250}
{doi: {{%
10\hspace{.1pt}\discretionary{.}{%
}{.}\hspace{.4pt}1145\discretionary{/}{%
}{/}3503250}}}


\bibitem{10.1145/3528223.3530127}
T.~M\"{u}ller, A.~Evans, C.~Schied, and A.~Keller.
\newblock Instant neural graphics primitives with a multiresolution hash encoding.
\newblock {\em ACM Trans. Graph.}, 41(4),  article no. 102,  15 pages, July 2022. \href{https://doi.org/10.1145/3528223.3530127}
{doi: {{%
10\hspace{.1pt}\discretionary{.}{%
}{.}\hspace{.4pt}1145\discretionary{/}{%
}{/}3528223\hspace{.1pt}\discretionary{.}{%
}{.}\hspace{.4pt}3530127}}}


\bibitem{10.1145/3450626.3459812}
T.~M\"{u}ller, F.~Rousselle, J.~Nov\'{a}k, and A.~Keller.
\newblock Real-time neural radiance caching for path tracing.
\newblock {\em ACM Trans. Graph.}, 40(4),  article no. 36,  16 pages, July 2021. \href{https://doi.org/10.1145/3450626.3459812}
{doi: {{%
10\hspace{.1pt}\discretionary{.}{%
}{.}\hspace{.4pt}1145\discretionary{/}{%
}{/}3450626\hspace{.1pt}\discretionary{.}{%
}{.}\hspace{.4pt}3459812}}}


\bibitem{https://doi.org/10.1111/cgf.13163}
H.~Saikia and T.~Weinkauf.
\newblock Global feature tracking and similarity estimation in time-dependent scalar fields.
\newblock {\em Computer Graphics Forum}, 36(3):1--11, 2017. \href{https://doi.org/10.1111/cgf.13163}
{doi: {{%
10\hspace{.1pt}\discretionary{.}{%
}{.}\hspace{.4pt}1111\discretionary{/}{%
}{/}cgf\hspace{.1pt}\discretionary{.}{%
}{.}\hspace{.4pt}13163}}}


\bibitem{NEURIPS2020_53c04118}
V.~Sitzmann, J.~Martel, A.~Bergman, D.~Lindell, and G.~Wetzstein.
\newblock Implicit neural representations with periodic activation functions.
\newblock In H.~Larochelle, M.~Ranzato, R.~Hadsell, M.~Balcan, and H.~Lin, eds., {\em Advances in Neural Information Processing Systems}, vol.~33, pp. 7462--7473. Curran Associates, Inc., 2020.

\bibitem{NEURIPS2019_b5dc4e5d}
V.~Sitzmann, M.~Zollhoefer, and G.~Wetzstein.
\newblock Scene representation networks: Continuous 3d-structure-aware neural scene representations.
\newblock In H.~Wallach, H.~Larochelle, A.~Beygelzimer, F.~d\textquotesingle Alch\'{e}-Buc, E.~Fox, and R.~Garnett, eds., {\em Advances in Neural Information Processing Systems}, vol.~32. Curran Associates, Inc., 2019.

\bibitem{Sun_2022_CVPR}
C.~Sun, M.~Sun, and H.-T. Chen.
\newblock Direct voxel grid optimization: Super-fast convergence for radiance fields reconstruction.
\newblock In {\em Proceedings of the IEEE/CVF Conference on Computer Vision and Pattern Recognition (CVPR)}, pp. 5459--5469, June 2022.

\bibitem{sun2025make}
J.~Sun, D.~Lenz, H.~Yu, and T.~Peterka.
\newblock Make the fastest faster: Importance mask for interactive volume visualization using reconstruction neural networks.
\newblock {\em arXiv preprint arXiv:2502.06053}, 2025.

\bibitem{10549835}
J.~Sun, X.~Xie, and H.~Yu.
\newblock Rmdncache: Dual-space prefetching neural network for large-scale volume visualization.
\newblock {\em IEEE Transactions on Visualization and Computer Graphics}, pp. 1--13, 2024. \href{https://doi.org/10.1109/TVCG.2024.3410091}
{doi: {{%
10\hspace{.1pt}\discretionary{.}{%
}{.}\hspace{.4pt}1109\discretionary{/}{%
}{/}TVCG\hspace{.1pt}\discretionary{.}{%
}{.}\hspace{.4pt}2024\hspace{.1pt}\discretionary{.}{%
}{.}\hspace{.4pt}3410091}}}


\bibitem{NEURIPS2020_55053683}
M.~Tancik, P.~Srinivasan, B.~Mildenhall, S.~Fridovich-Keil, N.~Raghavan, U.~Singhal, R.~Ramamoorthi, J.~Barron, and R.~Ng.
\newblock Fourier features let networks learn high frequency functions in low dimensional domains.
\newblock In H.~Larochelle, M.~Ranzato, R.~Hadsell, M.~Balcan, and H.~Lin, eds., {\em Advances in Neural Information Processing Systems}, vol.~33, pp. 7537--7547. Curran Associates, Inc., 2020.

\bibitem{tang2020deep}
D.~Tang, S.~Singh, P.~A. Chou, C.~Hane, M.~Dou, S.~Fanello, J.~Taylor, P.~Davidson, O.~G. Guleryuz, Y.~Zhang, et~al.
\newblock Deep implicit volume compression.
\newblock In {\em Proceedings of the IEEE/CVF conference on computer vision and pattern recognition}, pp. 1293--1303, 2020.

\bibitem{10.1145/3571735}
H.~Tang, L.~Ding, S.~Wu, B.~Ren, N.~Sebe, and P.~Rota.
\newblock Deep unsupervised key frame extraction for efficient video classification.
\newblock {\em ACM Trans. Multimedia Comput. Commun. Appl.}, 19(3),  article no. 119,  17 pages, Feb. 2023. \href{https://doi.org/10.1145/3571735}
{doi: {{%
10\hspace{.1pt}\discretionary{.}{%
}{.}\hspace{.4pt}1145\discretionary{/}{%
}{/}3571735}}}


\bibitem{tang2023ecnr}
K.~Tang and C.~Wang.
\newblock Ecnr: Efficient compressive neural representation of time-varying volumetric datasets.
\newblock {\em arXiv preprint arXiv:2311.12831}, 2023.

\bibitem{TANG2024103874}
K.~Tang and C.~Wang.
\newblock Stsr-inr: Spatiotemporal super-resolution for multivariate time-varying volumetric data via implicit neural representation.
\newblock {\em Computers \& Graphics}, 119:103874, 2024. \href{https://doi.org/10.1016/j.cag.2024.01.001}
{doi: {{%
10\hspace{.1pt}\discretionary{.}{%
}{.}\hspace{.4pt}1016\discretionary{/}{%
}{/}j\hspace{.1pt}\discretionary{.}{%
}{.}\hspace{.4pt}cag\hspace{.1pt}\discretionary{.}{%
}{.}\hspace{.4pt}2024\hspace{.1pt}\discretionary{.}{%
}{.}\hspace{.4pt}01\hspace{.1pt}\discretionary{.}{%
}{.}\hspace{.4pt}001}}}


\bibitem{teschner2003optimized}
M.~Teschner, B.~Heidelberger, M.~M{\"u}ller, D.~Pomerantes, and M.~H. Gross.
\newblock Optimized spatial hashing for collision detection of deformable objects.
\newblock In {\em Vmv}, vol.~3, pp. 47--54, 2003.

\bibitem{8727480}
K.-C. Wang, T.-H. Wei, N.~Shareef, and H.-W. Shen.
\newblock Ray-based exploration of large time-varying volume data using per-ray proxy distributions.
\newblock {\em IEEE Transactions on Visualization and Computer Graphics}, 26(11):3299--3313, 2020. \href{https://doi.org/10.1109/TVCG.2019.2920130}
{doi: {{%
10\hspace{.1pt}\discretionary{.}{%
}{.}\hspace{.4pt}1109\discretionary{/}{%
}{/}TVCG\hspace{.1pt}\discretionary{.}{%
}{.}\hspace{.4pt}2019\hspace{.1pt}\discretionary{.}{%
}{.}\hspace{.4pt}2920130}}}


\bibitem{weiss2022fast}
S.~Weiss, P.~Hermüller, and R.~Westermann.
\newblock Fast neural representations for direct volume rendering.
\newblock {\em Computer Graphics Forum}, 41(6):196--211, 2022. \href{https://doi.org/10.1111/cgf.14578}
{doi: {{%
10\hspace{.1pt}\discretionary{.}{%
}{.}\hspace{.4pt}1111\discretionary{/}{%
}{/}cgf\hspace{.1pt}\discretionary{.}{%
}{.}\hspace{.4pt}14578}}}


\bibitem{7348066}
W.~Widanagamaachchi, J.~Chen, P.~Klacansky, V.~Pascucci, H.~Kolla, A.~Bhagatwala, and P.-T. Bremer.
\newblock Tracking features in embedded surfaces: Understanding extinction in turbulent combustion.
\newblock In {\em 2015 IEEE 5th Symposium on Large Data Analysis and Visualization (LDAV)}, pp. 9--16, 2015. \href{https://doi.org/10.1109/LDAV.2015.7348066}
{doi: {{%
10\hspace{.1pt}\discretionary{.}{%
}{.}\hspace{.4pt}1109\discretionary{/}{%
}{/}LDAV\hspace{.1pt}\discretionary{.}{%
}{.}\hspace{.4pt}2015\hspace{.1pt}\discretionary{.}{%
}{.}\hspace{.4pt}7348066}}}


\bibitem{8031584}
W.~Widanagamaachchi, A.~Jacques, B.~Wang, E.~Crosman, P.-T. Bremer, V.~Pascucci, and J.~Horel.
\newblock Exploring the evolution of pressure-perturbations to understand atmospheric phenomena.
\newblock In {\em 2017 IEEE Pacific Visualization Symposium (PacificVis)}, pp. 101--110, 2017. \href{https://doi.org/10.1109/PACIFICVIS.2017.8031584}
{doi: {{%
10\hspace{.1pt}\discretionary{.}{%
}{.}\hspace{.4pt}1109\discretionary{/}{%
}{/}PACIFICVIS\hspace{.1pt}\discretionary{.}{%
}{.}\hspace{.4pt}2017\hspace{.1pt}\discretionary{.}{%
}{.}\hspace{.4pt}8031584}}}


\bibitem{10175377}
Q.~Wu, D.~Bauer, M.~J. Doyle, and K.-L. Ma.
\newblock Interactive volume visualization via multi-resolution hash encoding based neural representation.
\newblock {\em IEEE Transactions on Visualization and Computer Graphics}, 30(8):5404--5418, 2024. \href{https://doi.org/10.1109/TVCG.2023.3293121}
{doi: {{%
10\hspace{.1pt}\discretionary{.}{%
}{.}\hspace{.4pt}1109\discretionary{/}{%
}{/}TVCG\hspace{.1pt}\discretionary{.}{%
}{.}\hspace{.4pt}2023\hspace{.1pt}\discretionary{.}{%
}{.}\hspace{.4pt}3293121}}}


\bibitem{yang2025meta}
M.~Yang, K.~Tang, and C.~Wang.
\newblock Meta-inr: Efficient encoding of volumetric data via meta-learning implicit neural representation.
\newblock {\em arXiv preprint arXiv:2502.09669}, 2025.

\bibitem{yao2025visnerf}
S.~Yao, Y.~Lu, and C.~Wang.
\newblock Visnerf: Efficient multidimensional neural radiance field representation for visualization synthesis of dynamic volumetric scenes.
\newblock {\em arXiv preprint arXiv:2502.16731}, 2025.

\bibitem{NEURIPS2021_25e2a30f}
L.~Yariv, J.~Gu, Y.~Kasten, and Y.~Lipman.
\newblock Volume rendering of neural implicit surfaces.
\newblock In M.~Ranzato, A.~Beygelzimer, Y.~Dauphin, P.~Liang, and J.~W. Vaughan, eds., {\em Advances in Neural Information Processing Systems}, vol.~34. Curran Associates, Inc., 2021.

\bibitem{9203859}
Q.~Zhong, Y.~Zhang, J.~Zhang, K.~Shi, Y.~Yu, and C.~Liu.
\newblock Key frame extraction algorithm of motion video based on priori.
\newblock {\em IEEE Access}, 8:174424--174436, 2020. \href{https://doi.org/10.1109/ACCESS.2020.3025774}
{doi: {{%
10\hspace{.1pt}\discretionary{.}{%
}{.}\hspace{.4pt}1109\discretionary{/}{%
}{/}ACCESS\hspace{.1pt}\discretionary{.}{%
}{.}\hspace{.4pt}2020\hspace{.1pt}\discretionary{.}{%
}{.}\hspace{.4pt}3025774}}}


\bibitem{https://doi.org/10.1111/cgf.13399}
B.~Zhou and Y.-J. Chiang.
\newblock Key time steps selection for large-scale time-varying volume datasets using an information-theoretic storyboard.
\newblock {\em Computer Graphics Forum}, 37(3):37--49, 2018. \href{https://doi.org/10.1111/cgf.13399}
{doi: {{%
10\hspace{.1pt}\discretionary{.}{%
}{.}\hspace{.4pt}1111\discretionary{/}{%
}{/}cgf\hspace{.1pt}\discretionary{.}{%
}{.}\hspace{.4pt}13399}}}


\end{thebibliography}

\end{document}